\documentclass[12pt]{article}
\pdfoutput=1
\usepackage{float}
\usepackage{amsfonts}
\usepackage{amsmath}
\usepackage{bm}
\usepackage{url}
\usepackage{hyperref}
\usepackage{bbold}
\usepackage{slashed}
\usepackage{tikz}
\usepackage{graphicx}
\usepackage{epstopdf}
\usepackage{subfigure}
\usepackage{pgfplots}

\usepackage{amssymb}
\usepackage{color}
\usepackage{mathrsfs}
\usepackage{fancybox}
\usepackage{lipsum}
\usepackage{eurosym}
\usepackage{tcolorbox}
\usepackage{physics}
\usepackage{tensor}

\numberwithin{equation}{section}

\newcommand{\slr}{\text{SL}^+(2,\mathbb{R})}
\newcommand{\sltr}{\text{SL}(2,\mathbb{R})}
\newcommand{\slc}{\text{SL}(2,\mathbb{C})}
\newcommand{\dpi}{\mathcal{D}}
\newcommand{\mj}{\mathcal{J}}

\newcommand{\mo}{\mathcal{O}}

\newcommand{\cas}{\mathcal{C}}
\newcommand{\atbdy}{\rvert_{\text{bdy}}}
\newcommand{\gphi}{g^{-1}\partial_\phi g}
\newcommand{\gt}{g^{-1}\partial_t g}
\newcommand{\tj}[6]{ \begin{pmatrix}
   #1 & #2 & #3 \\
   #4 & #5 & #6 
  \end{pmatrix}}
  \newcommand{\sj}[6]{ \begin{Bmatrix}
   #1 & #2 & #3 \\
   #4 & #5 & #6 
  \end{Bmatrix}}
	\newcommand{\LL}{\scriptscriptstyle \text{L}}
	\newcommand{\RR}{\scriptscriptstyle \text{R}}
	
\title{Fine Structure of Jackiw-Teitelboim Quantum Gravity}
\begin{document}
\begin{titlepage}

\setcounter{page}{1} \baselineskip=15.5pt \thispagestyle{empty}

\vfil

${}$
\vspace{1cm}

\begin{center}
\def\thefootnote{\fnsymbol{footnote}}
\begin{changemargin}{0.05cm}{0.05cm} 
\begin{center}
{\Large \bf Fine Structure of Jackiw-Teitelboim Quantum Gravity}
\end{center} 
\end{changemargin}

~\\[1cm]
{Andreas Blommaert\footnote{\href{mailto:andreas.blommaert@ugent.be}{\protect\path{andreas.blommaert@ugent.be}}}, Thomas G. Mertens\footnote{\href{mailto:thomas.mertens@ugent.be}{\protect\path{thomas.mertens@ugent.be}}} and Henri Verschelde\footnote{\href{mailto:henri.verschelde@ugent.be}{\protect\path{henri.verschelde@ugent.be}}}}
\\[0.3cm]
\vspace{0.7cm}
{\normalsize { \sl Department of Physics and Astronomy,
\\[1.0mm]
Ghent University, Krijgslaan, 281-S9, 9000 Gent, Belgium}} \\
\vspace{0.5cm}

\end{center}


 \vspace{0.2cm}
\begin{changemargin}{01cm}{1cm} 
{\small  \noindent 
\begin{center} 
\textbf{Abstract}
\end{center} }
We investigate structural aspects of JT gravity through its BF description. In particular, we provide evidence that JT gravity should be thought of as (a coset of) the noncompact subsemigroup $\slr$ BF theory. We highlight physical implications, including the famous Plancherel measure $\sinh 2\pi\sqrt{E}$. Exploiting this perspective, we investigate JT gravity on more generic manifolds with emphasis on the edge degrees of freedom on entangling surfaces and factorization. It is found that the one-sided JT gravity degrees of freedom are described not just by a Schwarzian on the asymptotic boundary, but also include frozen $\slr$ degrees of freedom on the horizon, identifiable as JT gravity black hole states. Configurations with two asymptotic boundaries are linked to 2d Liouville CFT on the torus surface.
\end{changemargin}
 \vspace{0.3cm}
\vfil
\begin{flushleft}
\today
\end{flushleft}

\end{titlepage}

\newpage
\tableofcontents
\vspace{0.5cm}
\noindent\makebox[\linewidth]{\rule{\textwidth}{0.4pt}}
\vspace{1cm}

\addtolength{\abovedisplayskip}{.5mm}
\addtolength{\belowdisplayskip}{.5mm}

\def\plus{\raisebox{.5pt}{\tiny$+$\smpc}}

\addtolength{\parskip}{.6mm}
\def\spc{\hspace{1pt}}

\def\nspc{{\hspace{-2pt}}}
\def\ff{\rm\smpc f\smpc} 
\def\fff{\mbox{Y}}
\def\ww{{\rm w}}
\def\smpc{{\hspace{.5pt}}}

\def\zz{{\spc \rm z}}
\def\xx{{\rm x\smpc}}
\def\xxi{\mbox{\footnotesize \spc $\xi$}}
\def\jj{{\rm j}}
 \addtolength{\baselineskip}{-.1mm}

\renewcommand{\Large}{\large}

\setcounter{tocdepth}{2}
\addtolength{\baselineskip}{0mm}
\addtolength{\parskip}{.4mm}
\addtolength{\abovedisplayskip}{1mm}
\addtolength{\belowdisplayskip}{1mm}

\setcounter{footnote}{0}


\section{Introduction}
	
When considering models of two-dimensional gravity, the Jackiw-Teitelboim (JT) theory plays a privileged role \cite{Jackiw:1984je,Teitelboim:1983ux}:
\begin{equation}
\label{JTaction}
S[g,\Phi] = \frac{1}{16\pi G_2}\int d^2x \sqrt{-g}\, \Phi \left(R^{(2)}-\Lambda\right) + S_{\text{GH}}.
\end{equation}
It consists of a 2d metric $g_{\mu\nu}$, whose only physical degree of freedom is the Ricci scalar $R$, and a dilaton field $\Phi$. This model is the spherical dimensional reduction of pure 3d gravity with cosmological constant $\Lambda$ and as such, it is the closest one can get in two dimensions to a dynamical pure quantum gravity theory.\footnote{Recall that the Einstein-Hilbert action $S_{EH} \sim \int R$ is the Euler characteristic in 2d.} It also appears as the universal low-energy gravitational sector in SYK-type models \cite{KitaevTalks,Sachdev:1992fk,Polchinski:2016xgd,Jevicki:2016bwu,Maldacena:2016hyu,Jevicki:2016ito,randommatrix,Turiaci:2017zwd,Gross:2017hcz,Gross:2017aos,Das:2017pif,Das:2017wae,Berkooz:2018qkz,Berkooz:2018jqr}. \\
Being the spherical sector of 3d gravity, the JT model \eqref{JTaction} does not have any bulk propagating degrees of freedom, but it does have black hole solutions.\footnote{Propagating degrees of freedom can be introduced by coupling the system to an external matter sector as studied in e.g. \cite{Almheiri:2014cka,paper5,Mertens:2019bvy}. This will not be pursued here.} Furthermore, the JT action describes the dynamics of the near-horizon regime of near-extremal black holes. In that context, the zeroth order term $S_{\text{top}} \sim \Phi_0 \int d^2x \sqrt{-g}\,R +S_{\text{GH}} \sim \chi$ captures the ground state entropy $S_0$, whereas the remainder, at first order \eqref{JTaction}, captures the deviations from extremality. As such, pure JT \eqref{JTaction} only captures the deviations from extremality.\footnote{This has implications in that when we compute the black hole entropy, we will not capture the ground state entropy $S_0$, much like in \cite{lin}.} \\
When considering JT gravity \eqref{JTaction} on a manifold with a boundary, one finds that the dynamics is governed by Schwarzian quantum mechanics \cite{Jensen:2016pah,Maldacena:2016upp,Engelsoy:2016xyb}:\footnote{This is a primitive form of holography, of the same type as the Chern-Simons / WZW correspondence.}
\begin{equation}
\label{SSch}
S[f] = -C\int dt \, \left\{f,t\right\},
\end{equation}
with $\left\{f,t\right\} = \frac{f'''}{f'} - \frac{3}{2}\frac{f''^2}{f'^2}$, the Schwarzian derivative of $f$, the boundary time reparametrization. Schwarzian amplitudes can be explicitly computed and indeed exhibit virtual intermediate virtual black hole states \cite{schwarzian}, see also \cite{wittenstanford,Mandal:2017thl,Goel:2018ubv,Kitaev:2018wpr,Yang:2018gdb,Belokurov:2018aol}. We set $C=1/2$ from here on out.
\\~\\
Ever since the early work in the model \cite{Fukuyama:1985gg,Isler:1989hq,Chamseddine:1989yz,Jackiw:1992bw}, the JT action \eqref{JTaction} has been known to be identical to the action of a $\sltr$ BF theory, which in turn is the dimensional reduction of 3d $\sltr$ CS theory.\footnote{It should be noted that this identification was done at the classical level and locally, and that it does not guarantee the quantum equivalence of JT gravity and $\sltr$ BF. In particular the range of fields in the path integral depends explicitly on the group and not just on the algebra. In our case, we have to at least identify $g\in \sltr$ with $-g$ and the structure is reduced to P$\sltr$ $\simeq \sltr/\mathbb{Z}_2 \simeq $ SO(2,1). This modification will be left implicit here, and is relatively harmless. More impactful modifications are discussed shortly.} Away from the holographic boundary, this $ \sltr$ BF-model is describing the moduli space of flat $\sltr$ connections. The equivalence between JT gravity and its BF formulation is manifest in the first-order formulation. However, it is not immediate that the second-order and first-order formulation of gravity are equivalent quantum-mechanically in terms of path integration space. We can raise several important points with its relation to gravity. 
\begin{itemize}
\item A first important point is that metric invertibility is typically not imposed in the first-order (i.e. BF) formulation. It was shown in \cite{Verlinde:1989ua} to be related to picking the hyperbolic component of the moduli space of flat $\sltr$ connections. We will have more to say about this further on, and this is one of our motivations for restricting to the $\slr$ subsemigroup.
\item Next to this, there are two inequivalent choices of integration space over geometries that correspond to integrating over Teichm\"uller space $\mathcal{T}$ (the moduli space of flat hyperbolic $\sltr$ connections) or the moduli space of Riemann surfaces $\mathcal{M}$ (Teichm\"uller space modulo the mapping class group). Though equivalent on the disk, for higher genus surfaces we get different results using either $\mathcal{T}$ respectively $\mathcal{M}$. This is detailed in Appendices \ref{app:C} and  \ref{app:multi}.
\item Quantum gravity can be considered to include a summation over different bulk topologies, respecting the asymptotic structure. In this work, we choose to define the model by restricting to a predefined topology, mostly the disk and annulus topology.
\end{itemize}
Throughout this work, we \emph{choose} the path integration space for the bulk to correspond to the hyperbolic component of the moduli space of flat $\sltr$ connections, or Teichm\"uller space $\mathcal{T}$, of fixed topology.
\\~\\
In \cite{paper3} we made the claim that JT quantum gravity is in fact a $\slr$ BF theory, and not a $\sltr$ BF theory, with $\slr$ the subsemigroup of $\sltr$ obtained be restricting $\sltr$ matrices to matrices with all elements positive. In the first part of this work (section \ref{sect:subsemi}), we substantiate this claim. \\
BF theory for compact groups is understood rather well \cite{2dgt1,2dgt2}. JT gravity is different from this in a number of ways: the relevant group is noncompact, it is in fact not a group but a subsemigroup, and finally gravitational boundary conditions constrain the group theoretic degrees on the boundary resulting in a coset construction. We will deal with each of these issues one by one throughout sections \ref{sect:coset} and \ref{sect:subsemi}, gradually working our way up to JT gravity. This completes the precise BF formulation of JT gravity initiated in \cite{origins, paper3}. 
\\~\\
The remainder of this work is devoted to the study of JT gravity on more generic manifolds. The main focus is on JT gravity on a strip (Lorentzian) or equivalently an annulus (Euclidean), as this configuration is relevant for black hole physics. This is discussed in section \ref{sect:JTedge}. More general Euclidean topologies are discussed in Appendix \ref{app:multi}.
\\
In particular in section \ref{sect:JTedge} we explain how cutting manifolds assigns edge dynamics or JT edge modes to entangling surfaces, in the spirit of \cite{paper2}. The boundary surface can be made transparent, or equivalently the manifolds can be glued together by taking the trace in the extended Hilbert space associated with the edge degrees of freedom (see e.g. \cite{Buividovich:2008gq,Donnelly:2011hn,donnellywall,donnellyfreidel,wong,Fliss:2017wop,paper1,paper2,gluing1} and references therein). 
\\
As a byproduct we establish that the spectrum of JT gravity contains one-sided black hole states; unlike the Schwarzian theory which on its own is insufficient to describe the Hilbert space of one-sided JT black holes.\footnote{Although all correlation functions reduce to Schwarzian thermal calculations.} These states account for the Bekenstein-Hawking entropy in JT gravity, in the sense of the calculation in \cite{lin}.
\\~\\
Including edge modes then allows JT to factorize across horizons in the sense \eqref{tfdjt}, which is the sense in which generic gauge theories such as Maxwell factorize. Indeed, within a BF formulation of JT gravity, the factorizing structure \eqref{tfdjt} of the Hilbert space follows from basic group-theoretic properties. We highlight this structure in BF at the very beginning of this work in section \ref{sect:coset},\footnote{When this paper was nearing completion, a work of Donnelly and Wong \cite{Donnelly:2018ppr} appeared containing similar statements regarding the TFD in (quasi)-topological gauge theories.} and come back to this for JT gravity in section \ref{sect:JTedge}. This addresses one aspect of the factorization problem posed in \cite{harlowfactor,lin}. 
\\
It does not resolve all the subtleties though, as e.g. the JT spectrum is continuous without a volume-scaling divergence. This raises issues regarding a Hilbert space interpretation of such quantum systems \cite{wittenstanford,harlowfactor}, which are intrinsic to JT. To address this and other aspects of the factorization problem of \cite{harlowfactor}, one would have to consider a specific UV-ancestor of JT, like SYK, and find a discretized set of microstates. Whenever we use the word factorization throughout this work, we mean no more or no less than the property \eqref{tfdjt}.
\\
In any case, the pure states $\ket{k,s,\mathfrak{i}}$ in \eqref{tfdjt} play an important role in JT gravity and are worth studying.
\\~\\
As a warm-up for the JT edge mode story of section \ref{sect:JTedge} we consider compact group BF in section \ref{sect:BFedge}. Furthermore, we repeat the edge mode story for CS in Appendix \ref{app:CS} and compare the BF formulas of section \ref{sect:BFedge} with known formulas of 2d CFT. 
\\
Finally, in section \ref{s:twobouncor} we compute JT bilocal wormhole-crossing amplitudes and elaborate on an identification of these as a specific limit of Liouville torus amplitudes. 
\\~\\
A natural class of operator insertions in JT and BF are boundary-anchored Wilson lines. Generic correlation functions with Wilson lines inserted, possibly crossed, can be written down using a diagrammatic construction.\footnote{Bluntly, each Wilson line endpoint on the boundary circle gets a $3j$-symbol, each bulk Wilson line crossing gets a $6j$-symbol. The detailed rules are summarized in Appendix \ref{s:pgdiag} and their derivation can be found in \cite{origins, paper3}.} Though the emphasis in this work is not on such correlation functions, at several instances we will write down some amplitudes, with the goal of showing that the BF perspective on JT allows us to understand dynamics of JT quantum gravity on generic manifolds.

\section{Holography for Quantum Mechanics on Groups and Cosets}
\label{sect:coset}
We start this section with a review on how quantum mechanics on the group manifold $G$ appears when studying 2d BF theory on a disk \cite{origins,paper3}, with compact gauge group $G$. Later we generalize the boundary conditions to incorporate coset models for a subgroup $H \subset G$. Finally we discuss how to generalize to noncompact groups.

\subsection{Review: Compact Groups}
\label{sect:bf}
Consider BF theory on a disk with boundary labeled by $t$:
\begin{equation}
S[\chi,A]=\int_{\mathcal{M}} d^2x \Tr\left[\chi F\right]- \frac{1}{2} \int_{\partial \mathcal{M}} dt \Tr\left[\chi A_t\right].\label{BFaction}
\end{equation}
Variation of the action results in
\begin{equation}
\delta S[\chi,A] = (\text{bulk e.o.m.}) + \frac{1}{2} \int_{\partial \mathcal{M}}  dt\Tr\left[\chi \delta A_t - A_t\delta \chi\right],
\end{equation}
the boundary term can be dealt with by imposing:
\begin{equation}
\label{pogbdy}
A_t\atbdy=\chi\atbdy.
\end{equation}
Path integrating over $\chi$ forces $A =g^{-1}d g$, with $g$ periodic $g(t+\beta)=g(t)$ and we are left with the untwisted particle on a group action:
\begin{equation}
\boxed{S[g] = -\frac{1}{2} \int d t \Tr(\gt)^2,}\label{pog}
\end{equation}
studied e.g. in \cite{pg2,pg1}.\footnote{There is actually a redundancy for $g \sim V g$ for constant $V \in G$. This translates to a path integration space of $LG/G$ for the partition function. This modding by $G$ gives an additional factor of $1/\text{vol }G$ in the partition function (which we did not write) that strictly speaking foils a genuine Hilbert space interpretation of this path integral. We will interpret this factor as a contribution to the zero-temperature entropy as $e^{S_0}$ and dismiss it from here on out. See also appendix C of \cite{origins}. There will be an analogous subtlety for the non-compact JT case.} This theory will henceforth be refered to as quantum mechanics on the group manifold. More generally we can include a puncture in irrep $\lambda$ in the disk. Path integrating out $\chi$ now imposes a non-trivial holonomy on $A$: $A = g^{-1}d g + \lambda$. The result is the action:
\begin{equation}
S[g,\lambda]=-\int d t \Tr(\gt +\lambda)^2,\label{twistedpogaction}
\end{equation}
with partition function \cite{picken}:
\begin{equation}
Z(\beta,U_\lambda) = \sum_{R} \dim R \, \chi_R(U_\lambda) e^{-\beta \cas_R}, \qquad U_\lambda = e^{-2\pi \lambda},\label{twistedpog}
\end{equation}
in terms of the weight $\lambda \equiv \bm{\lambda} \cdot \mathbf{H}$, with $\mathbf{H}$ the Cartan generators. The Peter-Weyl theorem implies the Hilbert space of both BF on an interval and that of quantum mechanics on the group manifold consists of all matrix elements of all irreducible representations $R$ of $G$:
\begin{equation}
\mathcal{H} = \big\{\ket{R,a,b}, \quad a,b = 1\hdots \dim R\big\},
\end{equation}
with normalized coordinate space wavefunctions:
\begin{equation}
\label{basispw}
\left\langle g\right| \left.R,a,b\right\rangle = \sqrt{\text{dim }R}\, R_{ab}(g) = \sqrt{\text{dim }R}\left\langle R, a\right| g \left|R, b\right\rangle.
\end{equation}
One way of formulating this conclusion, is that a quantum particle on the group manifold can be written in terms of an emergent 2d spacetime. In this sense, this is a form of holography on the worldline (see also \cite{Janik:2018dll}), albeit one without propagating bulk degrees of freedom, in perfect analogy with the situation for 2d WZW CFTs.

\subsection{Factorization of the Thermofield Double}\label{s:factor}
\begin{figure}[h]
\centering
\includegraphics[width=0.8\textwidth]{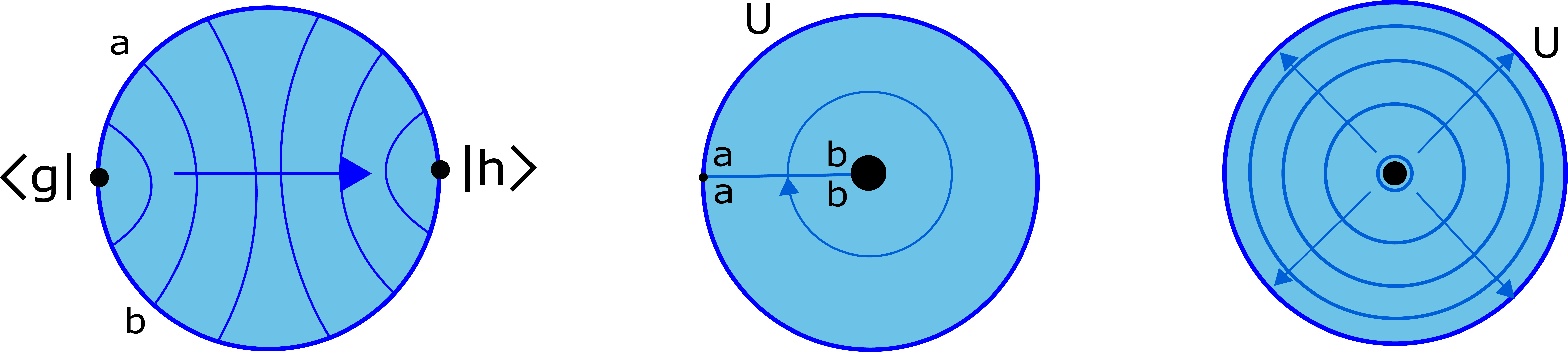}
\caption{Left: defect channel slicing of the amplitude where $gh^{-1} = U$. Middle: Angular slicing of the amplitude. Right: Circular slicing connecting the inner boundary and the outer.}
\label{IntroChannels}
\end{figure}

In \cite{paper3} we introduced several useful families of time-slicings of the BF disk. Next to the defect channel slicing (Figure \ref{IntroChannels} left), in this paper we introduce two more slicings that turn out to be very useful. These are an angular slicing of the disk, and a circular slicing (Figure \ref{IntroChannels} middle and right). The angular slicing is analogous to Schwarzschild time slicing in Euclidean signature. As we will be mostly interested in the Lorentzian continuation in this time coordinate, we will adhere to this slicing throughout most of this work.
The disk partition function can be computed in either of these channels:
\begin{align}
\label{pf}
Z(\beta,U)  &=  \sum_R \text{dim }R \, R_{ba}(h^{-1}) R_{ab}(g) e^{-\cas_R \beta} \nonumber \\
&= \sum_R \text{dim }R \, \int_G dg \, R_{ba}(g) R_{ab}(Ug^{-1}) e^{-\cas_R \beta} \nonumber \\
&= \sum_R \chi_R(\mathbf{1}) \, \chi_R(U) e^{-\cas_R \beta}.
\end{align}
The thermofield double (TFD) state is a semi-disk amplitude and can accordingly be calculated using either of these slicings. The defect channel slicing is most reminiscent of the definition of the TFD state as preparing the vacuum:
\begin{align}
&\begin{tikzpicture}[scale=1, baseline={([yshift=0cm]current bounding box.center)}]
\draw[thick,blue] (1.5,0.75) arc (360:180:1.5);
\draw[thick,blue] (-1.5,0.75) -- (1.5,0.75);
\draw[fill,black] (-1.5,0.75) circle (0.1);
\draw[fill,black] (1.5,0.75) circle (0.1);
\draw (1.2,0.95) node {\small \color{red}$a$};
\draw (-1.2,0.95) node {\small \color{red}$a$};
\draw (0,.25) node {\small \color{blue}$R$};
\draw (0,1) node {\small $g$};
\draw (0,-1) node {\small $\mathbf{1}$};
\end{tikzpicture} = \left\langle g\right|\left.\text{TFD}\right\rangle
\end{align}
The disk calculation results using \eqref{basispw} in:
\begin{equation}
\label{diska}
\bra{g}\ket{\text{TFD}} = \sum_{R,a,b} \bra{g}\ket{R,a,b}\bra{R,a,b}\ket{\mathbf{1}}e^{-\frac{\beta}{2}\cas_R} = \sum_{R} \text{dim R} \, \chi_R(g) \, e^{-\frac{\beta}{2}\cas_R},
\end{equation}
or
\begin{equation}
\ket{\text{TFD}} = \sum_{R,a} \sqrt{\text{dim R}}\, e^{-\frac{\beta}{2}\cas_R} \ket{R,a,a}.
\end{equation}
Consider now the wavefunction $\bra{g_1\cdot g_2}\ket{R,a,a}$ in combination with the defining property of representation matrices $R_{aa}(g_1\cdot g_2) = R_{ab}(g_1)R_{ba}(g_2)$. We find the factorization of the wavefunction:
\begin{equation}
\bra{g_1\cdot g_2}\ket{R,a,a}=\sum_b \frac{1}{\sqrt{\dim R}} \bra{g_1}\ket{R,a,b}\,\bra{g_2}\ket{R,b,a}.\label{statefactor}
\end{equation}
Using this, we can equivalently write the thermofield double state as:
\begin{align}
&\begin{tikzpicture}[scale=1, baseline={([yshift=0cm]current bounding box.center)}]
\draw[thick,blue] (1.5,0.75) arc (360:180:1.5);
\draw[thick,blue] (-1.5,0.75) -- (1.5,0.75);
\draw[fill,black] (-1.5,0.75) circle (0.1);
\draw[fill,black] (0,0.75) circle (0.1);
\draw[fill,black] (1.5,0.75) circle (0.1);
\draw (1.2,0.95) node {\small \color{red}$a$};
\draw (-1.2,0.95) node {\small \color{red}$a$};
\draw (0.2,0.95) node {\small \color{red}$b$};
\draw (-0.2,0.95) node {\small \color{red}$b$};
\draw (0,.25) node {\small \color{blue}$R$};
\draw (-0.75,1) node {\small $g_L$};
\draw (0.75,1) node {\small $g_R$};
\draw (0,-1) node {\small $\mathbf{1}$};
\end{tikzpicture} = \bra{g_L\otimes g_R}\ket{\text{TFD}}
\end{align}
Using \eqref{statefactor} we now find:
\begin{equation}
\boxed{
\ket{\text{TFD}}= \sum_{R,a,b}\, e^{-\frac{\beta}{2}\cas_R} \ket{R,a,b} \otimes \ket{R,a,b}}.\label{purification?}
\end{equation}
This corresponds to a state defined on the $t=0$ slice with a predefined bifurcation in two pieces $\mathcal{H}_L \otimes \mathcal{H}_R$. This formula is very suggestive and shows the purification of a thermal ensemble of states $\ket{R,a,b}$ associated with the submanifold obtained by cutting a two-sided geometry on the horizon. We will make this picture explicit in section \ref{sect:BFedge}, where we identify the states $\ket{R,b}$ as the edge states associated with the horizon.

\subsection{Cosets $G/H$}\label{s:cosets}
The boundary condition \eqref{pogbdy} can be generalized into\footnote{One can generalize this further by including sign changes as $A^a\atbdy=\pm\chi^a\atbdy$. These sign changes correspond to changing the signature of the bilinear form on the algebra $\mathfrak{g}$ at the boundary; this boils down to switching between different real forms of the complex algebra. The magnitude of the proportionality factor can be absorbed by a field redefinition.}
\begin{equation}
A^a_t \atbdy=\chi^a\atbdy, \qquad A^b_t \atbdy = \chi^b \atbdy = 0\label{bcchoice}
\end{equation}
for some subset of generators labeled $b$. This leads to a restricted particle on a group action:
\begin{equation}
S[g] = -\frac{1}{2} \int d t \left.\Tr(\gt)^2\right|_{\text{restricted}},
\end{equation}
We will focus on the case where the generators $\tau^b$ span a subalgebra $\mathfrak{h} \subset \mathfrak{g}$. The resulting theory then describes a particle on the right coset $G/H$. The extreme case of $H=G$ sets all boundary values of $\chi=0$ and removes all boundary dynamics: as a result the theory $G/G$ only contains topological data such as knots contained in the BF bulk. 
\\~\\
The Peter-Weyl theorem for groups $G$ is readily extended to right cosets $G/H$. Functions on the coset $G/H$ are restricted by right $H$-invariance: $\psi(g) = \psi(g\cdot H)$. In terms of the matrix element basis functions \eqref{basispw}, this leads to the constrained basis:
\begin{equation}
R_{a0}(g)= \left\langle R, a\right|g\left|R, 0\right\rangle = \left\langle R, a\right|g \cdot H \left|R, 0\right\rangle
\end{equation}
with right-states constrained by invariance under $H$ denoted by a label $0$: $H \left|R,0 \right\rangle = \left|R,0 \right\rangle$. For homogeneous spaces (to which we restrict from now on), there is only one such basis vector $\left|R,0 \right\rangle$ within each irrep $R$. Thus the Hilbert space is spanned by the orthonormal basis of so-called spherical functions:
\begin{equation}
\phi_{0a}^R(g) = \sqrt{\text{dim }R} R_{0a}(g),
\end{equation}
We can now directly write down the propagator on the coset manifold from $g=\mathbf{1}$ to $g=U$:
\begin{equation}
\label{twistpcoset}
Z_{G/H}(\beta,U) = \sum_R \phi_{0a}^R(\mathbf{1})\phi_{a0}^R(U) e^{-\cas_R \beta} = \sum_R \text{dim }R \, R_{00}(U) e^{-\cas_R \beta},
\end{equation}
\begin{figure}[h]
\centering
\includegraphics[width=0.8\textwidth]{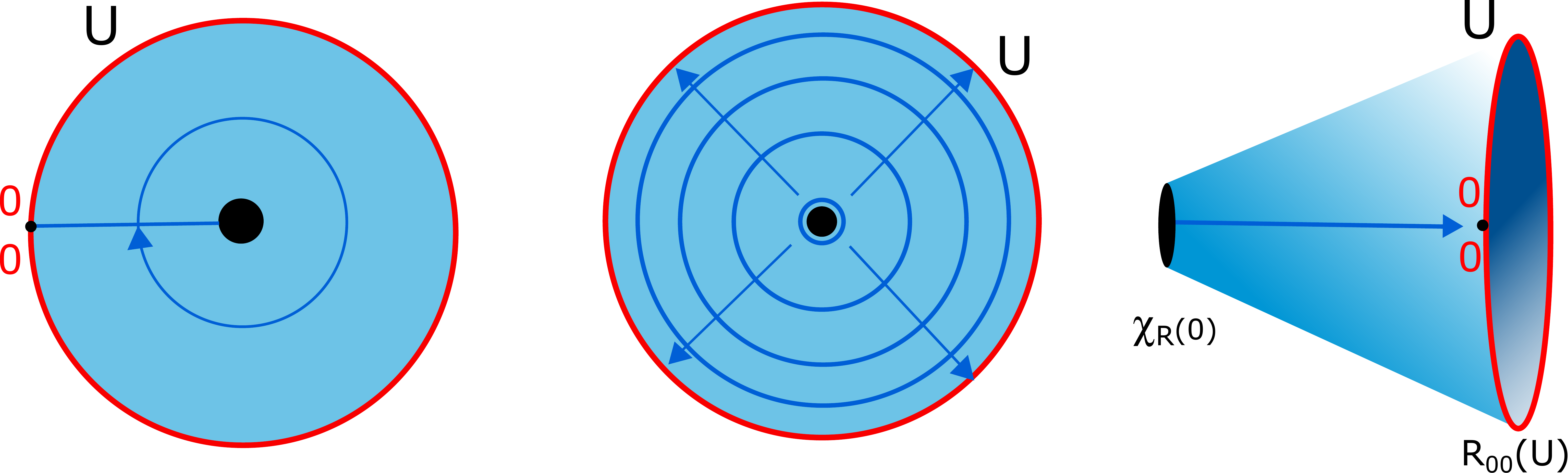}
\caption{Left: Angular slicing of the amplitude. Middle and Right: Circular slicing and annular region connecting the inner boundary and the outer, the latter projecting on the invariant indices. From hereon coset boundaries will always we depicted in red.}
\label{MoreChannels}
\end{figure}
As highlighted in Appendix \ref{app:slicing}, the angular slicing (Figure \ref{MoreChannels} left) in BF theory is manifestly equal to the boundary particle-on-a-coset evaluation. The second way of writing the amplitude in \eqref{twistpcoset} on the other hand is interpreted as closed channel propagation between initial and final states (Figure \ref{MoreChannels} middle and right). The matrix element 
\begin{equation}
R_{00}(g) \equiv \left\langle R, 0\right|g\left|R, 0\right\rangle
\end{equation}
is both left- and right- $H$-invariant and is called a zonal spherical function.
\\~\\
As shown in Appendix \ref{app:slicing}, regions in the bulk diagrams enclosed by Wilson lines are weighed by $\dim R$ reminiscent of inserting a complete set of wavefunctions of the parent $G$ theory. The deep interior does not know about the modding by $H$ and is insensitive to the choice of boundary conditions \eqref{bcchoice}. 
\\
Indeed: interior points come with free labels $a$, whereas boundary labels are constrained to be $0$. Accordingly, the $6j$-symbols that appear at the bulk crossing of Wilson lines are those of the parent group $G$. For JT gravity there is a similar scenario \cite{paper3}: the gravitational constraints are genuine boundary conditions and do not affect the theory in the deep bulk, as we discuss in section \ref{s:gravcos}.
\\~\\
As an illuminating example of a quantum particle on a coset manifold, take the two sphere $S^2 \simeq SU(2)/U(1)$. In this case, the full matrix element is the Wigner D-function, the spherical functions are the standard spherical harmonics, and the zonal spherical function is the Legendre function. We provided details and some more discussion in Appendix \ref{app:sphere}.
\\~\\
We end by remarking that cosets are quite numerous in the space of all manifolds, and the fact that we can directly generalize our conclusion from section \ref{sect:coset} to this case, is hence a vast expansion of the number of available models of this kind.

\subsection{Noncompact Groups}
\label{sect:noncompact}
Consider next quantum mechanics on a noncompact group manifold. The Peter-Weyl theorem (or equivalently the Plancherel decomposition) states how square integrable functions on the group manifold can be decomposed into representation matrix elements:
\begin{equation}
f(g)=\sum_{k,a,b}  c_{k,ab}\, R^k_{ab}(g),\quad \forall f\in L^2 (G).
\end{equation}
The difference with compact groups is that now continuous irrep labels $k$ will appear, as well as infinite-dimensional representations. The irrep matrix elements are orthogonal with respect to the Plancherel measure:
\begin{equation}
\int d g \, R^k_{a b}(g) R^{k'}_{c d}(g)^* = \frac{\delta(k-k')}{\rho(k)}\delta_{ac}\delta_{bd}.
\end{equation}
We read off the normalized eigenfunctions:
\begin{equation}
\phi^k_{a b}(g)=\sqrt{\rho(k)} R^k_{a b}(g).\label{eigenstates}
\end{equation}
The propagator on the group manifold is written down using these ingredients as:
\begin{equation}
\label{noncp}
Z_G(\beta,\lambda)=\int dk\, \phi^k_{a b}(g)\phi^k_{a b}(g\cdot U_\lambda)^*\,e^{-\beta\cas_k} = \int d k\, \rho(k) \, \chi_k (U_\lambda)\,e^{-\beta\cas_k}.
\end{equation}
In BF language, this is the amplitude for a disk-shaped region, so each such region is weighted with the Plancherel measure $\rho(k)$. For several irreps, including the unitary irreps of relevance in the Peter-Weyl decomposition, the representation space is infinite-dimensional. Its dimension is found as the character evaluated at the identity element. We will prove further on that this is also equal to the Plancherel measure:\footnote{For reader comfort, we have left several volume factors implicit, hence there is no contradiction between \eqref{charone} and the infinite dimensionality of the representation. We more carefully track these factors in Appendix \ref{app:reg} by relating finite-volume regularization to delta-regularization. It is the latter in which the Plancherel measure $\rho(k)$ is defined.}
\begin{equation}
 \chi_k(\mathbf{1}) \equiv \dim k=\rho(k),\label{charone}
\end{equation}
but for this we must first discuss non-compact cosets.
\\~\\
The propagator on coset manifolds $G/H$ with both $G$ and $H$ noncompact is well-understood and described in detail in \cite{coset}. It is the generalization of \eqref{twistpcoset}:
\begin{equation}
Z_{G/H}(\beta)=\int d k \rho_{G}(k)\,e^{-\beta \cas_k},\label{zfong}
\end{equation}
where $\rho_G(k)$ is the Plancherel measure on $G$ and where we used $R^k _{0 0}(\mathbf{1})=1$. Let's consider some instructive examples of this formula.

\begin{itemize}
\item $G=SL(2,\mathbb{C})$ and $H=SU(2)$. The resulting space is the Euclidean hyperbolic space $H_3^+$. The propagator on Euclidean AdS$_3$ is well-known \cite{David:2009xg}:
\begin{equation}
Z_{H_3^+}(\beta)=\int ds \, s^2 \,e^{-\beta \cas_s},
\end{equation}
where one indeed recognizes the $SL(2,\mathbb{C})$ Plancherel measure $\rho(s) = s^2$.

\item $G=\sltr$ and $H=U(1)$. The resulting space is the Euclidean hyperbolic plane $H_2^+$. The propagator on Euclidean AdS$_2$ is again well-known:
\begin{equation}
Z_{H_2^+}(\beta)=\int ds \, s\tanh(\pi s) \,e^{-\beta \cas_s},
\end{equation}
and we recover the $\sltr$ Plancherel measure $\rho(s) \sim s \tanh(\pi s)$.\footnote{Discrete representations of $\sltr$ are absent since discrete eigenmodes of the $H_2^+$ Laplacian do not exist.}

\item $G = G\times G$ and $H = G_{diag}$. This is the coset realization of the group $G$ itself.\footnote{The argument that there is only one state $\ket{k,0}$ for each irrep holds for this particular coset, see the discussion around (B.51) and (B.52) in \cite{coset}.}
For a direct product of groups $G=G_1\cross G_2$, the Plancherel measure is $\rho_G(\mu_1,\mu_2)=\rho_{G_1}(\mu_1)\cdot \rho_{G_2}(\mu_2)$, so:
\begin{equation}
\rho_{G\cross G}(\mu)=\rho_G(\mu)^2.
\end{equation}
Hence for the diagonal coset which is just the group, the partition function can be rewritten as:
\begin{equation}
Z_G(\beta)=\int dk\, \rho(k)^2\, e^{-\beta \cas_k}. \label{toproof}
\end{equation}
Comparing this equation with \eqref{noncp}, completes the proof of \eqref{charone}.
\end{itemize}
As a further example, in Appendix \ref{app:slc} we consider quantum mechanics on $\slc$.

\section{The Subsemigroup Structure of JT Gravity}
\label{sect:subsemi}
In this section, we build up towards describing JT gravity as a $\slr$ BF theory. The structure $\slr$ is a subsemigroup, consisting of $\sltr$ matrices with all positive entries:
\begin{equation}
\left(\begin{array}{cc}
a & b \\
c & d \\
\end{array}\right), \qquad ad-bc = 1, \quad a,b,c,d > 0.
\end{equation}
In sections \ref{sect:argue}, \ref{sect:argue3} and \ref{sect:argue2}, we gather evidence that this structure can indeed be identified with 2d JT gravity. \\
In section \ref{s:slrchar} we show that one can consistently describe quantum mechanics on the subsemigroup $\slr$. In section \ref{s:gravcos}, we work out the coset perspective on the JT disk amplitudes. 
In order to appreciate the difference between $\sltr$ and $\slr$, we present a short recap of the relevant representation theory in Appendices \ref{s:rep} and \ref{s:repsemi}.

\subsection{Evidence 1: Density of States and the Plancherel Measure}
\label{sect:argue}
\begin{figure}[h]
\centering
\includegraphics[width=0.55\textwidth]{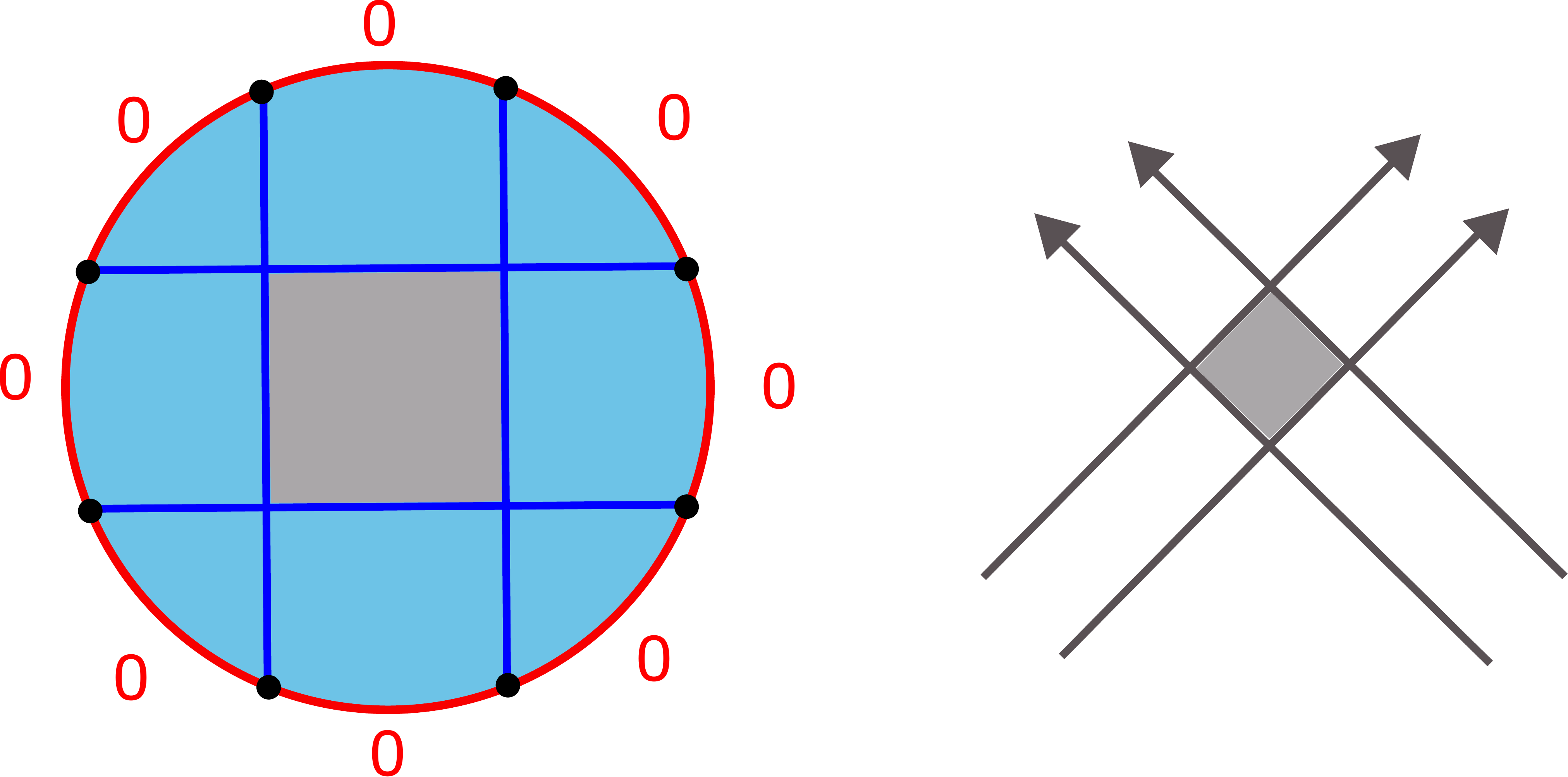}
\caption{Left: Configuration of crossing Wilson lines that enclose a bulk region. The coset projection (denoted by the $0$-symbol and the red-colored boundary) is not felt in the deep interior, but it is crucial to use the $\sinh 2\pi \sqrt{E}$ measure to agree with a semi-classical shockwave computation of the same topology (Right).}
\label{shockenclosed}
\end{figure}

Let us first present an argument in favor of the $\slr$ structure. For $\slr$ the Plancherel measure is $\sinh 2\pi \sqrt{E}$ \eqref{plmeasup} whereas for $\sltr$ the Plancherel measure is $\tanh \pi \sqrt{E}$ \eqref{plmeasu}. The former has a Cardy rise at large energies, consistent with the semi-classical Bekenstein-Hawking entropy formula, the latter doesn't. So an $\sltr$ BF theory will not result in a correct calculation of the black hole entropy \cite{lin}, as there are simply not enough states.\footnote{We will elaborate on the black hole states further on.}
\\~\\
Let us briefly touch on a second physical application for which it is pivotal that we describe JT gravity as a BF theory with Plancherel measure $\sinh 2\pi \sqrt{E}$, attributing this weight to each disk-shaped region. Recently, the semi-classical limit of the exact JT correlation functions was investigated in \cite{shocks}. Analyzing generic diagrams with crossing bilocal lines, the eikonal shockwave expressions were reproduced \cite{ss1,ss2}, where the corresponding shockwave diagram in real time is topologically the same as the crossing lines disk diagram. When performing such a calculation, it is crucial that each region in the Euclidean bulk carries a measure factor $\sinh 2\pi \sqrt{E}$, as these factors ultimately determine the saddle point that represents the mass of the original black hole on which the shockwaves propagate. \\
This is even more crucial for regions that are completely sealed off from the holographic boundary (Figure \ref{shockenclosed}), as no coset conditions are imposed at all for such region and the theory is sensitive to the full BF theory.

\subsection{Evidence 2: Hyperbolic Geometry}
\label{sect:argue3}
\begin{figure}[h]
\centering
\includegraphics[width=0.2\textwidth]{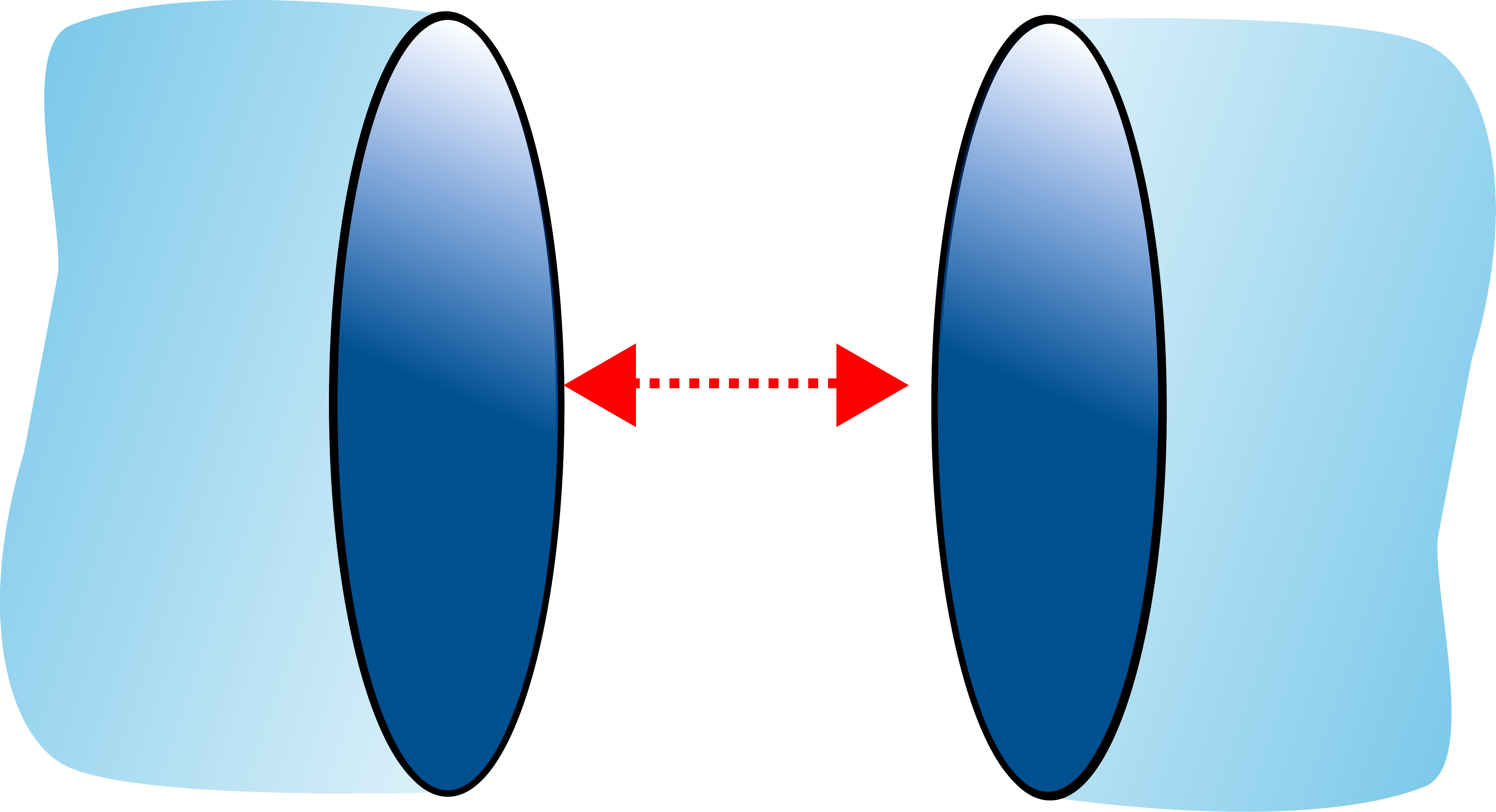}
\caption{Quantizing BF theory on a circle gives a complete basis by Peter-Weyl as the set of all characters of all unitary irreps. Gluing proceeds by using this basis.}
\label{HilbertCircle}
\end{figure}

\noindent A further argument in favor of $\slr$ can be made by thinking about more complicated geometries. In particular, when quantizing a BF-theory on a circle instead of an interval, the Hilbert space is spanned by the set of all class-functions on $G$, i.e. the characters of all unitary irreps. For a non-compact group with continuous irrep labels $k$ and $k'$, these satisfy the completeness relation:\footnote{In principle, an integral over twist angles is present in this equation. This is harmless for compact groups, see Appendix \ref{app:glueBF} for details. For the gravity case, the range of the twisting integral depends on the choice of Teichm\"uller space or the moduli space of Riemann surfaces, see Appendix \ref{app:glueJT}. The specific range is not important for the point we are making here.}
\begin{equation}
\label{gluu}
\int_{C} d\alpha \, \chi^k(\alpha)\chi^{k'}(\alpha^{-1}) = \delta(k-k'),
\end{equation}
to be used when gluing two tubes together (Figure \ref{HilbertCircle}). Such a relation holds equally well for a subsemigroup as $\slr$.

\noindent The integral \eqref{gluu} ranges over the subgroup of all conjugacy class elements $C$ of the group $G$. For a compact group, this is the maximal torus mod Weyl $T/W$. For a non-compact group, the situation is not so simple. In the case of $\sltr$, the set of conjugacy class elements splits in elliptic $\left|\text{Tr}g\right| < 2$, parabolic $\left|\text{Tr}g\right| = 2$ and hyperbolic classes $\left|\text{Tr}g\right| > 2$, and one has to sum over the three classes as well. Restricting further to $\slr$ where all elements are positive numbers, the constraint $ad-bc=1$ combined with positivity rules out the elliptic and parabolic class. Indeed, since $b,c > 0$ we must have $ad > 1$ and hence $\text{Tr}g = a+d > a + a^{-1} > 2$, as was to be shown. The parabolic class represented by the identity element is located with measure zero at the bottom of the hyperbolic class. We hence have $C = \left\{\text{hyperbolic class elements}\right\}$. \\
Moreover, it is known how the different conjugacy classes work geometrically in JT gravity. Elliptic class states correspond to conical defects, whereas hyperbolic class states correspond to smooth tubes \cite{Mertens:2019tcm}.\footnote{The parabolic class generates a cusp infinitely far away and can be viewed as a degenerate case.} If one is interested in smooth 2d geometries, in particular with a non-singular (invertible) metric, then one has to restrict to the hyperbolic class. Indeed, the component of the moduli space of flat $\sltr$ connections that is related to gravity, is the so-called hyperbolic component where all tubes have hyperbolic holonomies. The above argument illustrates that restricting to $\slr$ makes immediate contact with non-singular gravity, and in particular gives a path integration space ranging only over non-singular metrics. \\
In fact, even though we lack a true proof, we believe that the hyperbolic component of the moduli space of flat $\sltr$ connections is to be identified with the moduli space of flat $\slr$ connections. We provide some arguments for this in Appendix \ref{app:moduli}.

\subsection{Evidence 3: Limits of 3d Gravity and Quantum Groups}
\label{sect:argue2}
Next, we elaborate on a deeper structural reason for the group-theoretic $\slr$ structure of JT gravity. \\
Jackiw-Teitelboim gravity is unambiguously defined as a suitable dimensional reduction of 3d gravity. The dynamics of 3d gravity is governed in essence by the Virasoro modular bootstrap, which in turn is governed by the representation theory of the quantum group $\text{SL}_q^+(2,\mathbb{R})$. This was discussed in detail by Ponsot and Teschner in \cite{Ponsot:1999uf,Ponsot:2000mt}. By taking suitable limits of their formulas we end up uniquely with the representation theory of $\slr$.
\\~\\
In discussing the harmonic analysis on the quantum group $\text{SL}_q^+(2,\mathbb{R})$ in the context of the Virasoro modular bootstrap, Ponsot and Teschner write down the following Plancherel decomposition:\footnote{In fact this Plancherel decomposition was announced without proof by Ponsot and Teschner, and proven only later in the mathematics literature \cite{Ip}.}
\begin{equation}
    L^2 (\text{SL}_q^+(2,\mathbb{R})) \simeq \int_\oplus d\mu(P) \, \mathcal{P}_P \otimes \mathcal{P}_P\label{planqaunt},
\end{equation}
where $\mathcal{P}_P$ are the self-dual representations of $U_q(\mathfrak{sl}(2,\mathbb{R}))$, $d\mu(P)$ is the Plancherel measure on SL$_q^+(2,\mathbb{R})$ and $P>0$. Explicitly, the measure reads:
\begin{align}
d\mu(P) = d P \, \abs{S_b(2\alpha)}^2, \qquad \alpha = Q/2 + iP,
\end{align}
with 
\begin{equation}
\left|S_b(2\alpha)\right|^2 = S_0^{\, P} = 4 \sinh(2\pi Pb)\sinh(2\pi P/b),
\end{equation}
where we recognize the Virasoro modular $S$ matrix element $S_0^{\, P}$. In the classical limit $b\to 0$, with $P=bk$, this becomes the Sklyanin measure:
\begin{equation}
S_0^{\, P} \, \to \, k \sinh 2\pi k ,\label{sklyanin}
\end{equation}
which is just the Plancherel measure on $\slr$. The objects appearing on the r.h.s. in \eqref{planqaunt} are viewed more naturally as representations of the modular double $U_q(\mathfrak{sl}(2,\mathbb{R})) \otimes U_{\tilde{q}}(\mathfrak{sl}(2,\mathbb{R}))$. The classical limit $q\to 1$ of these representations does not yield a double copy of the classical group $\sltr$, instead the representations are \emph{self-dual}, and form a basis of functions on $\slr$ \cite{Ip}. Hence the classical limit of \eqref{planqaunt} is just the Plancherel decomposition of $\slr$:
\begin{equation}
\label{plusplanch}
\boxed{L^2 (\text{SL}^+(2,\mathbb{R})) \simeq \int_{0}^{\infty}dk \, k \sinh 2\pi k  \,  \mathcal{P}_k \otimes  \mathcal{P}_k}.
\end{equation}
Note that no discrete representations are present. The Plancherel decomposition \eqref{plusplanch} is to be read as the statement that the matrix elements $K^{++}_{s_1 s_2}(g)$ \eqref{matgen} of $\slr$ are complete in $\slr$ in the sense that:
\begin{align}
f(g) = \sum_{k,s_1,s_2}c_{k,s_1s_2} K^{++}_{s_1 s_2}(g), \qquad \forall f \in L^2 (\text{SL}^+(2,\mathbb{R})),
\end{align}
for uniquely determined expansion coefficients $c_{k,s_1s_2}$, with the associated orthonormality
\begin{equation}
\int d g \, K^{++}_{s_1 s_2}(g) K^{++}_{s_3 s_4}(g)^* = \frac{\delta(k-k')\delta(s_1-s_3)\delta(s_2-s_4)}{k \sinh 2\pi k },
\end{equation}
and completeness relation:
\begin{equation}
\int ds_1ds_2 \int dk \, k \sinh 2\pi k \, K^{++}_{s_1 s_2}(g_2) K^{++}_{s_1 s_2}(g_2)^* = \delta(g_1-g_2).
\end{equation}
As a consistency check on the limiting procedure from \eqref{planqaunt} to \eqref{plusplanch}, recall from Appendix \ref{s:repsemi} the $\slr$ gravitational wavefunction:
\begin{equation}
R^k(\phi) = e^\phi K_{2ik}(e^\phi),\label{whittaker}
\end{equation}
which is the mixed parabolic matrix element of the Cartan element $\phi$. In the mathematics literature, this is the so-called Whittaker function (or coefficient) \cite{Jacquet,Schiffmann,Hashizume1,Hashizume2}. The JT result \eqref{whittaker} matches with the classical limit $b\to0$ of the Whittaker function of $U_q(\mathfrak{sl}(2,\mathbb{R})) \otimes U_{\tilde{q}}(\mathfrak{sl}(2,\mathbb{R}))$ \cite{Kharchev:2001rs}.
\\~\\
When considering out-of-time ordered correlation functions in JT gravity, $6j$-symbols of $\slr$ pop up \cite{paper3}. Alternatively, these $6j$ symbols are obtained as the classical limit $b\to 0$ of the braiding matrices of Virasoro conformal blocks. The fusion matrices of Virasoro are given as $6j$-symbols of the quantum group SL$_q^+(2,\mathbb{R})$.\footnote{A very nice discussion on this can be found in \cite{verlindejacksonlauren}.} As a consistency check, the orthogonality relation of the quantum $6j$ symbols \cite{verlindejacksonlauren}:
\begin{equation}
\int d\mu(P) \sj{K_1}{L_1}{P}{K_2}{L_2}{Q}_q\sj{K_1}{L_1}{P}{K_2}{L_2}{R}_q=\frac{1}{S_0^P}\delta(Q-R),
\end{equation}
is taken in the $b\to 0$ limit to \eqref{sklyanin}:
\begin{equation}
\int d p \, p\sinh 2\pi p \sj{k_1}{l_1}{p}{k_2}{l_2}{q}\sj{k_1}{l_1}{p}{k_2}{l_2}{r}=\frac{\delta(q-r)}{q \sinh 2\pi q}.\label{6jortho}
\end{equation}
Within JT gravity, gravitational Wilson lines can be uncrossed in the bulk at no cost. This can be proven directly in the path integral before initiating an explicit calculation \cite{paper3}. The above formula which includes the $6j$-symbols that appear at bulk crossings of Wilson lines in JT, expresses precisely this operation, \emph{given that} we work with a BF theory whose Plancherel decomposition is precisely \eqref{plusplanch}. So on top of identifying the $6j$-symbols as those of $\slr$, \eqref{6jortho} also proves that the Plancherel decomposition of the BF theory associated to JT gravity is precisely \eqref{plusplanch}.
\\~\\
A related point is that in \cite{schwarzian,shocks} Schwarzian OTO correlators were obtained by applying the braiding R-matrix in 2d Virasoro CFT for each line crossing. The double-scaling Schwarzian limit then demonstrated that each such procedure generates an additional momentum integral, with the $k_i \sinh(2\pi k_i)$ measure accompanying it. This includes regions that end up being completely enclosed in the interior of the bulk.

\subsection{Quantum Mechanics on $\slr$}
\label{s:slrchar}
Motivated by the previous subsections, we will now prove that the particle on the subsemigroup $\slr$ or equivalently $\slr$ BF on a disk is a mathematically consistent model. The contents of this section build on some $\slr$ representation theory summarized in Appendix \ref{s:repsemi}. The consistency hinges on the fact that the $\slr$ manifold is a submanifold of the $\sltr$ manifold.
\subsubsection*{Partition Function}
The particle on $\slr$ is defined by the path integral:
\begin{equation}
\int_{\slr} \left[\dpi \phi\right] \left[\dpi \gamma_+\right] \left[\dpi \gamma_-\right] \exp{-\int_0^\beta d t \left((\partial_t \phi)^2 + e^{-2\phi}\partial_t \gamma_+ \partial_t\gamma_-\right)},
\end{equation}
on the thermal manifold $g(t+\beta)=g(t)$ and constrained to the $\slr$ patch $\gamma_-,\gamma_+>0$ \eqref{Gauss}. Within a Hamiltonian context, we obtain the propagator (or twisted partition function) on the $\slr$ manifold:
\begin{equation}
Z(\beta,g,U_\lambda g)=\int d k \int d\alpha d\beta\, \rho(k)\ e^{-\beta\cas_k}\, K^{++}_{\alpha\beta}(U_\lambda g)K^{++}_{\alpha\beta}(g)^*.\label{semigroupprop}
\end{equation}
Here, $\alpha$ and $\beta$ label the hyperbolic basis of $\slr$. Because we are considering propagation on the $\slr$ submanifold, obviously $g$ and $U_\lambda$ are restricted to be positive. The matrix elements of $\slr$ are a subset of the hyperbolic basis matrix elements of $\sltr$:
\begin{equation}
\mathbf{K}(g) = \left(\begin{array}{cc}
K^{++}(g) & K^{+-}(g) \\
K^{-+}(g) & K^{--}(g) \\
\end{array}\right),
\end{equation}
with composition property $\mathbf{K}(g_1 \cdot g_2) = \mathbf{K}(g_1) \cdot \mathbf{K}(g_2)$ and inverse $\mathbf{K}(g^{-1}) = \mathbf{K}(g)^{-1}$. Using the explicit expressions for the matrix elements \cite{vilenkin,VK}, one readily finds
\begin{equation}
K^{+-}_{\alpha\beta}(g)=0, \quad g\in\slr \label{slrvanish}
\end{equation}
Similarly, the matrix representation can be shown to be unitary:\footnote{This is explicitly demonstrated in Appendix \ref{app:uni}.}
\begin{equation}
K^{++}_{\beta\alpha}(g)^*=K^{++}_{\alpha\beta}(g)^{-1}\label{orthogonal}
\end{equation}
For $g$ positive, the property \eqref{slrvanish} can be used to show that group composition of $\sltr$ implies
\begin{equation}
K^{++}_{\alpha\beta}(g)^{-1}=K^{++}_{\alpha\beta}(g^{-1}), \quad g \in \slr,
\end{equation}
and hence:\footnote{Notice here that it is crucial that $\slr$ is not just a semigroup, but a \emph{sub}semigroup of $\sltr$. In particular the embedding of $\sltr$ allows us to give meaning to $g^{-1}$ for $g$ positive. Elements of $\slr$ do have an inverse, but it lies outside of $\slr$.}
\begin{equation}
K^{++}_{\alpha\beta}(g)^*=K^{++}_{\beta\alpha}(g^{-1}), \quad g \in \slr \label{crucial}
\end{equation}
Using \eqref{slrvanish}, one furthermore proves that the following property holds:
\begin{equation}
K^{++}(h)\cdot K^{++}(g)=K^{++}(h\cdot g), \quad h \in \slr, g \in \sltr \label{crucial2}
\end{equation}
for \emph{any} $g\in\sltr$. Putting the pieces together we get
\begin{equation}
\Tr K^{++}(U_\lambda g)\cdot  K^{++}(g)^* =\Tr K^{++}(U_\lambda) = \chi_{k}^+(U_\lambda).
\end{equation}
Hence the propagator on the $\slr$ manifold becomes:
\begin{equation}
\label{slprpf}
Z^+ (\beta,\lambda)=\int d k\,\rho(k)\, \chi_{k}^+(U_\lambda) \,e^{-\beta\cas_k}.
\end{equation}
Notice that we recover the fact that the $\slr$ manifold is homogeneous, simply because the $\sltr$ manifold is.
\\~\\
Let's now give an explicit expression of the characters, exploiting its embedding within $\sltr$. 
The $\sltr$ character $\chi_{k}(U_\lambda) = \Tr K^{++}(U_\lambda)  + \Tr K^{--}(U_\lambda) $
Using formulas (9) and (10) on p358 of \cite{vilenkin} one finds $K^{--}(\lambda) = K^{++}(\lambda)$, and hence $\chi_{k}(U_\lambda) =2 \chi_{k}^{+}(U_\lambda) $.\footnote{In fact we can use \cite{vilenkin} to prove a more generic property. The general character of $\sltr$ can be rewritten as:
\begin{equation}
\chi_\mu(g) = \chi^+_\mu(g)+\chi^+_\mu(e\cdot g\cdot e), \quad g\in \slr, \label{chslrsltrgen}
\end{equation}
with $e=diag(-1,1)$. The action of $e$ on wavefunctions $f_\mu(x)$ defined on the positive axis is: $e\cdot f_\mu(x)=f_\mu(-x)$, effectively mapping $\mathbb{R}^+$ to $\mathbb{R}^-$ and $K^{++}$ to $K^{--}$. Explicitly for the relevant wavefunctions we obtain $\bra{-x}\ket{s}=e^{\pi s}\bra{x}\ket{s}$ and $\bra{s}\ket{-x}=e^{-\pi s}\bra{s}\ket{x}$ where we used $-1 = e^{i\pi}$ since we cannot go through the branch cut. Writing the character as $\chi^+_\mu(e\cdot g\cdot e)=\int ds\, \bra{s}e\cdot g\cdot e\ket{s}$, inserting a completeness relation in the $x$-basis and using the above properties one finds that 
\begin{equation}
\chi^+_\mu(e\cdot g\cdot e)=\chi^+_\mu(g).\label{charslplus}
\end{equation}
Using this in \eqref{chslrsltrgen} and again dropping an irrelevant factor $2$ we obtain
\begin{equation}
\chi_\mu(g)=\chi_\mu^+(g),
\end{equation}
for all positive $g$.} The net factor $2$ is irrelevant and the appropriate finite characters for $\sltr$ are\footnote{See Appendix \ref{ss:ambiguities}.}
\begin{equation}
\chi_{k}^+(U_\lambda) = \cos 2\pi k\lambda.
\end{equation}
Equation \eqref{slprpf} can then be written more explicitly as:
\begin{equation}
\boxed{Z^+ (\beta,\lambda)=\int_{0}^{+\infty} dk\,k \sinh 2\pi k \, \cos 2\pi k \lambda \,e^{-\beta k^2}.}
\end{equation}
The vacuum character on the other hand is the Plancherel measure by \eqref{charone}:
\begin{equation}
\chi_{k}^+(\mathbf{1}) = \rho(k) = k \sinh 2\pi k .
\end{equation}
So the partition function of a particle on $\slr$ is:
\begin{equation}
Z^+ (\beta) = \int_{0}^{+\infty} dk\,\left(k \sinh 2\pi k \right)^2\, e^{-\beta k^2}.
\end{equation}

\subsubsection*{Correlation Functions}
\begin{figure}[h]
\centering
\includegraphics[width=0.7\textwidth]{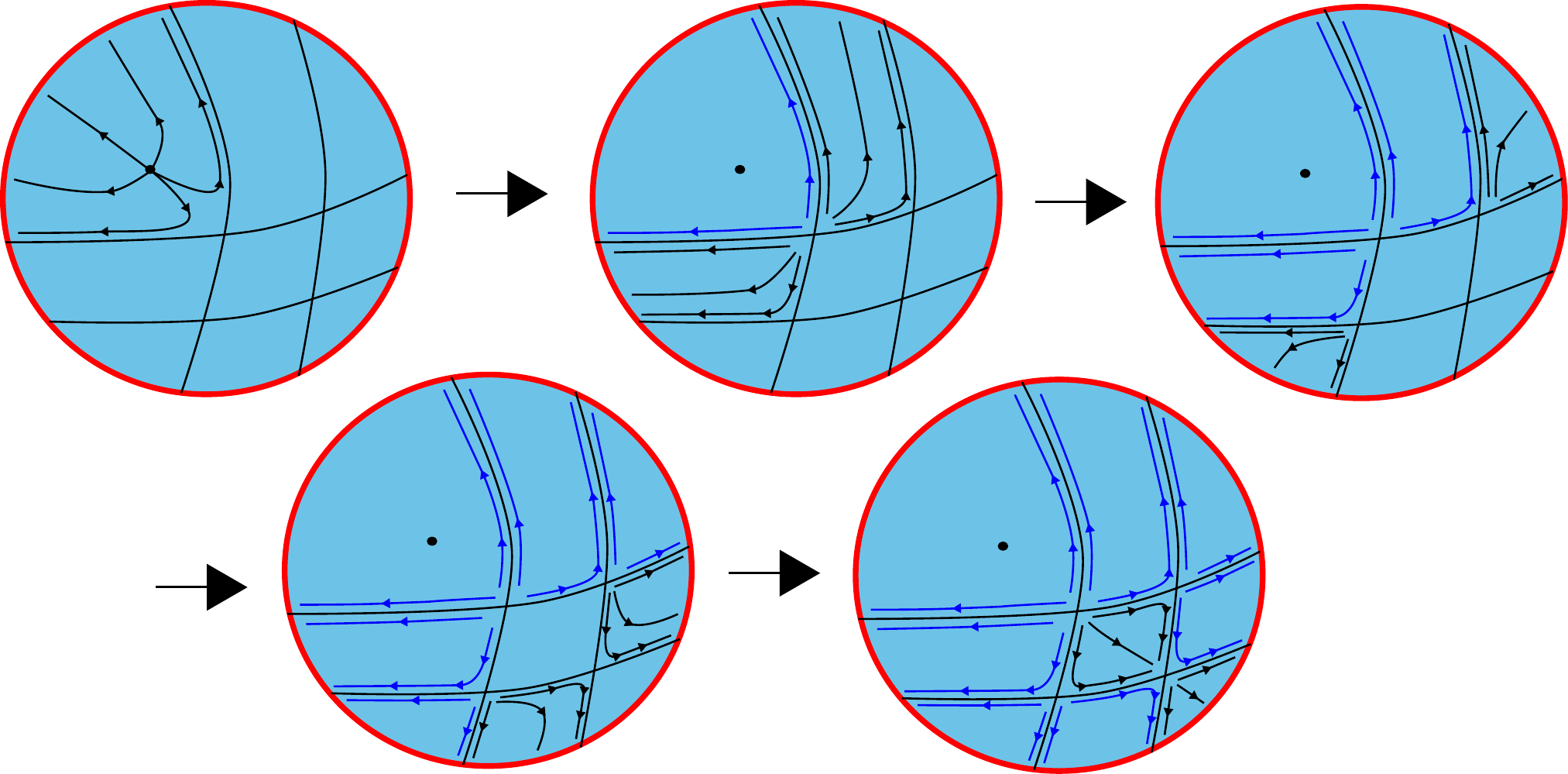}
\caption{We evolve a set of oriented Cauchy slices (black) through the disk. In this way, an orientation is associated to each of the boundaries of the smaller disks (blue) that allows for an $\slr$ BF calculation in each of these disks. The black dot represents the horizon.}
\label{fig:cauchylorentzian}
\end{figure}

We can now use the methodology of \cite{paper3} to calculate a generic $\slr$ disk correlation function, decomposing the full amplitude into propagators and $3j$-symbols.\footnote{This deconstruction is of similar spirit as that of higher genus string amplitudes into tubes and three-holed spheres.} This decomposition can also be appreciated by starting solely with the boundary theory and realizing that this immediately gives a particular bulk slicing of the amplitude, the coset slicing. We provide details on this argument in Appendix  \ref{app:slicing}. 
\\~\\
By \eqref{plusplanch}, a complete set of states of $\slr$ BF theory is given by the semigroup element states $\ket{g}$ with $g\in\slr$ resulting in the resolution of the identity:
\begin{equation}
\mathbf{1} = \int_{\slr} d g \ket{g}\bra{g}.\label{complete}
\end{equation}
Amplitudes of $\slr$ BF including several Wilson line insertions can be constructed as usual by cutting the manifold into disk-shaped regions, inserting completeness relations \eqref{complete} on the edges of the regions, calculating the amplitude for each disk-like region with fixed $g_i$ on the boundaries, and then gluing the disk back together including the external Wilson lines by performing integrals over $g_i$ of the type:
\begin{equation}
\int dg \, K^{++}_{s_1s_4}(g) K^{++}_{s_2s_5}(g) K^{++}_{s_3s_6}(g)^{*} = \tj{\mu_1}{\mu_2}{\mu_3}{s_1}{s_2}{s_3}\tj{\mu_1}{\mu_2}{\mu_3}{s_4}{s_5}{s_6},
\end{equation}
where we used the crucial property \eqref{crucial}. On the right hand side one recognizes the vertex functions of interest as the $\slr$ (hyperbolic) $3j$ symbols.
\\~\\
There is still the question of mathematical consistency of this calculation to be answered. For $\slr$, within each disk-like region, the calculation only works as explained around \eqref{semigroupprop} if the disk can be written as Hamiltonian propagation from \emph{positive} group elements to other positive group elements.\footnote{Otherwise $\sltr$ representation theory is required in contradiction with the ansatz that a consistent truncation to $\slr$ BF theory exists.} Positivity of a group element along a certain line requires the choice of an orientation on this line. As illustrated for example in Figure \ref{fig:cauchylorentzian}, this is accomplished by choosing a set of \emph{oriented} Cauchy surfaces within the disk.

\subsection{Constrained Asymptotic States}\label{s:gravcos}
The Schwarzian theory dual to JT gravity on a disk, can be viewed as quantum mechanics on a particular coset of $\slr$, inherited from the coset constraints to obtain 2d Liouville CFT from $\sltr$ WZW CFT \cite{berooguri,Balog:1990mu, Dijkgraaf:1991ba}. It is instructive to see that we can obtain the JT disk amplitudes from this coset construction using the results of section \ref{s:cosets}.
\\~\\
Explicitly, gravitational disk wavefunctions are associated with the parabolic state $\ket{\mathfrak{i}_+}$ defined in Appendix \ref{s:repsemi} to satisfy $J_+\ket{\mathfrak{i}_+}=i\ket{\mathfrak{i}_+}$ \cite{paper3,Dijkgraaf:1991ba}. In terms of functions $f$ on $\slr$, the condition is
\begin{equation}
\label{cosconstr}
f(g \cdot h_+(\gamma)) = e^{-\gamma} f(g), \qquad \forall \gamma \in \mathbb{R}^+, \quad g \in \slr.
\end{equation}
Unlike in section \ref{sect:coset}, this does not define functions invariant under some subgroup; rather covariant functions are studied. This modification does not alter any of the results of section \ref{sect:coset} though. The JT disk partition function is hence calculated in the angular slicing of Figure \ref{IntroChannels} middle as \eqref{twistpcoset}:
\begin{equation}
Z(\beta) = \int d k \int_{-\infty}^{+\infty}ds \, \phi^k_{\mathfrak{i}s}(g) \phi^k_{\mathfrak{i}s}(g)^* e^{-\beta k^2},
\end{equation}
with 
\begin{equation}
\phi^k_{\mathfrak{i}s}(g) = \sqrt{k \sinh 2\pi k} R_{\mathfrak{i}s}(g),
\end{equation}
a basis for the gravitational coset. Indeed, the functions $R_{s's}(g) = \left\langle s'\right| g \left|s\right\rangle$ are complete in $L^2(\slr)$. Of these, only those linear combinations of the form
\begin{equation}
\left|\mathfrak{i}_+\right\rangle = \int_{-\infty}^{+\infty}ds \, \left\langle s\right|\left.\mathfrak{i}_+\right\rangle \left|s\right\rangle,
\end{equation}
fulfill the gravitational constraints \eqref{cosconstr}. The Hilbert space can be written in the form:
\begin{equation}
\mathcal{H} = \Big\{\left|k,s,\mathfrak{i}\right\rangle, \quad s \in \mathbb{R}, \quad J_+\ket{k,\mathfrak{i}} = i\ket{k,\mathfrak{i}}\Big\}
\end{equation}
or in the dual group basis as the states $\ket{\phi,\gamma_-}$. The Schwarzian states $\ket{k,\mathfrak{i},\mathfrak{i}}$ respectively $\ket{\phi}$ used in \cite{harlowfactor,paper3,lin,schwarzian} live on the defect slices of Figure \ref{IntroChannels} left.

\section{Edge States of BF Theory}
\label{sect:BFedge}
Next, we will explain the precise nature of the edge degrees of freedom in Jackiw-Teitelboim gravity that appear at entangling surfaces (or black hole horizons). To obtain these edge dynamics we follow the logic of \cite{paper2}.
\\
As earlier, we start by focusing on compact BF theory; the generalization to JT gravity becomes straightforward with the previous section in mind.

\subsection{Edge Dynamics from the Path Integral}
The correct way to split the BF Lorentzian path integral of a surface $\Sigma$ in two pieces $L$ and $R$ proceeds by introducing a functional delta constraint as in \cite{paper2} (Figure \ref{gluing}):
\begin{figure}[h]
\centering
\includegraphics[width=0.5\textwidth]{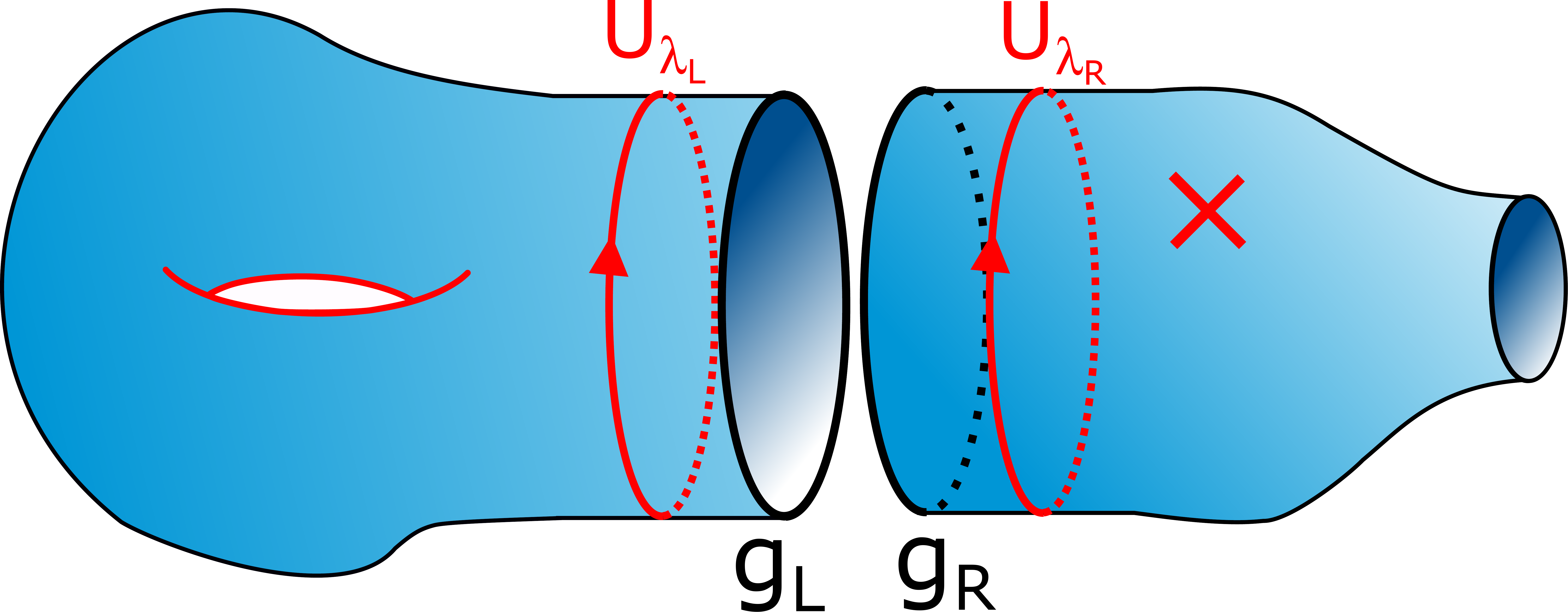}
\caption{Gluing two BF sectors along one boundary in terms of two particle-on-a-group models $g_L$ and $g_R$.}
\label{gluing}
\end{figure}
\begin{align}
\int \frac{\left[\dpi A_L \dpi A_R\right]}{\text{vol }G}[\dpi \chi_L \dpi \chi_R]\exp\left(iS[A_L,\chi_L]\right)\delta(A_L-A_R\atbdy)\delta(\chi_L-\chi_R\atbdy) \exp(iS[A_R,\chi_R]).\label{gluebf}
\end{align}
Integrating out $\chi_L$ and $\chi_R$, forces the connections to be flat in the bulk of $L$ and $R$ and the path integral over $A_L$ (and $A_R$) is reduced to a path integral over independent boundary group element configurations on all boundaries of $\Sigma$ as well as on the gluing boundaries.\footnote{Depending on the topology there may or may not also be an integral over topologically nontrivial flat connections.} The path integral over $A$ in general also includes an integral over holonomies $\int A = U_\lambda$  along the gluing boundary. For example, if $L$ is a disk, the holonomy is fixed, but if $L$ is an annulus, the holonomy is an additional degree of freedom to be integrated over.
\\
Explicitly, localization on flat connections results in $A_{L}\atbdy = d g_{L} \, g_{L}^{-1} + \lambda_L$ and $A_{R}\atbdy = d g_{R} \, g_{R}^{-1} + \lambda_R$. In the path integral \eqref{gluebf}, the functional delta becomes:
\begin{equation}
\delta(A_L-A_R\atbdy)=\delta\left(d g_L \, g_L^{-1}- d g_R \, g_R^{-1}\right) \, \delta(\lambda_L-\lambda_R),\label{delta}
\end{equation}
and two twisted particle on a group actions pop up associated with the gluing surface (one for $L$ and one for $R$). The action on the left boundary, is minus the right one, as the orientation of the boundary surface in $L$ respectively $R$ is opposite.\footnote{This descends from the parity transformation on the Chern-Simons action taking $k\to -k$ to flip the orientation.} As a result, the two actions cancel when we enforce the functional delta constraints and set $\lambda_R = \lambda_L \equiv \lambda$:\footnote{The final equality uses that $Z_{S^2} = 1$. There are several ways to argue for this. Performing the double-scaling large $k$ limit on the Chern-Simons partition function on $S^2 \times S^1$ is trivial since \cite{jones}
\begin{equation}
Z^{CS}(S^2 \times S^1) = 1.
\end{equation}
Alternatively, the volume of the moduli space of flat gauge connections on $S^2$ is trivial:
\begin{equation}
\int \frac{\left[\mathcal{D}\chi \right] \left[\mathcal{D}A_\mu \right]}{\text{vol }G} e^{i S_{BF}} = \int \frac{\left[\mathcal{D}g \right] }{\text{vol }G_{\partial}}  = 1,
\end{equation}
since there is only 1 gauge orbit on $S^2$.}
\begin{align}
\label{split}
\hdots \int \frac{[\dpi g_L] [\dpi g_R]}{\text{vol }G_{\partial}}\exp\left(iS[g_L,\lambda]\right)\delta(g_L-g_R)\exp\left(-iS[g_R,\lambda]\right) \hdots  &=  \hdots \frac{\int [\dpi g]} {\text{vol }G_{\partial}} \hdots\nonumber\\ &= \hdots 1 \hdots.
\end{align}
The dots represent the other degrees of freedom in $L$ and $R$ that are irrelevant for this argument.
This procedure consistently glues the submanifolds together.
\\~\\
Notice that the argument of the functional delta in \eqref{delta} is just the current density on the boundary, so it can be read as $\delta(\mj_L-\mj_R)$. The theory associated with the submanifold $R$ only is obtained from \eqref{split} as in \cite{paper2} by dropping all reference to $L$:
\begin{equation}
Z_R=\int d\lambda_R\int [\dpi g_R]\exp\left( i S[g_R,\lambda_R]\right) = \int[\dpi \mj_R]\int_{A_t \atbdy = \chi \atbdy = \mj_R} [\dpi A_R][\dpi\chi_R]\exp(iS[A_R,\chi_R]). \label{edgedynamics}
\end{equation}
As shown by the second equality, this formula can be interpreted as the path integral on the right manifold sourced by a boundary current $\mj_R$, including an additional path integral over the boundary charges $\mj_R$ to account for the edge degrees of freedom, in the spirit of \cite{paper2}. In canonical language, this means there is an extended Hilbert space that accounts for edge states on the dividing surface. The gluing condition $\delta(\mj_L-\mj_R)$ acts as a Gupta-Bleuler constraint that extracts the physical subsector from the extended Hilbert space.\footnote{See \cite{wong} for similar statements on edge states in CS.} The path-integral over $\mj_L= \mj_R$ glues the manifolds together. 

\subsection{Two-Boundary Models}\label{twobdybf}
\begin{figure}[h]
\centering
\includegraphics[width=0.85\textwidth]{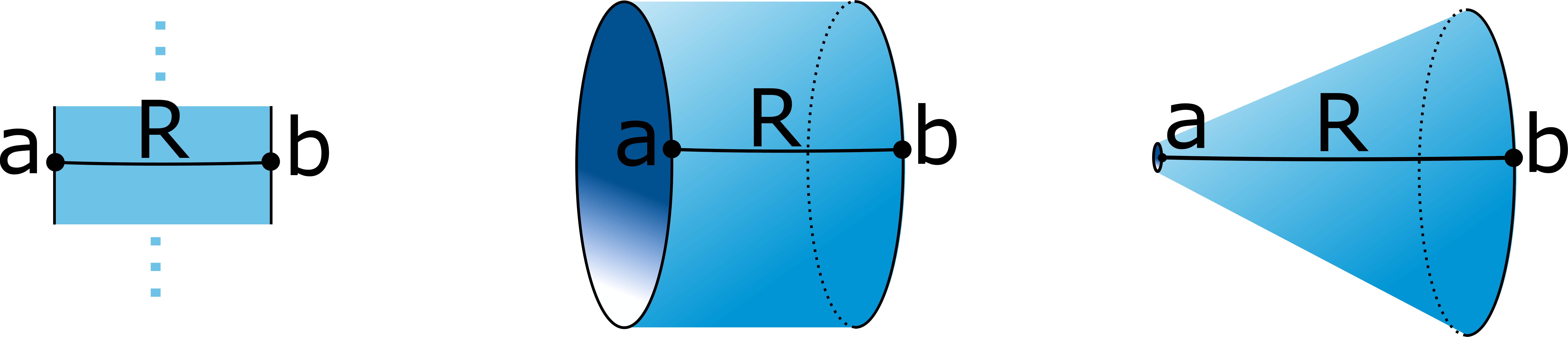}
\caption{Left: Hilbert space of BF on $I$. Middle: Thermal Cylinder. Right: Thermal disk obtained by shrinking the left boundary by redshift.}
\label{Wlannulus}
\end{figure}

\begin{figure}[h]
\centering
\includegraphics[width=0.5\textwidth]{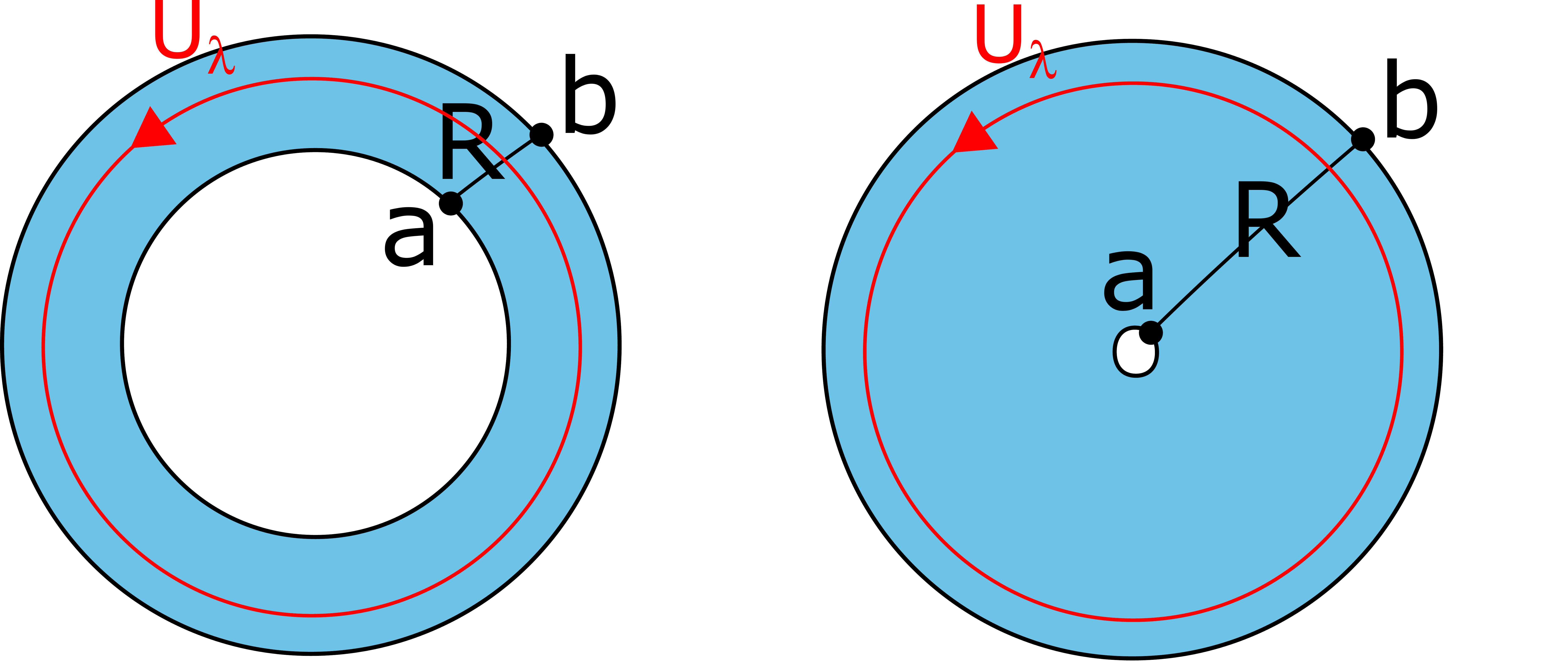}
\caption{Left: Annulus. Right: Closing of the inner hole to form a disk.}
\label{Wlannulus2}
\end{figure}

\begin{figure}[h]
\centering
\includegraphics[width=0.25\textwidth]{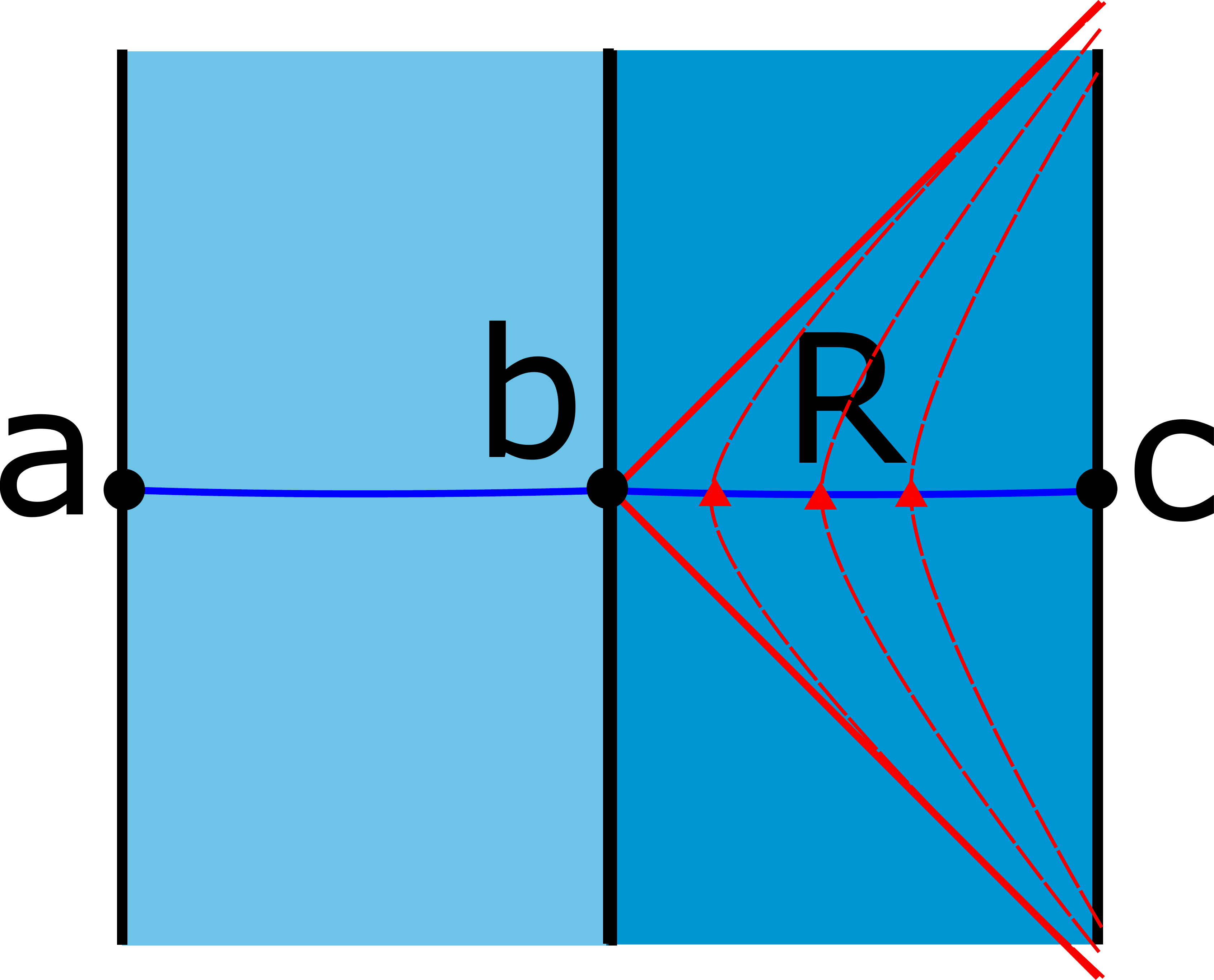}
\caption{Splitting an interval in two pieces using the modular Hamiltonian.}
\label{annulusredshift}
\end{figure}

As an application of the above, and as a preparation for the gravity case, we will show how to split a spatial interval in two pieces. 
\\
Consider first the BF model on a Lorentzian strip $I$. The Euclidean configuration associated with this setup is $I\cross S^1$ with two circular boundaries that break topological invariance. This manifests itself as the dependence of the path integral on a choice of metric / einbein on the boundary curves, through its circumferences $\beta_L$ respectively $\beta_R$ (Figure \ref{Wlannulus}). Flatness of $F=0$ implies $A = d g g^{-1} + \lambda$ where $\lambda$ is an unspecified holonomy: the time circle is not contractable hence $\lambda$ is a physical degree of freedom of the theory to be integrated over. Via the usual argument, the action for this configuration only depends on large values of $g$.\footnote{Bulk profiles of $g$ are redundant. In particular, $g\rvert_\partial$, with two disconnected boundary components in this scenario.} We obtain the path integral for this configuration as:
\begin{equation}
Z = \int [\dpi A][\dpi \Phi] e^{- S[A,\Phi]} = \sum_{\lambda}\int [\dpi g_L][\dpi g_R] e^{-S[g_L,\lambda] - S[g_R,\lambda]}.\label{za1}
\end{equation}
This could have been obtained along the lines of \eqref{edgedynamics} by cutting a tubular neighbourhood out of some generic manifold.
The result of this Euclidean path integral is:
\begin{equation}
Z(\beta_L,\beta_R)=\int d\lambda Z(\beta_L,\lambda)Z(\beta_R,\lambda),\label{za}
\end{equation}
where one recognizes the twisted particle on a group partition functions \eqref{twistedpog}.\footnote{We provide a more technical account on gluing the disks together, emphasizing the path integration space, in Appendix \ref{app:glueBF}.} Writing this out using orthogonality of the finite characters\footnote{This is the classical limit $k\to\infty$ of $S$-matrix unitarity in 2d CFT.}
\begin{equation}
\sum_\lambda \chi_R(U_\lambda)\chi_{R'}(U_\lambda)^* = \delta_{R,R'},\label{charortho}
\end{equation}
this becomes (Figure \ref{Wlannulus2}):
\begin{equation}
Z(\beta_L,\beta_R)=\sum_{R}\left(\dim R  \, e^{-\beta_L \cas_R}\right)\left(\dim R \,  e^{-\beta_R \cas_R}\right). \label{zib}
\end{equation}
\\~\\
Two interesting limits are the thermal cylinder where we take $\beta_L = \beta_R$ and the disk obtained by $\beta_L=0$. They are shown in Figure \ref{Wlannulus}. \\
\noindent For the thermal case, \eqref{zib} implies that the spectrum of the theory consists of the states $\ket{R,a,b} \equiv \ket{R,a}\otimes \ket{R,b}$ and the Hamiltonian is $ H_L +  H_R$. Due to the Peter-Weyl theorem, the Hilbert space of 2d BF on an interval is indeed given by these states, to be interpreted as open strings with one endpoint on each boundary (Figure \ref{Wlannulus} left). \\
The latter case $\beta_L = 0$ comes into play when constructing the thermofield double from the Rindler Hilbert space or equivalently when computing vacuum entanglement entropy of an interval with an adjacent interval. As shown by the modular flow in Figure \ref{annulusredshift}, the particle on a group on the inner boundary is \emph{frozen} and does not contribute to the modular Hamiltonian: $K=\beta H_R$.
We recover the disk amplitude: 
\begin{equation}
Z(\beta) = \sum_\lambda Z_L(0,\lambda) Z_R(\beta,\lambda) = \sum_R (\dim R) ^2 \, e^{-\beta \cas_R},\label{zob}
\end{equation}
which includes a sum over edge modes, and is comparable to \eqref{edgedynamics}. The edge degrees of freedom associated with the horizon or inner boundary are identified as the states $\ket{R,a}$. The precise microstate $\ket{R,a}$ contributes zero energy and does not affect any of the bulk observations a right-observer would perform, which translates to the fact that the correlation functions in a pure microstate $\ket{R,a} \otimes \mathcal{H}_R$ are independent of $a$.
\\~\\
Formula \eqref{zob} is a consistency check: including the correct edge degrees of freedom to a one-sided theory ensures that the trace in the Rindler Hilbert space equals the thermal disk path integral. Graphically, summing over edge degrees of freedom $a$ stuffs the hole in the annulus (Figures \ref{Wlannulus} and \ref{Wlannulus2} right). This proves the claims made around \eqref{purification?}. From the above we can directly purify the density matrix to re-obtain the thermofield double state:
\begin{equation}
\label{tfd}
\left|\text{TFD}\right\rangle = \sum_{R,a,b}\, e^{-\frac{\beta}{2} \cas_R} \ket{R,a,b} \otimes \ket{R,a,b}.
\end{equation}
The conclusion here is that whereas \eqref{diska} and \eqref{tfd} describe the same state, only the latter makes manifest the factorization of the theory, as it can be directly read as a purification of the Rindler thermal density matrix, which crucially includes an edge sector on the horizon.

\section{Edge States of JT Gravity}
\label{sect:JTedge}
In this section we generalize the BF discussion of the previous section to JT gravity. We consider two different two-boundary models. There is a distinction to be made between a holographic boundary, where gravitational constraints are to be imposed, and entanglement boundaries where no such constraints are imposed.
\\~\\
First we discuss a configuration with two holographic boundaries. After that, we consider one holographic boundary and one entangling boundary, which describes a one-sided black hole configuration.

\subsection{Wormhole States}\label{s:worm}
\begin{figure}[h]
\centering
\includegraphics[width=0.25\textwidth]{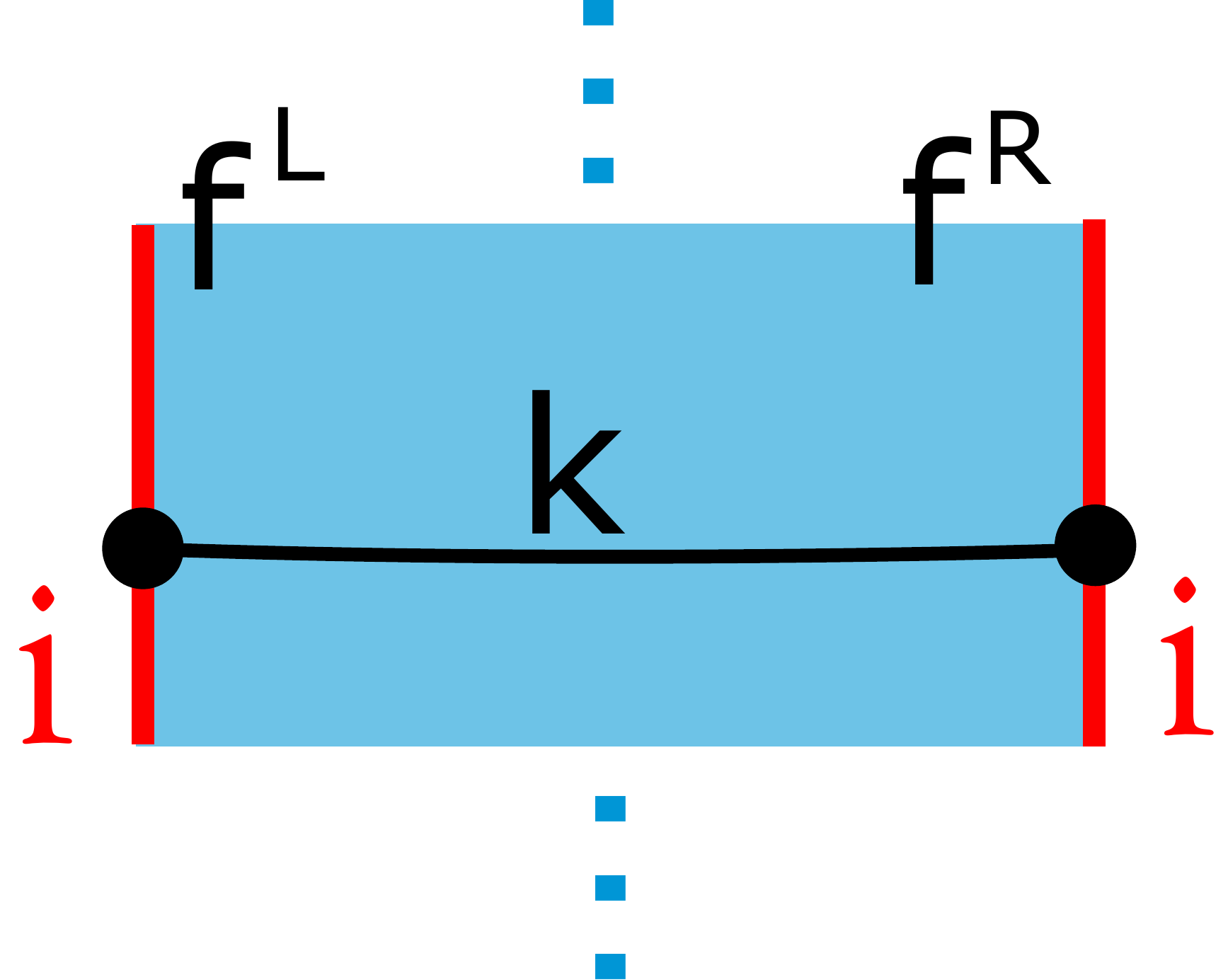}
\caption{Hilbert space of JT gravity with two Schwarzian (holographic) boundaries. In path integral language, the degrees of freedom are given by separate left and right reparametrizations $f^{\LL}(\tau)$ and $f^{\RR}(\tau)$.}
\label{twoschw}
\end{figure}

Consider first Jackiw-Teitelboim gravity between two holographic (Schwarzian) boundaries, $L$ and $R$, on which the gravitational boundary conditions are to be enforced \cite{Coussaert:1995zp,Fitzpatrick:2016mtp,Gonzalez:2018enk, paper3}:
\begin{equation}
\left.A\right|_{\partial \mathcal{M}} = iJ^- - \frac{T(\tau)}{2}  iJ^+,\label{gravcon}
\end{equation}
in terms of a dynamical function $T(\tau)$ and the generators \eqref{gener}. These boundary conditions act by constraining the boundary theory from a particle on $\slr$ to the Schwarzian theory (Figure \ref{twoschw}) \cite{paper3}, in terms of the time reparametrizations $f^{\LL}$ and $f^{\RR}$ of the left- respectively right holographic boundary, defined as:
\begin{equation}
T_{L,R}(\tau) \equiv \left\{\tanh\frac{\pi}{\beta_{L,R}}\lambda f^{\LL,\RR}(\tau),\tau\right\}.
\end{equation}
The Hilbert space of this gravitational coset system is of the form $\ket{k,\mathfrak{i},\mathfrak{i}}$, as we will demonstrate.

\noindent The thermal path integral for this configuration is the analogue of \eqref{za1} and includes an integral over conjugacy class elements (or orbits) $\lambda$:
\begin{align}
Z(\beta_L,\beta_R) &= \int d\lambda \int [\dpi f^{\LL}]  e^{-S[f^{\LL},\lambda]}\int [\dpi f^{\RR}]e^{ - S[f^{\RR},\lambda]} \nonumber \\
&= \int d\lambda\, Z(\beta_L,\lambda) Z(\beta_R,\lambda),\label{zsch2}
\end{align}
with the twisted Schwarzian action
\begin{equation}
S[f,\lambda] = - \frac{1}{2}\int_{0}^{\beta}d\tau \, \left\{\tanh\left(\frac{\pi}{\beta}\lambda f(\tau)\right),\tau\right\} = -\frac{1}{2} \int_{0}^{\beta}d\tau \, \left(\left\{f,\tau\right\} - \frac{2\pi^2}{\beta^2}\lambda^2 \dot{f}^2\right),\label{sch}
\end{equation}
where $f(\tau + \beta) = f(\tau) + \beta$, and the twisted Schwarzian partition function \cite{origins,schwarzian}:
\begin{equation}
Z(\beta,\lambda)=\int_{0}^{+\infty} dk \, S_\lambda^{\, k} \, e^{-\beta k^2}.\label{zsch}
\end{equation} 
Explicitly to derive this one simply takes the Schwarzian double-scaling limit of a Virasoro character $\chi_\lambda(\tau)$ \cite{origins,schwarzian}.\footnote{This can be interpreted as the Schwarzian limit of a brane system in 2d Liouville CFT with one FZZT and one ZZ brane for $\lambda \neq 0$, or a ZZ-ZZ system for $\lambda = 0$.} The Virasoro modular $S$-matrices in this limit are given by
\begin{align}
S_\lambda^{\, k} &=\cos 2\pi \lambda k, \label{svir} \\
S_0^{\,k} &= k \sinh 2\pi k. \label{plancherelvir}
\end{align}
Using $S$-matrix unitarity 
\begin{equation}
\int d\lambda S_k^{\,\lambda} S_\lambda^{\,k'} = \int_{0}^{+\infty} d\lambda \, \cos 2\pi k\lambda \cos 2\pi k'\lambda = \frac{1}{4}\, \delta(k-k'),
\end{equation}
\eqref{zsch2} is rewritten into a form that makes manifest the content of the Hilbert space of the theory:
\begin{equation}
\boxed{Z(\beta_L,\beta_R) = \int_{0}^{+\infty}dk \, e^{-k^2(\beta_L+\beta_R)}}\label{pfann}
\end{equation}
We deduce that only the constrained states  $\ket{k,\mathfrak{i},\mathfrak{i}}$ make up the Hilbert space of this theory. We will call these the \emph{wormhole states} of JT gravity.
\\
The states in this Hilbert space are labeled in the same way as in the defect channel slicing of Figure \ref{IntroChannels} left and as in the Hilbert space of a single Schwarzian theory \cite{schwarzian}; as in each of these scenarios we are considering a Cauchy surface connecting two constrained boundaries (Figure \ref{twoschw}).


\subsection{Black Hole States}
\label{s:jt}
The question arises how Jackiw-Teitelboim gravity behaves away from the asymptotic boundary. Does it behave as an unconstrained $\slr$ BF theory or does it still feel the constraints? In particular, when cutting a manifold in the sense of \eqref{split}, do we get Schwarzian actions on the gluing boundaries or particle on $\slr$ actions? 
\\
For cosets in section \ref{s:cosets}, we found that interior regions are insensitive to the constraints and behave as if they are part of the parent $G$ theory. We provided arguments in section \ref{s:gravcos} that the gravitational theory should be viewed as a specific example of a coset model. This suggests the edge dynamics of JT on a gluing surface is that of a particle on $\slr$.
\\~\\
Following the logic around Figure \ref{annulusredshift}, the edge theory is frozen on the horizon. Using the twisted $\slr$ partition function $Z^{+}(\beta,\lambda)$ from \eqref{slprpf}, we can write:
\begin{equation}
Z(\beta)=\int d\lambda\, Z^{+}(0,\lambda) Z(\beta,\lambda).\label{zbh1}
\end{equation}
The finite characters of $\slr$ \eqref{charslplus} are $\chi_k^+(U_\lambda)=\cos 2\pi k \lambda$. Notice that these are identical to the classical $b\to 0$ limit of the Virasoro $S$-matrix \eqref{svir} appearing in $Z(\beta,\lambda)$.\footnote{\label{fn:smatrix}What we have proven here is a non-compact generalization of a well-known result. Consider the modular $S$-matrices associated to two compact groups $G$ and $G/H$ for $G$ and $H$ compact. It is an elementary result that these are identical $S^{G}=S^{G/H}$. In particular this carries through in the classical double-scaling $k\to\infty$ limit to:
\begin{equation}
\chi^{G}_R(U) = \chi^{G/H}_{R}(U)
\end{equation}}
This means we can use $\slr$ character orthogonality to rewrite \eqref{zbh1} as:
\begin{equation}
Z(\beta)=\int_{0}^{+\infty} d k \Big(k\sinh 2\pi k\Big) \left(e^{-\beta k^2}\right)=\Tr e^{-\beta H}.\label{zbhsch}
\end{equation}
From this one finds the spectrum of states as $\ket{k,s,\mathfrak{i}} \equiv \ket{k,s}\otimes\ket{k,\mathfrak{i}}$, with $s$ a hyperbolic $\slr$ label as introduced in Appendix \ref{s:repsemi}. The result \eqref{zbhsch} is the JT disk amplitude \eqref{zsch}, proving that we have included precisely the correct edge states by postulating a particle on $\slr$ lives on the entangling surface.\footnote{Including a frozen Schwarzian on the horizon, we would end up with $Z(\beta)=\int d k \, e^{-\beta k^2}$. This is not the JT disk amplitude so edge degrees of freedom would not have been taken into account correctly.}
\\
In the context of Section \ref{s:factor}, this is just the statement that an $\slr$ representation matrix factorizes using its defining property as 
\begin{equation}
R^k_{\mathfrak{i}\mathfrak{i}}(g_1\cdot g_2)=\int d s \, R^k_{\mathfrak{i} s}(g_1)R^k_{s \mathfrak{i}}(g_2),
\end{equation}
hence
\begin{equation}
\ket{k,\mathfrak{i},\mathfrak{i}}=\frac{1}{\sqrt{V k\sinh 2\pi k}}\int d s \ket{k,\mathfrak{i},s}\otimes \ket{k,\mathfrak{i},s}.\label{statesfactorjt}
\end{equation}
So the Hartle-Hawking calculation already illustrates the edge states should be the states $\ket{k,s}$, and this is confirmed by \eqref{zbhsch}.\footnote{To distill the volume prefactor $V$ that properly normalizes $\ket{k,\mathfrak{i},\mathfrak{i}}$, a more careful treatment is needed relating finite-volume regularization to delta regularization in this context. This is performed in appendix \ref{app:reg}. Relatedly, the trace over these hyperbolic labels also includes an additional volume factor:
\begin{equation}
    \Tr(\dots) =\int_0^\infty \frac{d k}{V} \int_{-\infty}^\infty ds \bra{k,s,\mathfrak{i}}(\dots)\ket{k,s,\mathfrak{i}}.\label{trace}
\end{equation}
These volume factors can all be traced back to the $\sltr$ modding in the symplectic Schwarzian path integral. It is intrinsic to \emph{all} BF-theories (and their 3d Chern-Simons ancestors): a similar $G$-modding appears in that context for the particle on group \eqref{pog} path integral, and the TFD \eqref{tfd} secretly has a similar $1/\sqrt{\text{Vol }G}$ as \eqref{tfdjt}. The appearance of these volume factors have been subject to critique \cite{wittenstanford,harlowfactor}, hindering a genuine Hilbert space interpretation of such symplectic path integrals. 
}
\\~\\
From \eqref{zbhsch} we can directly write down the purification of the thermal ensemble:\footnote{Its norm is indeed the Schwarzian partition function $Z$, when using $\delta(k-k) = V_C$ and $\text{dim k} = \frac{V}{V_C} \rho(k)$, as explained in more detail in appendix \ref{app:reg}.}
\begin{equation}
\boxed{\ket{\text{TFD}}= \frac{1}{\sqrt{V}}\int_{0}^{+\infty} d k\, \int_{-\infty}^{+\infty} d s\, e^{-\frac{\beta}{2}k^2} \ket{k,s,\mathfrak{i}}\otimes \ket{k,s,\mathfrak{i}}. }\label{tfdjt}
\end{equation}
This is the sense in which we can think of JT gravity states as factorizing across surfaces.
\\
The Von Neumann entropy of the thermal state was calculated in \cite{lin} and gives the Bekenstein-Hawking entropy in the limit where the bulk is classical. In writing \eqref{tfdjt}, we have pinpointed the gravitational states responsible for this entropy, so the conclusion is that the states $\ket{k,s}\otimes \ket{k,\mathfrak{i}}$ can be interpreted as \emph{black hole states} or one-sided states of JT gravity.
\\~\\
It has been argued \cite{harlowfactor} that JT gravity does not factorize across a horizon, and this factorization problem can be decomposed into several subproblems.
\begin{itemize}
\item 
Firstly, gravity experiences non-local constraints that hamper a direct factorization across a surface. This happens in much the same way as Maxwell theory with its Gauss-law constraint. For Maxwell however, it is well-known how to address this issue: one introduces an extended Hilbert space and gluing condition, basically allowing Wilson lines to split across the surface.\footnote{See for example \cite{donnellywall,paper1}.} The price to pay is the introduction of edge degrees of freedom, charges in the Maxwell case. Since JT was written in terms of a (non-Abelian) gauge theory, we have provided here the analogue of this argument for JT gravity.
    \item 
		These additional horizon degrees of freedom, captured by the $s$-index in \eqref{tfdjt}, are not represented at the holographic boundary.\footnote{Given the degree of freedom $f$, one has no information whatsoever on the precise microstate underlying this state. Relatedly, it was observed in \cite{Kourkoulou:2017zaj} that the pure states in SYK are all described by the same Schwarzian action and no distinction can be made between them within this low-energy regime.} This is a rephrasing of the statement that Schwarzian dynamics is capturing thermodynamics, not microphysics. That these horizon degrees of freedom are not localized on the asymptotic boundary, illustrates that this is indeed not a microscopic realization of the AdS/CFT correspondence. 
    \item As mentioned in the introduction, JT gravity \eqref{JTaction} does not capture the extremal (or zero-temperature) entropy $S_0$ of some parent microscopic theory.\footnote{In case of integral $e^{S_0}$, one can incorporate this in principle by adding an additional (energy-independent) degeneracy $e^{S_0}$ label to each state.} Strictly speaking, such factors hamper a direct Hilbert space interpretation of the symplectic thermal path integrals. The volume factors $V$ that we tracked in the above formulas can be treated in the same vein, and interpreted as contributions to $S_0$, similarly to the way it worked for the BF model with compact group. \\
		Furthermore, the spectrum is continuous so no discrete microstates (as in e.g. the D1-D5 system) exist. Formula \eqref{tfdjt} and its interpretation should be read taking into account these caveats: we have found a description of the states that yield the black hole entropy, but there is no hope for a genuine discrete counting problem within JT gravity, as is expected from the very get-go for such pure gravity theories, see also \cite{Martinec:1998wm}. Upon embedding within a full-fledged holographic UV theory, these horizon states $s$ are expected to be the IR-limit of the dynamical and fundamental degrees of freedom, with gauge theory and gravity emerging from these more microscopic degrees of freedom \cite{Harlow:2015lma}.\footnote{A wormhole-threading Wilson line is only factorizable upon introducing horizon degrees of freedom in such a way that in the low-energy effective field theory, this replacement makes no difference for correlation functions. However, one has access to all possible horizon charges to facilitate this with no information for the low-energy observer on which charge was actually used: these can be thought of as labeling the different states that count the entropy.}
\end{itemize}
Using \eqref{statesfactorjt} we can rewrite the TFD state of JT gravity \eqref{tfdjt} in terms of wormhole states as:
\begin{equation}
\ket{\text{TFD}}=\int_{0}^{+\infty} dk \, \sqrt{k \sinh 2\pi k } \, e^{-\frac{\beta}{2} k^2}\ket{k,\mathfrak{i},\mathfrak{i}}   .\label{tfdjtnonfactor}
\end{equation}
This is the form that appeared in the literature \cite{harlowfactor,lin}, where factorization is not manifest.\footnote{Projecting it onto a $g$-eigenstate, one writes:
\begin{equation}
\label{diskb}
\left\langle g=\varphi\right|\left.\text{TFD}\right\rangle = \int dk \, k \sinh 2\pi k \, R_{k,00}(\varphi) \, e^{-\frac{\beta}{2} k^2} = \int dk  \, k \sinh 2\pi k\, e^{\varphi} K_{2ik}(e^{\varphi}) \, e^{-\frac{\beta}{2} k^2} .
\end{equation}
The group variable $\varphi = -d$ can be geometrically interpreted as a bulk length parameter between both sides, as shown in \cite{Yang:2018gdb}. This is a direct geometric interpretation of the abstract group variable.}

\section{Two-Boundary Correlation Functions}\label{s:twobouncor}
Let us return to the situation with two asymptotic boundaries discussed in section \ref{s:worm}. In this setup, we encounter a new type of Wilson line operators with endpoints on different boundaries.\footnote{Such operators are $\sltr$ covariant under $\sltr_\text{L}$ or $\sltr_\text{R}$ separately, but invariant under only the diagonal combination. W.r.t. each boundary, these operators are of the form of those discussed in Appendix D of \cite{schwarzian}, which were analyzed in terms of KZ equations.} In the dual boundary theory one is led to studying correlators of the type:
\begin{equation}
\label{wormcorr}
\int_{0}^{+\infty}d\lambda \, \int [\dpi f^{\LL}][\dpi f^{\RR}] \mathcal{O}_{\lambda,\text{LR}}^{\ell}(\tau_1,\tau_2) \hdots e^{-S[f^{\LL},\lambda] - S[f^{\RR},\lambda]}
\end{equation}
for one or more bilocal operators connecting both boundaries $\mathcal{O}_{\lambda,\text{LR}}^{\ell}(\tau_1,\tau_2)$. In the BF formulation of JT gravity this is easy. But let us first give a more precise holographic expression for the bulk crossing Wilson line $\mathcal{O}_{\lambda,\text{LR}}^{\ell}(\tau_1,\tau_2)$. After integrating out $\chi$, we find:
\begin{equation}
\label{kaplan}
\mathcal{P}e^{\int_{z_i}^{z_f} A(z) dz} \quad \longrightarrow \quad \left(\frac{\dot{F^{\LL}}(\tau_1) \dot{F^{\RR}}(\tau_2)}{(F^{\LL}(\tau_1)-F^{\RR}(\tau_2))^2}\right)^\ell
\end{equation}
where $\left\{F^{\LL, \RR},\tau\right\} = T_{L,R}(\tau)$, for possibly different time reparametrizations $F^{\LL}$ and $F^{\RR}$ at the endpoints. The proof can be found in Appendix \ref{app:kaplan}.\footnote{In earlier work \cite{paper3}, we demonstrated this for a Wilson line with both endpoints on the same boundary (with hence $F^{\LL} = F^{\RR}$), where the Wilson line could be deformed to lie entirely within the boundary. This proof no longer holds for bulk-crossing Wilson lines, or for Wilson lines encircling punctures such as those discussed in Appendix A of \cite{paper3}.} Performing a final reparameterization to the variables used in the action \eqref{sch} $F^{\LL, \RR} = \tanh(\frac{\pi}{\beta_{LR}} \lambda f^{\LL, \RR})$, we find:
\begin{equation}
\label{bilocalworm}
\mathcal{O}_{\lambda,\text{LR}}^{\ell}(\tau_1,\tau_2) = \left(\frac{\dot{f^{\LL}}(\tau_1) \dot{f^{\RR}}(\tau_2)}{\sinh(\pi \lambda \left(\frac{f^{\LL}(\tau_1)}{\beta_L} - \frac{f^{\RR}(\tau_2)}{\beta_R}\right))^2}\right)^{\ell}.
\end{equation}
Let us emphasize that the two asymptotic boundary model discussed here is very different from the TFD. The model, unlike the TFD, has two independent clocks $f^{\LL}$ and $f^{\RR}$ running on each of its boundaries, reflected in the separate temperatures $\beta_L$ and $\beta_R$.\footnote{The annulus amplitude contains two separate boundary theories at finite temperature simultaneously, whereas the TFD configuration is only thermal upon tracing out half of the theory. The two sides of the TFD state are mirror images of one another and hence it takes as many degrees of freedom to describe the dynamical clock for a TFD configuration than for a single-sided configuration.}
\\
Symmetries of the model \eqref{wormcorr} include independent time shifts $f^{\LL}(\tau) \to f^{\LL}(\tau + a_1)$, $f^{\RR}(\tau)\to f^{\RR}(\tau + a_2)$ on both boundaries. The independence of both boundary times shows that the amplitude for a single such bulk crossing Wilson line will be time-independent: the time $t_L$ of an incoming pulse in the $L$ system learns the $L$ observer nothing about the time $t_R$ at which the pulse left the $R$ system. 
\\~\\
As an application of the BF perspective on JT gravity let us write down two single Wilson line correlation functions in this model.
\begin{align}
&\begin{tikzpicture}[scale=0.8, baseline={([yshift=0cm]current bounding box.center)}]
\draw[thick] (0,0) circle (0.7);
\draw[thick] (0,0) circle (2);
\draw[thick,blue] (0.7,0) -- (2,0);
\draw[fill,black] (0.7,0) circle (0.1);
\draw[fill,black] (2,0) circle (0.1);
\draw (1.35,0.35) node {\small \color{blue}$\ell$};
\draw (0.9,0.3) node {\small $s$};
\draw (0.9,-0.3) node {\small $s$};
\draw (1.8,0.35) node {\small $\mathfrak{i}$};
\draw (1.8,-0.25) node {\small $\mathfrak{i}$};
\draw (0,1.35) node {\small $k$};
\draw (-0.25,0) node {\small $\beta_L$};
\draw (-2.35,0) node {\small $\beta_R$};
\end{tikzpicture}
\hspace{1.5cm}
&\begin{tikzpicture}[scale=0.8, baseline={([yshift=0cm]current bounding box.center)}]
\draw[thick] (0,0) circle (0.7);
\draw[thick] (0,0) circle (2);
\draw[thick,blue] (2,0) arc (270:138:1);
\draw[fill,black] (1.22,1.6) circle (0.1);
\draw[fill,black] (2,0) circle (0.1);
\draw (1.4,0.8) node {\small $k_2$};
\draw (0.8,1.5) node {\small $\mathfrak{i}$};
\draw (1.25,1.25) node {\small $\mathfrak{i}$};
\draw (1.7,0.35) node {\small $\mathfrak{i}$};
\draw (1.7,-0.25) node {\small $\mathfrak{i}$};
\draw (0,1.35) node {\small $k_1$};
\draw (0.8,0.55) node {\small \color{blue} $\ell$};
\draw (-0.25,0) node {\small $\beta_L$};
\draw (-2.35,0) node {\small $\beta_R$};
\end{tikzpicture}
\end{align}
Taking a particle on $\slr$ on the inner boundary, we find the correlator for a single bilocal straddling the annulus:
\begin{equation}
\label{annslr}
\left\langle \mathcal{O}_{\text{LR}}^{\ell}(\tau_L,\tau_R)\right\rangle = \frac{1}{Z}\int d k k\sinh 2\pi k \int ds \tj{k}{\ell}{k}{\mathfrak{i}}{0}{\mathfrak{i}}\tj{k}{\ell}{k}{s}{0}{s} e^{-k^2(\beta_L+\beta_R)} = \delta_{\ell,0}.
\end{equation}
The special case of $\beta_L = 0$ can be interpreted as a Wilson line stretching from the holographic boundary to the black hole horizon, the fact that the resulting amplitude vanishes is a manifestation of the fact that bulk operators do not couple to horizon degrees of freedom \cite{paper1,paper2}. Taking the inner boundary to be the Schwarzian instead, we find:\footnote{We used the known expression for the Schwarzian $3j$-symbol:
\begin{equation}
\tj{k}{\ell}{k}{\mathfrak{i}}{0}{\mathfrak{i}}^2 = \frac{\Gamma(\ell)^2\Gamma(\ell\pm 2ik)}{\Gamma(2\ell)}.
\end{equation}}
\begin{equation}
\label{annwl}
\left\langle \mathcal{O}_{\text{LR}}^{\ell}(\tau_L,\tau_R)\right\rangle = \frac{1}{Z}\int dk k\sinh 2\pi k \frac{\Gamma(\ell)^2\Gamma(\ell\pm 2ik)}{\Gamma(2\ell)}e^{-k^2 (\beta_L+\beta_R)},
\end{equation}
the numerator indeed reduces to \eqref{pfann} in the $\ell \to 0$ limit. Continuing to real-time is trivial and in terms of the time-ordered and anti-time-ordered two-point correlators $G_\ell^{\pm}(t_{L},t_R) \equiv \left\langle \mathcal{O}_{\text{LR}}^{\ell,\pm}(t_L,t_R)\right\rangle$, one readily has:
\begin{equation}
G_\ell^+(t_{L},t_R) - G_\ell^-(t_{L},t_R)= \left\langle \left[\mathcal{O}_L^\ell(t_L),\mathcal{O}_R^\ell(t_R)\right]\right\rangle = 0.
\end{equation}
The dual spacetime is connected since the correlator \eqref{annwl} is non-zero, but no communication can occur between both boundaries.
\\~\\
Similarly, a Wilson line correlator with both endpoints on the same boundary, taking again $\slr$ on the inner boundary, is:
\begin{equation}
\label{annslr2}
\left\langle \mathcal{O}^{\ell}(\tau_1,\tau_2)\right\rangle = \frac{1}{Z}\int d k_1^2 \sinh 2\pi k_1 \int dk_2^2 \sinh 2\pi k_2 \, \tj{k_1}{\ell}{k_2}{\mathfrak{i}}{0}{\mathfrak{i}}^2 e^{-k_1^2(\beta_L + \beta_R-\tau_{21})}e^{-k_2^2\tau_{21}}.
\end{equation}
Taking the infinite redshift $\beta_L \to 0$ limit, the result is the same as the Schwarzian disk computation, demonstrating exterior observables are insensitive to the precise microphysics in the edge sector. This conclusion is readily generalized to arbitrary correlation functions, and is qualitatively the same conclusion as that obtained in dynamical theories such as Maxwell in arbitrary dimensions \cite{paper1}. Taking instead a Schwarzian to live on the inner boundary, one finds
\begin{equation}
\label{annschw}
\left\langle \mathcal{O}^{\ell}(\tau_1,\tau_2)\right\rangle = \frac{1}{Z}\int d k_1 \int d k_2 k_2\sinh 2\pi k_2 \frac{\Gamma(\ell \pm i k_1 \pm i k_2)}{\Gamma(2\ell)}e^{-k_1^2(\beta_L + \beta_R -\tau_{21})}e^{- k_2^2 \tau_{21}}.
\end{equation}
These kinds of computations can be readily generalized to multi-boundary Euclidean JT configurations, we provide an example in Appendix \ref{app:multi}.

\subsubsection*{A Liouville Perspective}\label{s:liou}
\begin{figure}[h]
\centering
\includegraphics[width=0.5\textwidth]{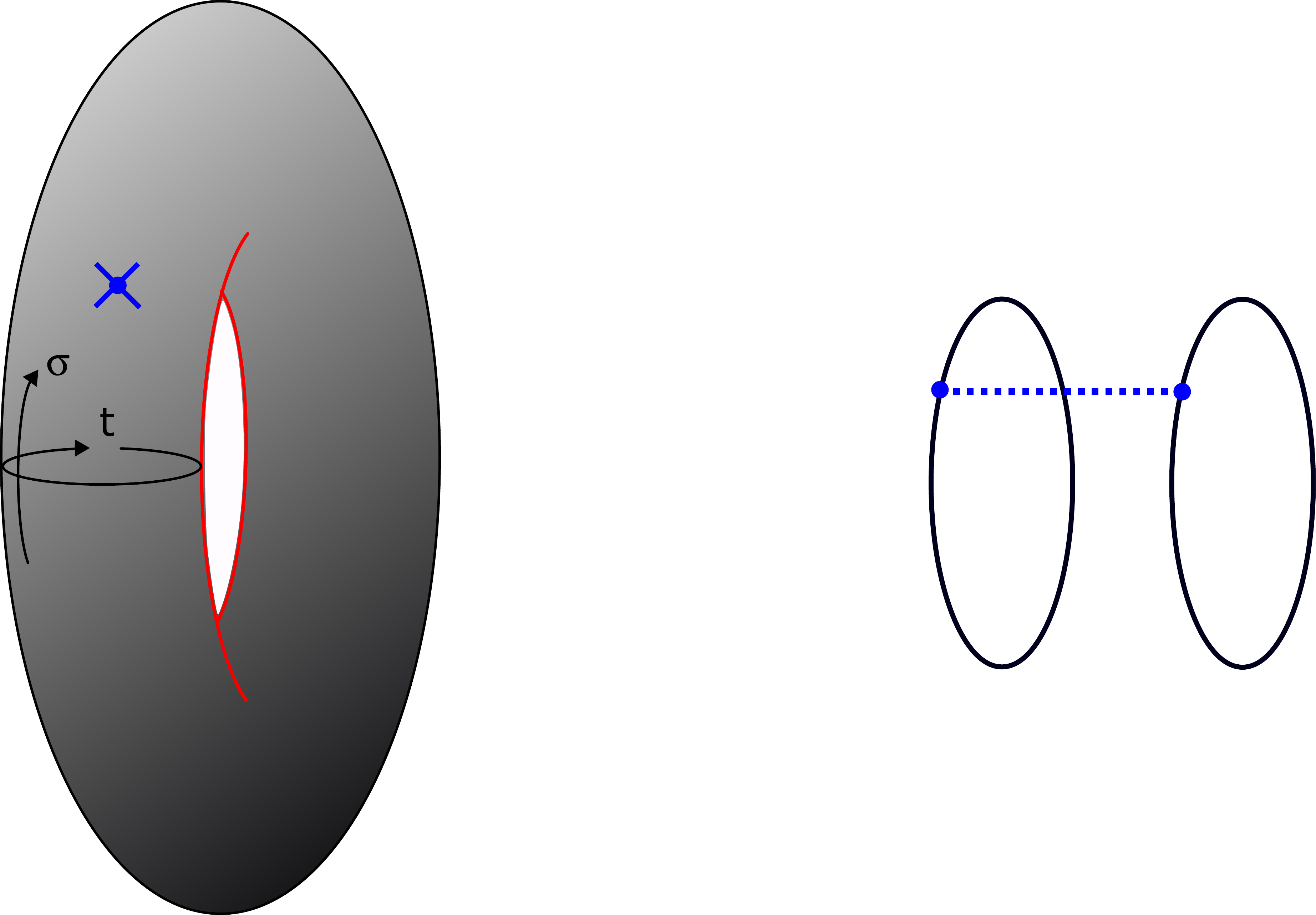}
\caption{Left: Liouville torus one-point function. Right: In the double scaling limit we end up with two Schwarzian-like systems, one from the holomorphic and one from the anti-holomorphic sector. They interact through the integral over $\lambda$ and operator insertions.}
\label{wormhcorr}
\end{figure}

We demonstrate here that as an alternative to the BF calculations, JT correlators of the type \eqref{wormcorr} can alternatively be obtained by taking the Schwarzian double-scaling limit of Liouville CFT on a torus surface. Insertions of Liouville primary vertex operators then correspond to the Schwarzian wormhole-crossing bilocals \eqref{bilocalworm}. This is a direct generalization of the argument used in \cite{schwarzian, origins} where Schwarzian disk correlators were obtained by taking the Schwarzian double-scaling limit on Liouville on the cylinder between ZZ-branes.
\\~\\
The Liouville torus partition function is well-known \cite{Seiberg:1990eb}:
\begin{equation}
\label{Lvpf}
Z(\tau) = V_\phi \int_{0}^{+\infty} dP \left|\chi_P(\tau)\right|^2, \qquad \chi_P(\tau) = \frac{q^{P^2}}{\eta(\tau)}, \qquad q=e^{2\pi i \tau},
\end{equation}
and is identical to that of a 2d free boson due to the KPZ scaling law \cite{Knizhnik:1988ak}.\footnote{The volume factor $V_\phi$ is interpreted as the length of the $\phi$-direction, when interpreting Liouville theory as the target space in string theory. We drop it here for convenience, but it can be tracked more carefully as the zero-mode twist of the fields $f^{\LL}$ and $f^{\RR}$ introduced in \eqref{GNtrans}. It can equally be explained in the Schwarzian limit by the precise gluing measure one uses when gluing disks together. We explore this further in Appendix \ref{app:liou}.} It famously contains only the continuous Virasoro primaries at 
\begin{equation}
h= Q^2/4 + P^2, \qquad P \in \mathbb{R}^+, \qquad Q=b+b^{-1}, \qquad c = 1 + 6Q^2,
\end{equation}
with the vacuum $h = 0$ being left out, a well-known argument against a gravity dual of Liouville CFT. It is modular invariant since
\begin{equation}
\int_{0}^{+\infty} d P\, S_{P_1}^{\,  P} \, S_P^{\, P_2} = \int_{0}^{+\infty} d P \cos(4\pi P P_1)\cos(4\pi P P_2) = \frac{1}{2} \, \delta(P_1-P_2).
\end{equation}
We will reproduce this partition function \eqref{Lvpf} from the Liouville path integral perspective, by deconstructing it into Virasoro coadjoint orbits. Consider the phase space Liouville path integral on the torus surface:
\begin{equation}
Z(\tau) = \int \left[\mathcal{D}\phi\right]\left[\mathcal{D}\pi_\phi\right]e^{\int dt \int d\sigma \left(i \pi_\phi \dot{\phi} - \mathcal{H}(\phi,\pi_\phi)\right)},
\end{equation}
with the Liouville Hamiltonian:
\begin{equation}
\label{HamLiou}
\mathcal{H}(\phi,\pi_\phi) = \frac{1}{8\pi b^2}\left(\frac{\pi_\phi^2}{2} + \frac{\phi_\sigma^2}{2} + e^{\phi}\right).
\end{equation}
We perform the following field redefinition from $(\phi, \pi_\phi)$ into $(f^{\LL},f^{\RR})$:\footnote{This is a slight variant of the one first introduced by Gervais and Neveu in a canonical framework \cite{Gervais:1981gs,Gervais:1982nw,Gervais:1982yf,Gervais:1983am}, see also \cite{Henneaux:1999ib,origins}. Notation: $f' \equiv \partial_\sigma f$.}
\begin{align}
\label{GNtrans}
e^{\phi} &= -2 \frac{f^{\LL '} f^{\RR '}}{\sinh\left(\frac{f^{\LL}-f^{\RR}}{2}\right)^2}, \\
\label{GNtrans2}
\pi_\phi &= \frac{f^{\LL ''}}{f^{\LL '}} - \frac{f^{\RR ''}}{f^{\RR '}}- \coth\left(\frac{f^{\LL}-f^{\RR}}{2}\right)(f^{\LL '}+f^{\RR '}),
\end{align}
in terms of fields $f^{\LL}$ and $f^{\RR}$, which are quasiperiodic in the sense:
\begin{equation}
\label{bcl}
f^{\LL}(\sigma + 2 \pi,t) = f^{\LL}(\sigma,t) + 2\pi\lambda, \quad f^{\RR}(\sigma + 2 \pi,t) = f^{\RR}(\sigma,t) + 2\pi\lambda,
\end{equation}
with $\lambda$ labeling orbits or conjugacy class elements.\footnote{One can appreciate the appearance of this extra parameter $\lambda$ by noting that \eqref{GNtrans} and \eqref{GNtrans2} describe periodic Liouville fields $\phi$ and $\pi_\phi$ for any value of $\lambda$. This parameter should hence be included in the phase space description of the theory. This is analogous to what happens in compact WZW theories \cite{Falceto:1992bf, origins, Henneaux:1999ib}.} The path integral over $\phi$ and $\pi_\phi$ is replaced by a path integral over $f^{\LL}$ and $f^{\RR}$ as well as an integral over $\lambda$, since $\lambda$ labels physically inequivalent configurations:
\begin{equation}
\int \left[\mathcal{D} \phi \right]\left[\mathcal{D} \pi_\phi\right] \quad \to \quad \int \left[\mathcal{D} f^{\LL} \right]\left[\mathcal{D} f^{\RR} \right]\int_{0}^{+\infty} d\lambda,
\end{equation}
with the unit measure on the space of conjugacy class elements (see Appendix \ref{ss:ambiguities}). The Jacobian in this transformation follows from the Pfaffian of the symplectic form. It was computed in this setup explicitly in \cite{origins} and will not be written explicitly here. The Hamiltonian \eqref{HamLiou} is transformed into:\footnote{There is a renormalization effect here that should be found by treating the Liouville determinant more carefully. We have effectively set $c = 6/b^2$, which is the classical result. Tracking this effect more carefully will not bother us here, as we are interested in the Schwarzian double scaling limit that includes $c\to +\infty$.}
\begin{align}
\label{hamil}
\mathcal{H} = -\frac{c}{24\pi}\left\{\tanh\frac{f^{\LL}}{2},\sigma\right\} -\frac{c}{24\pi} \left\{\tanh\frac{f^{\RR}}{2},\sigma\right\}.
\end{align}
Rescaling the fields as $f^{\LL} \to \lambda f^{\LL}$ and $f^{\RR} \to \lambda f^{\RR}$, one finds that the Liouville path integral \eqref{Lvpf} decomposes into a diagonal sum (integral) over coadjoint orbit actions:\footnote{There is a common $U(1)$ redundancy in the field redefinition \eqref{GNtrans} and \eqref{GNtrans2}, $f^{\LL, \RR} \to f^{\LL, \RR} + \alpha$, so the integration space is
\begin{equation}
I = \frac{\text{diff} S^1_{\text{L}} \otimes \text{diff} S^1_{\text{R}}}{U(1)}.
\end{equation}}
\begin{equation}
\label{newpi}
Z(\tau) = \int_{0}^{+\infty} d\lambda \int_I \, \left[\mathcal{D} f^{\LL}\right] \left[\mathcal{D} f^{\RR} \right] \, e^{-S[f^{\LL}]-S[f^{\RR}]},
\end{equation}
with \cite{alekseev1,alekseev2}
\begin{equation}
\label{newgeoma}
S[f] = \int dt \int_{-\pi}^{\pi} d\sigma  \left(i\left[\frac{c}{48\pi}\frac{\dot{f}}{f'}\left(\frac{f'''}{f'}- 2 \left(\frac{f''}{f'}\right)^2 \right) - b_0 \dot{f}f'\right] - \frac{c}{12\pi}\left\{\tanh \frac{\lambda f}{2},\sigma\right\}\right),
\end{equation}
and the orbit parameter $b_0 = \left(\frac{2\pi }{\beta} \frac{c}{24\pi} \lambda\right)^2$.
\\~\\
In the double-scaling Schwarzian limit of interest, one takes the central charge $c \sim 1/b\to +\infty$ along with the circumference in the $t$-direction to go to zero, keeping the product fixed (for more details see \cite{schwarzian, origins}). This eliminates the $ \pi_\phi \dot{\phi}$ term in the action (the term in square brackets in \eqref{newgeoma}), and leaves only the Hamiltonian \eqref{hamil}. Setting $\sigma \to \tau$, this reduces precisely to \eqref{zsch2}.\footnote{In \cite{origins}, this system was studied between ZZ-branes. The latter are dealt with with the doubling trick, combining the $f^{\LL}$- and $f^{\RR}$- degrees of freedom into a single periodic field $F$, directly reproducing the Virasoro vacuum character. Changing branes amounts to changing the character to any Liouville primary of interest.} It furthermore follows that the field redefinition \eqref{GNtrans} maps Liouville vertex operators $e^{2\ell\phi}$ to Wilson lines stretched between the two asymptotic boundaries \eqref{bilocalworm}.
\\~\\
We conclude that the  Schwarzian limit of Liouville torus correlation functions compute correlation functions of the type \eqref{wormcorr}. The two Schwarzian sectors interact indirectly through modular invariance of the torus, and directly by bilocal operator insertions (Figure \ref{wormhcorr}).
\\~\\
An immediate check is on the partition function itself. The Liouville torus partition function \eqref{Lvpf} reduces to the JT gravity partition function \eqref{pfann} in the Schwarzian limit.\footnote{Note that the absence of the $\sinh$-measure here is in direct unison with the flat measure on Liouville theory itself.}
\\~\\
For the correlation function \eqref{annwl}, one can use the $q\to 0$ Schwarzian double-scaling limit of the torus conformal block expansion of the one-point function ($V_\ell(z,\bar{z}) = e^{2 \ell \phi}$) \cite{Fateev:2009me,Hadasz:2009db,Hadasz:2009sw}:
\begin{align}
\left\langle V_\ell \right\rangle_{\tau} &= \sum_{\text{primaries}}\left\langle h_s \right| V_\ell \left| h_s \right\rangle \left|\mathcal{F}_s(q)\right|^2, \qquad h_s = Q^2/4 + P_s^2, \qquad q = e^{2\pi i \tau} \nonumber \\
&= \int \frac{dP_s}{2i}\left|q^{-P_s^2/4}\eta(q)^{-1}H_{P_s,\ell}(q)\right|^2 C(-P_s,\ell,P_s)
\end{align}
As in \cite{schwarzian}, setting $P_s = b k_s$, the block $H_{P_s,\ell}(q)$ reduces in this limit to the primary propagation and the DOZZ coefficient $C(-P_s,\ell,P_s)$ then precisely yields \eqref{annwl}. The independence of the correlator on the bilocal times $\tau_1$ and $\tau_2$ originates in this language from the independence on the location of the Liouville primary vertex operator $V_\ell$. Generalizations to multiple such insertions is then straightforward using Liouville techniques by inserting complete sets of Liouville states and reducing all conformal blocks to primary propagation as in \cite{schwarzian}. Within our choice of variables, the torus conformal blocks are graphically:
\begin{align}
&\begin{tikzpicture}[scale=0.5, baseline={([yshift=0cm]current bounding box.center)}]
\draw[thick] (0,0) circle (1.5);
\draw[thick,blue] (1.5,0) -- (3.5,0);
\draw[fill,black] (1.5,0) circle (0.1);
\draw[fill,black] (3.5,0) circle (0.1);
\draw (2.5,1) node {\small $\ell$};
\draw (0,2) node {\small $k$};
\end{tikzpicture}
&\begin{tikzpicture}[scale=0.5, baseline={([yshift=0cm]current bounding box.center)}]
\draw[thick] (0,0) circle (1.5);
\draw[thick,blue] (1.5,0) -- (3.5,0);
\draw[thick,blue] (-1.5,0) -- (-3.5,0);
\draw[fill,black] (1.5,0) circle (0.1);
\draw[fill,black] (3.5,0) circle (0.1);
\draw[fill,black] (-3.5,0) circle (0.1);
\draw (2.5,1) node {\small $\ell_1$};
\draw (-2.5,1) node {\small $\ell_2$};
\draw (0,2) node {\small $k_1$};
\draw (0,-2) node {\small $k_2$};
\end{tikzpicture}
\hspace{2cm}
\hdots 
\end{align}
Notice that calculating bilocals with endpoints on the \emph{same} asymptotic boundary seems to be impossible within the Liouville language. In that respect, the BF formulation of JT gravity developed above and in \cite{origins,paper3} is more versatile.

\section{Discussion}
We summarize the main lessons learned about the BF structure of JT gravity:
\begin{itemize}
    \item JT quantum gravity is precisely equal to an $\slr$ BF theory with coset boundary constraints. The ubiquitous $\sinh 2\pi \sqrt{E}$ density of states in the theory is simply the Plancherel measure of $\slr$. For almost all purposes, neither the fact that $\slr$ is noncompact, nor the fact that it is only a subsemigroup affect any of the diagrammatic rules for constructing BF amplitudes. The gravitational boundary conditions can be viewed as a coset construction in the BF language.
    \item In Appendix \ref{app:multi} it is explained how to calculate JT gravity amplitudes on manifolds with handles or multiple boundaries. One goes about this by isolating punctured disks with Schwarzian boundaries from the remaining amplitude, using known results for both\footnote{The calculation of the punctured disks follows from this work and \cite{paper3}, the topological amplitude were discussed in \cite{dijkgraafwitten}.} and then gluing the pieces back together.\footnote{This was simultaneously investigated in more detail in \cite{sss2}.} An important subtlety arises in these calculations that can be tracked back to the noncompactness of the group. Depending on the chosen integration space of geometries, BF calculations on manifolds with handles or more than two boundaries may diverge \cite{dijkgraafwitten}. In particular, on such higher genus surfaces, the volumes of Teichm\"uller space $\mathcal{T}$ diverge. To obtain a finite result one should mod by the mapping class group and integrate over the moduli space of Riemann surfaces $\mathcal{M}$ \cite{dijkgraafwitten}. On the disk, which was the main interest of this work, these are identical. We detail some of this story in Appendices \ref{app:C}.\footnote{It is amusing to note that both integration spaces over geometries can seemingly be reached when we think of JT gravity as arising as the low energy limit of Liouville on the same bulk surface. Quantum Liouville theory as we know it from CFT is like a quantum theory for Teichm\"uller space \cite{techner}, whilst the Liouville theory that pops up in the minimal string (see for example in \cite{sss2}) is more like a quantum theory for the moduli space of Riemann surfaces $\mathcal{M}$. The latter is dual to a matrix model \cite{sss2}, the former is not. A discussion on quantum Liouville on the disk and how JT gravity on the disk arises in a nearly-classical limit is coming soon \cite{paper6}.}
\end{itemize}
Whereas we believe we have amassed convincing evidence in favor of $\slr$, it would be good if more could be acquired. 
\\~\\
In the second part of this work we investigated edge dynamics and entanglement in JT gravity. Let us summarize the results.
\begin{itemize}
    \item By cutting the JT path integral on a given manifold we learned that an $\slr$ quantum mechanics lives on all entangling boundaries, whereas the asymptotic boundaries are described by Schwarzian quantum mechanics.
    \item From the perspective of a Rindler observer, the $\slr$ quantum mechanics on the horizon is frozen due to infinite redshift. Its degrees of freedom can be used to represent the JT black hole (or one-sided) states and account for the Bekenstein Hawking entropy \cite{lin}. Alternatively these new degrees of freedom simply arise in the factorization of a BF state on an interval into smaller intervals.\footnote{See also the very recent work \cite{Donnelly:2018ppr}.} The extended Hilbert space associated with the resulting subregion includes the edge states \cite{paper1,paper2,donnellywall,donnellyfreidel} or black hole states. We emphasize again that this is a description of the relevant states, but does not constitute what one would call a microscopic counting of the black hole entropy starting with a discrete counting problem. This is a problem beyond the reach of pure gravity.
\end{itemize}
Finally, we discussed JT gravity on a manifold with two Schwarzian boundaries, where the full path integral of the system can be written in terms of Schwarzian quantum mechanics on both boundaries. The resulting theory is identical to the double-scaling Schwarzian limit of the full Liouville path integral. This identification is strengthened by the fact that amplitudes of wormhole-crossing Wilson lines match with the double-scaling limit of Virasoro torus conformal block expansions. Besides providing an alternative perspective on JT amplitudes, this provides the torus conformal block literature \cite{Fateev:2009me,Hadasz:2009db,Hadasz:2009sw} with an interesting limit, and connects it to the SYK literature.
\\
This may come as somewhat of a surprise. Though Virasoro coadjoint orbit models are the building blocks of 3d quantum gravity, the role of full-fledged Liouville theory in 3d quantum gravity is less clear \cite{3dgrav1,3dgrav2,maloneywitten,Barnich:2017jgw,jensen}. However, in the double-scaling limit, full Liouville CFT is relevant for two-sided geometries.
\\~\\
We end with some speculation about entanglement and black hole entropy in 3d pure gravity. We saw in Appendix \ref{app:CS} that the partition function for CS theory in a Rindler wedge $\times S^1$ was just calculating the solid torus amplitude $\chi_0(S\cdot \beta)$. Accordingly, to compute the partition function for 3d gravity (which consists roughly of two copies of $\sltr$ CS of opposite chirality), we would naively write:
\begin{equation}
\left|\chi_0(S \cdot \beta)\right|^2,
\end{equation}
in terms of the Virasoro vacuum character. The resulting density of states is $\rho(\lambda,\lambda') = \tensor{S}{_0^\lambda} \tensor{S}{_0^{\lambda'}}$, which is the expression written down in \cite{McGough:2013gka} and which matches the semiclassical BTZ black hole entropy. A Hilbert space interpretation in terms of one-sided states along the lines of \eqref{tfdfac} is less obvious. 
For compact cosets $G/H$ the conclusion would be that a frozen $G$ WZW model lives on the horizon and accounts for the edge states. The precise statement in the gravity case certainly deserves further study.

\section*{Acknowledgements}
We thank L. Iliesiu, E. Mazenc and G.J. Turiaci for discussions. AB and TM gratefully acknowledge financial support from Research Foundation Flanders (FWO Vlaanderen).

\appendix
\section{BF Amplitudes}
\label{s:pgdiag}
We review the Feynman rules for correlation functions of boundary-anchored Wilson lines in BF \cite{origins,paper3}.

\begin{itemize}
\item Draw a disk with the Wilson line insertions. 
\item Each disk-shaped region is assigned an irrep $R_i$, and contributes a weight $\dim R_i$. A label $m_i$ denoting eigenvalues of a maximal set of commuting generators is assigned to each boundary segment. One sums over these labels $R_i$ and $m_i$ to obtain the amplitude.
\item Each boundary segment carries a Hamiltonian propagation factor proportional to the length $L_i$ of the relevant segment $i$ (depending on the chosen einbein). Each intersection of an endpoint of a Wilson line with the boundary is weighted with a $3j$-symbol.
\begin{align}
\label{frules}
\begin{tikzpicture}[scale=0.8, baseline={([yshift=-0.1cm]current bounding box.center)}]
\draw[thick] (-0.2,0) arc (170:10:1.53);
\draw[fill,black] (-0.2,0.0375) circle (0.1);
\draw[fill,black] (2.8,0.0375) circle (0.1);
\draw (3.4, 0) node {\footnotesize $\tau_1$};
\draw (-0.7,0) node {\footnotesize $\tau_2$};
\draw (1.25, 1.6) node {\footnotesize $\textcolor{red}{m}$};
\draw (1.25, 0.5) node {\footnotesize $R$};
\draw (5, 0) node {$\raisebox{10mm}{$\qquad\qquad= \exp{- (\tau_2-\tau_1) \mathcal{C}_R},$}$};
\end{tikzpicture}
\end{align}
\begin{align}
\begin{tikzpicture}[scale=1, baseline={([yshift=-0.1cm]current bounding box.center)}]
\draw[thick] (-.2,.9) arc (25:-25:2.2);
\draw[fill,black] (0,0) circle (0.08);
 \draw[thick,blue](-1.5,0) -- (0,0);
\draw (.5,-1) node {\footnotesize $\textcolor{red}{m_2}$};
\draw (.5,1) node {\footnotesize$\textcolor{red}{m_1}$};
\draw (-0.3,.3) node {\footnotesize$\textcolor{red}{m}$};
\draw (-1,.3) node {\footnotesize$ \color{blue}R$};
\draw (-1,.8) node {\footnotesize$ R_1$};
\draw (-1,-.5) node {\footnotesize$ R_2$};
\draw (3,0.1) node {$\raisebox{-10mm}{$\ =\  \, \tj{R_1}{R_2}{R}{m_1}{m_2}{m}.$}$}; \end{tikzpicture}\ \label{pg3j}
\end{align}
\item A Wilson line crossing in the bulk comes with a $6j$-symbol of the group.
\begin{align}
\label{crossing}
\ \begin{tikzpicture}[scale=1, baseline={([yshift=0cm]current bounding box.center)}]
\draw[thick,blue] (-0.85,0.85) -- (0.85,-0.85);
\draw[thick,blue] (-0.85,-0.85) -- (0.85,0.85);
\draw[dotted,thick] (-0.85,-0.85) -- (-1.25,-1.25);
\draw[dotted,thick] (0.85,0.85) -- (1.25,1.25);
\draw[dotted,thick] (-0.85,0.85) -- (-1.25,1.25);
\draw[dotted,thick] (0.85,-0.85) -- (1.25,-1.25);
\draw (1.5,0) node {\scriptsize $R_4$};
\draw (-1.5,0) node {\scriptsize $R_2$};
\draw (-.75,.33) node {\scriptsize \color{blue}$R_A$};
\draw (.78,.33) node {\scriptsize \color{blue}$R_B$};
\draw (0,1.5) node {\scriptsize  $R_1$};
\draw (0,-1.5) node {\scriptsize $R_3$};
\end{tikzpicture}~~\raisebox{-3pt}{$\ \ \  = \ \ \sj{R_B}{R_1}{R_4}{R_A}{R_3}{R_2}$}~~~
\end{align}

\end{itemize}
\subsection{Coset Slicing}
\label{app:slicing}
\begin{figure}[h]
\centering
\includegraphics[width=0.98\textwidth]{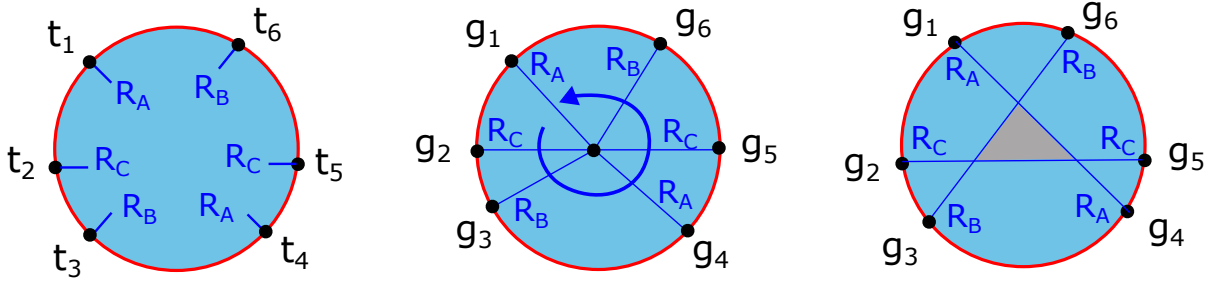}
\caption{Left: Particle-on-a-group evaluation of a six-point function of three bilocals inserted at times $t_i$, $i=1\hdots 6$. Middle: Bulk BF angular slicing corresponding to the same computation. Right: Different representation of the same amplitude, where now the bulk BF diagram contains an enclosed region that does not know about the coset constraints.}
\label{PoGpaper4}
\end{figure}

We demonstrate next that the slicing of coset models can be identified with angular slicing in the BF model directly.
\\~\\
In \cite{paper3} we computed a generic correlation function directly within the particle-on-a-group model by inserting complete sets of states in between all operator insertions. E.g. for three bilocals, 
\begin{equation}
\text{Tr}\left[e^{-(\beta-t_{61}) H} \mathcal{O}_A e^{-t_{12}H}\mathcal{O}_B e^{-t_{23}H}\mathcal{O}_Ce^{-t_{34}H}\mathcal{O}_Ae^{-t_{45}H}\mathcal{O}_Be^{-t_{56}H} \mathcal{O}_C \right],
\end{equation}
one inserts complete sets of $\mathbf{1} = \int dg_i \, \left|g_i\right\rangle\left\langle g_i\right|$, $i=1\hdots 6$ in between all legs of operators, followed by complete sets of $\left|R,a,b\right\rangle$ to diagonalize the Hamiltonian propagation factors $e^{-t_{ij}H}$. The computation can then be manifestly identified with a computation in BF in angular slicing (Figure \ref{PoGpaper4} left and middle) \cite{paper3}. 
This identification immediately extends to coset constructions. Denoting a coset element as $x \in G/H$, the completeness relation on $G/H$ can be rewritten as:
\begin{equation}
\mathbf{1} = \int_{G/H} dx \left|x\right\rangle\left\langle x\right| = \frac{1}{\text{Vol }H}\int_{G} dg \left|g\right\rangle\left\langle g\right|.
\end{equation}
Introducing complete sets in coordinate space $x$ like this, we can use precisely the same construction as above to get the generic correlator.
\\~\\
In \cite{paper3}, we explained that these pie-shaped bulk diagrams may be freely deformed into diagrams with enclosed regions (see e.g. Figure \ref{PoGpaper4} right). In particular, it can be shown that enclosed interior regions obtained in this manner are to be weighted with $\dim R$ coming from the $G$ parent theory; the interior of the disk does not know about the modding by $H$.

\subsection{Examples}
To illuminate the more abstract discussion of section \ref{sect:coset} we work out two examples.

\subsubsection{Quantum Mechanics on $SU(2)/U(1)$}
\label{app:sphere}
As an instructive example that is interesting in its own right, we consider the right coset of $SU(2)$ by $U(1)$ that yields the 2-sphere $S^2$. The $SU(2)$ manifold can be parameterized by Euler angles ($\theta,\phi,\psi$):
\begin{equation}
\label{Euler}
g =  e^{i \frac{\phi}{2}\sigma_3} e^{i \frac{\theta}{2}\sigma_1} e^{i \frac{\psi}{2}\sigma_3} = \left(\begin{array}{cc}
\cos\frac{\theta}{2}e^{\frac{i}{2}(\phi+\psi)} & i\sin\frac{\theta}{2}e^{\frac{i}{2}(\phi-\psi)} \\
-i\sin\frac{\theta}{2}e^{-\frac{i}{2}(\phi-\psi)} & \cos\frac{\theta}{2}e^{-\frac{i}{2}(\phi+\psi)} \\
\end{array}\right),
\end{equation}
with $\sigma_i, \, i=1,2,3$ the Pauli matrices. 

Choosing $A^{3}\atbdy = \chi^3 \atbdy = 0$ and $A^{1,2}\atbdy=\chi^{1,2}\atbdy$, we obtain the Lagrangian:
\begin{equation}
\left.\Tr(\gt)^2\right|_{\text{restricted}} = (g^{-1}\partial_t g)^{1} (g^{-1}\partial_t g)^{1} + (g^{-1}\partial_t g)^{2} (g^{-1}\partial_t g)^{2} = -\frac{1}{2}\left(\partial_t^2 \theta + \sin(\theta)^2 \partial_t^2 \phi\right),
\end{equation}
which is the action of a particle on $S^2 \simeq SU(2)/U(1)$.
\\~\\
The partition function is \eqref{twistpcoset}:
\begin{equation}
Z_{S^2}(\beta) = \sum_j (2 \text{j} +1)  \, e^{- \beta \text{j}(\text{j}+1)},
\end{equation}
which indeed matches the spectrum of the rigid rotor quantum mechanical system. The matrix elements $R_{ab}(g)$ of $SU(2)$ are given by the Wigner D-functions $D^j_{m,m'}(\theta,\phi,\psi)$. For each irrep, there is precisely one state right-invariant under $J_3$: the $m=0$ state. The spherical function basis therefore consists of the spherical harmonics:
\begin{equation}
\left\langle \theta,\phi\right| \left.j ,m,0 \right\rangle = \bra{j, m} g(\theta,\phi) \ket{j, 0} = Y^j_m (\theta,\phi),
\end{equation}
and the zonal spherical function is the Legendre function:
\begin{equation}
\left\langle \theta \right| \left.j ,0,0 \right\rangle = \bra{j, 0} g(\theta,\phi) \ket{j, 0} = P_j(\cos\theta).
\end{equation}
Using these, we can e.g. write down the correlator with a single boundary-anchored Wilson line:
\begin{align}
&\begin{tikzpicture}[scale=1, baseline={([yshift=0cm]current bounding box.center)}]
\draw[thick,red] (0,0) circle (1.5);
\draw[thick,blue] (-1.5,0) -- (1.5,0);
\draw[fill,black] (-1.5,0) circle (0.1);
\draw[fill,black] (1.5,0) circle (0.1);
\draw (0,1.7) node {\small \color{red}$0$};
\draw (1.2,0.2) node {\small \color{red}$m$};
\draw (-1.2,0.2) node {\small \color{red}$\bar{m}$};
\draw (0,-1.7) node {\small \color{red}$0$};
\draw (-2,0) node {\small $\tau_2$};
\draw (2,0) node {\small $\tau_1$};
\draw (0,.25) node {\small \color{blue}$j$};
\draw (0,1) node {\small $j_1$};
\draw (0,-1) node {\small $j_2$};
\end{tikzpicture} = \Big\langle \mathcal{O}_{j,m\bar{m}}(\tau_1,\tau_2)\Big\rangle \nonumber \\ 
&= \delta_{m0}\delta_{\bar{m}0}\sum_{j_1,j_2}(2j_1+1)(2j_2+1) e^{-\tau_{21}j_1(j_1+1)}e^{-(\beta-\tau_{21})j_2(j_2+1)} \tj{j_1}{j}{j_2}{0}{0}{0}\tj{j_1}{j}{j_2}{0}{0}{0}.\label{pg2pt}
\end{align}
Slicing this amplitude using Cauchy surfaces with both endpoints on the outer boundary, requires using the $R_{00}(\theta)$ zonal spherical functions. Using the angular slicing where only one endpoint touches the boundary, requires using spherical functions $R_{i0}(\theta,\phi)$ instead. Formula \eqref{pg2pt} is obtained using the well-known identities:
\begin{alignat}{2}
&\int_{0}^{\pi} d\theta \sin(\theta) \, P_{j_1}(\cos\theta) P_{j_2}(\cos\theta) P_{j_3}(\cos\theta) &&= \tj{j_1}{j_2}{j_3}{0}{0}{0}^2,\\
&\int_{0}^{\pi} d\theta \sin(\theta) \, Y^{j_1}_{m_1} (\theta,\phi) Y^{j_2}_{m_2} (\theta,\phi) Y^{j_3}_{m_3} (\theta,\phi) &&=  \tj{j_1}{j_2}{j_3}{0}{0}{0} \tj{j_1}{j_2}{j_3}{m_1}{m_2}{m_3}.
\end{alignat}
As explained above, regions that are in the deep interior and closed off from the boundary, see the full $SU(2)$ BF model with matrix elements the Wigner D-functions. 
\\~\\
We can give a complementary perspective on this by looking at the Casimir differential equation. The left- and right regular representation (realization) of the $\mathfrak{su}(2)$ algebra in Euler angles \eqref{Euler}, found by imposing $\hat{D}^a_L g = \tau^a g$ and $\hat{D}^a_R g = g \tau^a$ is given by the sets of differential operators:
\begin{alignat}{4}
&i\hat{D}^1_L &&= \cos\phi {\partial_\theta} + \frac{\sin\phi}{\sin \theta}{\partial_\psi} - \frac{\sin\phi}{\tan\theta}{\partial_\phi}, \qquad &&i\hat{D}^1_R &&= \cos\psi {\partial_\theta} + \frac{\sin\psi}{\sin \theta}{\partial_\phi} - \frac{\sin\psi}{\tan\theta}{\partial_\psi}, \nonumber \\
&i\hat{D}^2_L &&= -\sin\phi {\partial_\theta} + \frac{\cos\phi}{\sin \theta} {\partial_\psi} - \frac{\cos\phi}{\tan\theta}{\partial_\phi}, \qquad &&i\hat{D}^2_R &&= \sin\psi {\partial_\theta} - \frac{\cos\psi}{\sin \theta} {\partial_\phi} + \frac{\cos\psi}{\tan\theta}{\partial_\psi}, \\
&i\hat{D}^3_L &&= {\partial_\phi}, \qquad &&i\hat{D}^3_R &&= {\partial_\psi}. \nonumber
\end{alignat}
The $\mathfrak{su}(2)$ Casimir equation is then directly found as
\begin{align}
\left(\partial_\theta^2 + \cot \theta \partial_\theta + \frac{1}{\sin\theta^2}\left(\partial_\phi^2 -2\cos\theta \partial_\phi \partial_\psi + \partial_\psi^2\right) \right) D^j_{m,m'}(\theta,\phi,\psi) = j (j+1) D^j_{m,m'}(\theta,\phi,\psi),
\end{align}
solved by the Wigner D-functions $D^j_{m,m'}(\theta,\phi,\psi)$. Setting $J^3_R = \hat{D}^3_R= 0$, one finds
\begin{align}
\left(\partial_\theta^2 + \cot \theta \partial_\theta + \frac{1}{\sin\theta^2}\partial_\phi^2\right) Y^l_m (\theta,\phi) = j (j+1) Y^j_m (\theta,\phi),
\end{align}
in terms of the spherical harmonics $Y^j_m (\theta,\phi)$. Additionally setting $J^3_L = \hat{D}^3_L = 0$, one finds
\begin{align}
\left(\partial_\theta^2 + \cot \theta \partial_\theta \right)  P_j(\cos\theta) = j (j+1)  P_j(\cos\theta),
\end{align}
solved in terms of the Legendre functions $P_j(\cos\theta)$. This process of imposing the coset conditions $J^3_R = 0$ and $J^3_L = 0$ is the direct analogue of the gravitational / Liouville constraints discussed in Appendix F of \cite{paper3}. The left- and right-regular representation operators act on the bra, respectively the ket of the matrix element $R_{ab}(g) \equiv \left\langle R,a\right| g \left|R,b\right\rangle$.

\subsubsection{Quantum Mechanics on $\slc$}
\label{app:slc}
As a second instructive example we consider a particle on $\slc$. From \eqref{toproof} we obtain the partition function:
\begin{equation}
Z_{\slc}(\beta)=\int ds \, s^4 \,e^{-\beta (s^2+1/4)}.
\end{equation}
To obtain a basis of the representation, one conventionally diagonalizes the generator $J^3 = m$, or after Fourier transforming to a continuous 2-sphere of labels $(x,\bar{x})$:\footnote{See e.g. \cite{Teschner:1997ft,Teschner:1997fv,Kutasov:1999xu}.}
\begin{equation}
\psi^j_{m,\bar{m}}(g) = \int_{\mathbb{C}} d^2x x^{j+m} \bar{x}^{j+\bar{m}} \psi^j(x,\bar{x},g)
\end{equation}
Within this basis, inserting a single boundary-anchored Wilson line $\mathcal{O}^{s}_{x,\bar{x}}(\tau_1,\tau_2)$ of $\slc$ gives the correlator:
\begin{equation}
\int_{0}^{+\infty}ds_1 ds_2 \int_{\mathbb{C}} d^2x_1 d^2 x_2 \, s_1^4 s_2^4 \, \tj{s_1}{s}{s_2}{x_1}{x}{x_2}\tj{s_1}{s}{s_2}{\bar{x}_1}{\bar{x}}{\bar{x}_2}e^{-s_1^2 \tau_{12}^2 - s_2^2(\beta-\tau_{12})^2},
\end{equation}
where
\begin{align}
\left|\tj{s_1}{s}{s_2}{x_1}{x}{x_2}\right|^2 =& \left|x_2-x\right|^{2(j_2+j-j_1)}\left|x_2-x_1\right|^{2(j_2+j_1-j)}\left|x-x_1\right|^{2(j+j_1-j_2)}\nonumber\\ &\frac{\Gamma(-j_1-j-j_2-1)\Gamma(j_2-j-j_1)\Gamma(j-j_1-j_2)\Gamma(j_1-j-j_2)}{\Gamma(-2j_1-1)\Gamma(-2j-1)\Gamma(-2j_2-1)},
\end{align}
are the well-known $3j$-symbols of $\slc$ \cite{Kerimov:1978wf}, identifiable as conformal three-point functions as recently discussed in \cite{Gadde:2017sjg}.

\section{Moduli Space of Flat $\slr$ Connections}
\label{app:moduli}
\begin{figure}[h]
\centering
\includegraphics[width=0.35\textwidth]{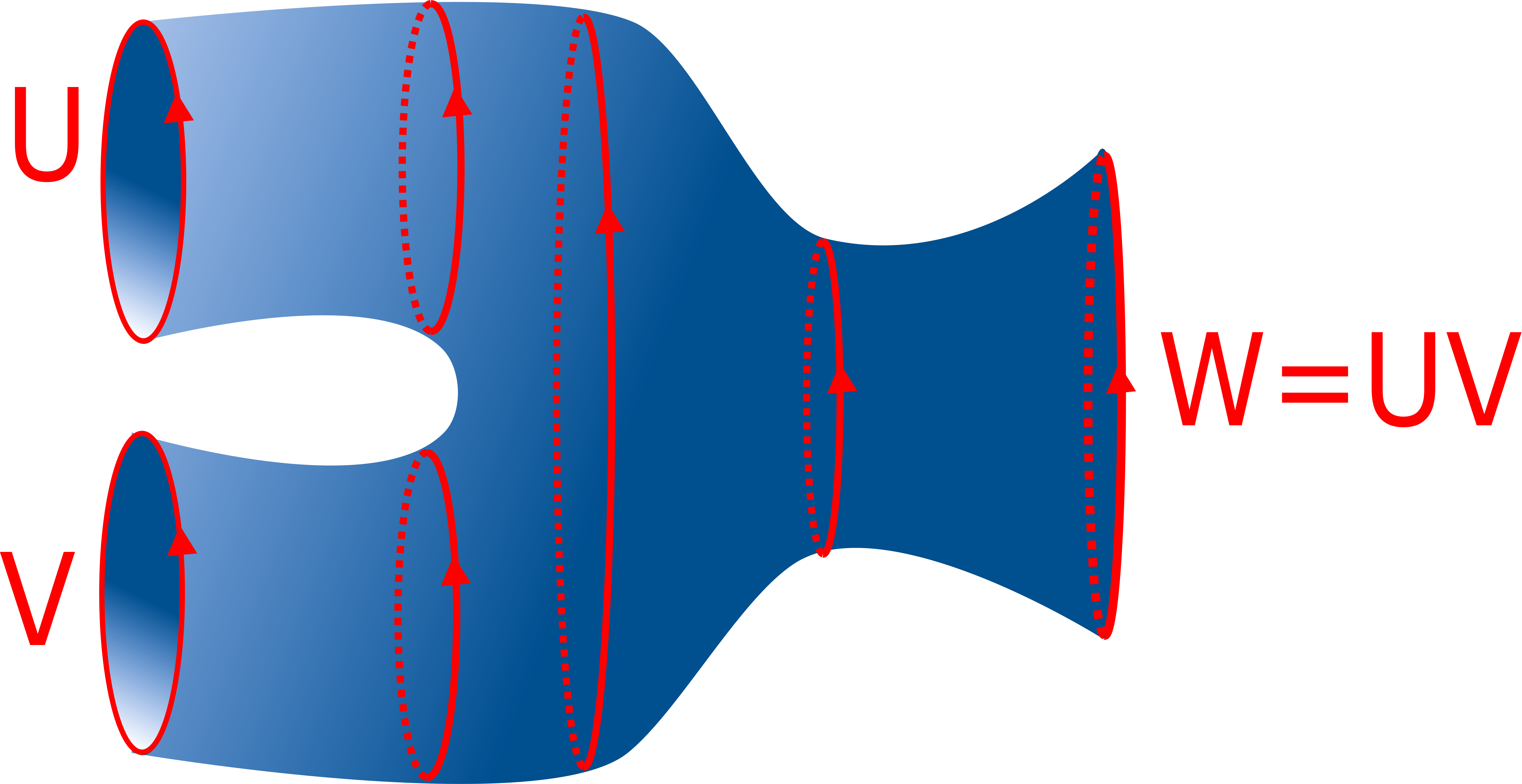}
\caption{Choice of oriented slices on a three-holed sphere, with holonomies $U$, $V$, $W$ restricted by $W= UV$.}
\label{3holesphere}
\end{figure}

We present some arguments here that that the component of the moduli space of flat $\sltr$ connections relevant for hyperbolic geometry, can be identified with the moduli space of flat $\slr$ connections. 
\\~\\
The component of moduli space of flat $\sltr$ connections that can be identified with hyperbolic geometry, are those connections with hyperbolic monodromy around each closed geodesic. As explained in subsection \ref{sect:argue3}, $\slr$ has only hyperbolic conjugacy classes. Therefore, the moduli space of flat $\slr$ connections is a subset of Teichm\"uller space $\mathcal{T}$.
\\~\\
The question is whether this map is also surjective: can we find for each point in Teichm\"uller space a flat $\slr$ connection with the corresponding monodromies? 
\\
We do not have a complete proof for this and test it only in a specific example. Consider the three-holed sphere (Figure \ref{3holesphere}).
The set of flat connections on a surface $\Sigma$ is $A= dg \,g^{-1}$, with holonomies $\mathcal{P} \exp \oint_{\mathcal{C}_i} A = U_i \in G$ around each cycle. An element in the moduli space of flat connections is given in terms of the values of these holonomies around each topologically supported cycle. In other words, the moduli space is $\text{Hom}(\pi_1(\Sigma) \to G) / G$, where an overall $G$-conjugation is modded out. \\
For the example at hand, we have three boundary holonomies $U$, $V$, $W$ satisfying $W= UV$. In $\slr$ it is important to check if one can choose all holonomies $\in \slr$ consistently, corresponding to a choice of oriented slices, while still satisfying the constraint $W=UV$. This is readily realized (Figure \ref{3holesphere}). The question is now whether we can span the set of boundary lengths $(a,b,c) \in (\mathbb{R}^+)^3$, given the constraint $W=UV$.\footnote{Ignoring this constraint, it would be readily true because $\slr$ has all hyperbolic conjugacy classes.} Given any choice of $a,b,c$, let $c$ be the largest one of these. Then we use the overall $G$-conjugation to choose the holonomies as, following \cite{Stanford:2019vob} and corresponding to the specific choice of slices of Figure \ref{3holesphere}:
\begin{equation}
U = \left(\begin{array}{cc}
e^{a/2} & \kappa \\
0 & e^{-a/2} \\
\end{array}\right), \quad V = \left(\begin{array}{cc}
e^{-b/2} & 0 \\
1 & e^{b/2} \\
\end{array}\right),
\end{equation}
with boundary lengths $a$ and $b$ respectively. Hence
\begin{equation}
W = UV = \left(\begin{array}{cc}
e^{(a-b)/2} + \kappa & \kappa e^{b/2}\\
e^{-a/2} & e^{-(a-b)/2} \\
\end{array}\right).
\end{equation}
All of these matrices $\in \slr$ if $\kappa \geq 0$. Given these $U$ and $V$ $\in \slr$, we can reach any boundary length $c$ for $W$. Indeed:
\begin{equation}
2\cosh \frac{c}{2} = 2\cosh\left(\frac{a-b}{2}\right) + \kappa,
\end{equation}
and for any given $a$ and $b$, we can adjust $\kappa \geq 0$ to obtain the prescribed value of the third boundary length $c$.\footnote{For this to work, we need the information that $c \geq a,b$, which implies $c\geq \left|a-b\right|$.}
\\~\\
Any higher-genus Riemann surface can be decomposed into three-holed spheres glued together. This gluing allows the introduction of a relative twist which in this language is the 1-parameter centralizer of the hyperbolic holonomy matrix for each of the $3g-3+b$ geodesic gluing cycles. We can imagine using the above computation then as a basis for a general proof. \\
It is furthermore interesting and reassuring to note that in the mathematics literature, a deep link between positivity properties of monodromy matrices and the hyperbolic/Hitchin component of the moduli space has been uncovered, see e.g. \cite{Fock:2006,Guichard:2018}.
\\~\\
Note that this complete set of monodromies required when gluing surfaces together is unrelated to the type of defects we can insert into the surface. We can add for example conical singularities (elliptic monodromies) in the surface, but they do not appear in gluing integrals. This is in direct analogy to Liouville or quantum Teichm\"uller theory.

\section{From Finite-volume to Delta-regularization}
\label{app:reg}
In the main text, we have been slightly cavalier on the overall volume-factors appearing in our formulas. In this appendix, we track these factors more carefully. We focus on non-compact groups with a continuous set of irreps (such as the continuous irreps of $\sltr$). We will deal with these representations by relating the finite-volume regularization with the delta-regularization, the first well-suited to develop physical intuition, while the latter is mathematically rigorous and links back to the Plancherel measure.
\\~\\
The volume-regularized Schur orthogonality relation
\begin{equation}
\label{Schur}
\int_G dg \, R^k_{ab}(g)R^{k'}_{cd}(g^{-1}) = V \frac{\delta_{kk'}}{\dim k}\delta_{ad}\delta_{bc},
\end{equation}
is transformed into the delta-regularized version:
\begin{equation}
\label{pldef}
\int_G dg \, R^k_{ab}(g)R^{k'}_{cd}(g^{-1}) = \frac{\delta(k-k')}{\rho(k)}\delta_{ad}\delta_{bc},
\end{equation}
related by the formal equality 
\begin{equation}
\frac{\dim k}{V}\, \delta(k-k') = \delta_{kk'}\,\rho(k).\label{formal}
\end{equation}
From \eqref{pldef}, we can also read off the delta-normalized wavefunctions as
\begin{equation}
\left\langle g\right|\left.k,a,b\right\rangle = \sqrt{\rho(k)}R^k_{ab}(g).
\end{equation}
Tracing over the indices in \eqref{Schur}, one finds the character orthogonality in the form:
\begin{equation}
\int_G dg \, \chi^k(g)\chi^{k'}(g^{-1}) = \delta(k-k')\frac{\dim k}{\rho(k)}.
\end{equation}
Restricting to the subgroup of conjugacy class elements $C$, one has instead
\begin{equation}
\label{gluealpha}
\int_{C} d\alpha \, \chi^k(\alpha)\chi^{k'}(\alpha^{-1}) = \delta(k-k'),
\end{equation}
identifying $\dim k/V = \rho(k)/\delta(k-k)$, and hence we have formally $\delta(k-k) = V_C$ as the volume of the space of conjugacy class elements. This is the formal translation of the fact that the space of irrep labels and the space of conjugacy class elements are Fourier duals to each other. Hence:
\begin{equation}
\frac{\dim k}{V}=\frac{\rho(k)}{V_C}.
\end{equation}
The summation over irreps has to transform contragrediently to the Kronecker delta,\footnote{Explicitly: $\sum_R \delta_{RR'} = 1 = \int dk \delta(k-k')$.} so using \eqref{formal}, we find the continuous replacement of the sum:
\begin{equation}
\sum_R\to V_C \int d k.
\end{equation}
Let us apply these equations to some concrete situations.
\begin{itemize}
\item
The Schwarzian partition function, represented as the path integral over $L(G/H)/G \equiv \frac{\text{diff }S^1}{\sltr}$, is 
\begin{equation}
\frac{1}{V}\sum_R \text{dim }R e^{-\beta \mathcal{C}_R} \quad \to\quad \int dk \rho(k) e^{-\beta \mathcal{C}_k}.
\end{equation}
\item
The twisted Schwarzian partition function, $L(G/H)/T \equiv \frac{\text{diff }S^1}{U(1)}$, with holonomy $\alpha$, is
\begin{equation}
\label{schtwi}
\frac{1}{V_C}\sum_R \chi_k(\alpha) e^{-\beta \mathcal{C}_R} \quad \to\quad \int dk \, \chi_k(\alpha) \, e^{-\beta \mathcal{C}_k}.
\end{equation}
Two such twisted partition functions are glued by using \eqref{gluealpha}, and give $\int dk e^{-\beta \mathcal{C}_k}$. \\
Taking $\alpha = \mathbf{1}$, one finds\footnote{Note that for $G = \mathbb{R}$, $V=V_C$ and $\rho(k)=1$ indeed. For other groups $V/V_C$ diverges, as one expects.}
\begin{equation}
\boxed{
\chi_k(\mathbf{1}) = \dim k = \frac{V}{V_C} \rho(k)},
\end{equation}
which is the more precise way of stating \eqref{charone}.
\item The twisted partition function of a particle on $\slr$, LG/T, or alternatively the propagator on $\slr$ between the point $\mathbf{1}$ and $\alpha$, is given by
\begin{equation}
\frac{1}{V}\sum_R \text{dim R} \chi_R(\alpha) e^{-\beta \mathcal{C}_R} \quad \to\quad \int dk \rho(k) \chi_k(\alpha) e^{-\beta \mathcal{C}_k}.
\end{equation}
Gluing this to \eqref{schtwi}, one indeed finds back $ \int dk \rho(k) e^{-\beta \mathcal{C}_k}$.
\item
The partition function of a particle on $\slr$, LG/G, is then
\begin{equation}
\frac{1}{V}\sum_R (\text{dim R})^2 e^{-\beta \mathcal{C}_R} \quad \to\quad \frac{V}{V_C}\int dk \rho(k)^2 e^{-\beta \mathcal{C}_k},
\end{equation}
formally divergent for any noncompact non-abelian group, due to the volume factors appearing upfront. Nonetheless, they are multiplicative prefactors in this language, and we can divide them out to define a sensible model.
\end{itemize}

\section{Gluing Measures}\label{app:C}
In \textbf{section \ref{app:glueBF}}, we provide some details on gluing and twists in BF that were left implicit in the main text in favor of readability.\\
In \textbf{section \ref{app:glueJT}} we argue that a similar story in JT gravity is the backbone of the difference between two possible integration spaces: Teichm\"uller space, or the moduli space of Riemann surfaces. This is the crux of the discrepancy between two different formulas for the annulus partition function in JT gravity: the Liouville or Teichm\"uller-inspired equations \eqref{pfann} or \eqref{Lvpf} in the current work, versus equation (127) of \cite{sss2}.\\
In \textbf{section \ref{app:liou}} we detail this story in the language of Liouville CFT on a torus.\\
In \textbf{section \ref{ss:ambiguities}} we discuss the measure on the space of conjugacy class elements.

\subsection{Twists in Compact BF}
\label{app:glueBF}
\begin{figure}[H]
\centering
\includegraphics[width=0.7\textwidth]{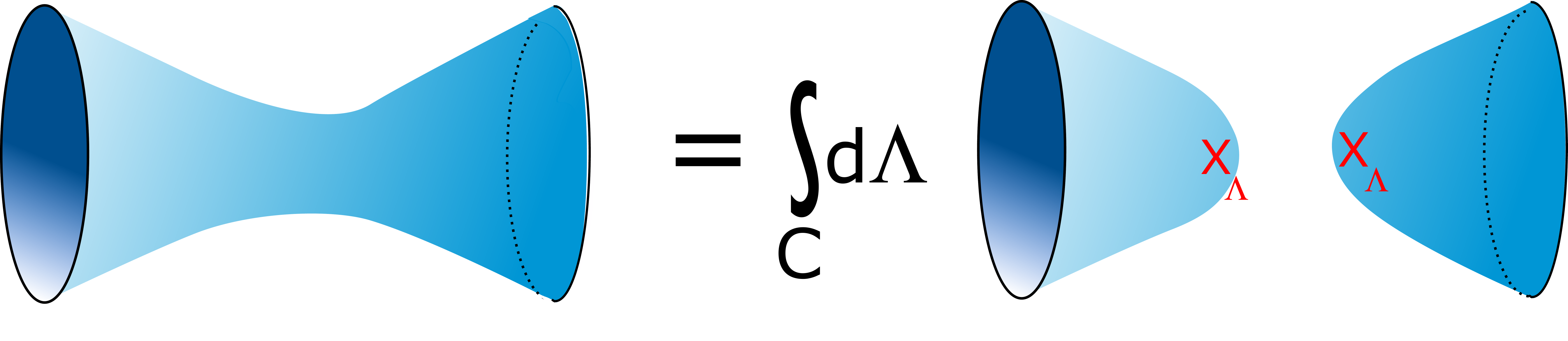}
\caption{Left; Annulus amplitude in compact BF theory. The description in terms of two particle on group models contains a common $G$-gauge symmetry, acting diagonally on both boundary actions. Right: Schematic decomposition in terms of twisted partition functions, glued together by integrating over $\Lambda \in C$.}
\label{WilsonCyl}
\end{figure}

Consider the annulus amplitude in BF theory for a compact Lie group $G$ (Figure \ref{WilsonCyl} left).

\noindent As we studied in section \ref{twobdybf}, the dynamics of this model is captured by two particle-on-group models, one on each boundary circle, with a common $G$-redundancy, and glued using the holonomy variable $M \in G$. 
\\
The BF path integral is a phase space path integral, for which we can write the symplectic integration space in several ways \cite{mooreseibergelitzurschwimmer}:
\begin{equation}
\int_G d M \, \frac{LG_M \times LG_M}{G_M} \,\, \simeq \,\, \left|G\right| \int_C d\Lambda \, \, \frac{LG_\Lambda \times LG_\Lambda}{Z(\Lambda)} \frac{1}{\left|Z(\Lambda)\right|} \,\, \simeq \,\, \left|G\right| \int_C d\Lambda \,  \frac{LG_\Lambda}{Z(\Lambda)} \times \frac{LG_\Lambda}{Z(\Lambda)} \times \frac{Z(\Lambda)}{\abs{Z(\Lambda)}}.\label{intspace}
\end{equation}
Here, $M$ denote the monodromies: $g(t+\beta) = M \cdot g(t)$. There is a diagonal gauge redundancy in description:
\begin{equation}
g_{L}(t) \sim G \cdot g_{L}(t),\quad g_{R}(t) \sim G \cdot g_{R}(t), \quad M \sim G M G^{-1}.
\end{equation}
Part of this gauge is fixed by only integrating over conjugacy class elements $\Lambda\in C$ covering all physically inequivalent $M$.\footnote{
This integral can be written as a sum over all irreps, using the fact that there exists an isomorphism between $C$ and the irreps of the group as $\Lambda = e^{-\frac{2\pi}{k} \lambda}$ \cite{mooreseibergelitzurschwimmer}. Here, $\lambda$ denote the weight vectors $\lambda = \bm{\lambda} \cdot \bm{H}$ as $k\to +\infty$.
} This leaves only the diagonal redundancy:
\begin{equation}
    g_{L}(t) \sim h \cdot g_{L}(t),\quad g_{R}(t) \sim h \cdot g_{R}(t), \quad \Lambda \sim h \Lambda h^{-1}=\Lambda,
\end{equation}
composed of elements $h$ that commute with $\Lambda$ i.e. that act within a conjugacy class. In other words, $h$ spans the centralizer $Z(\Lambda)$ of $\Lambda$.\footnote{
For compact $G$, the subgroup of conjugacy class elements $C$ can be identified with the maximal torus $T$ mod Weyl $W$. Each element $\in T/W$ by definition commutes with $T$ and so at the very least $T \subseteq Z(\Lambda)$. Some conjugacy class elements though contain an enlarged centralizer, e.g. the identity element has an increased centralizer $Z(\mathbf{1}) = G$.}
\\
In the second equality, we wrote out the fact that fixing the diagonal gauge is equivalent to fixing the gauge of both $g_R(t)$ and $g_L(t)$ and separately integrating over a twist variable $h\in Z(\Lambda)$ that represents the off-diagonal transformation:
\begin{equation}
    g_L(t)\to g_L(t),\quad g_R(t)\to h\cdot g_R(t).
\end{equation}
The set of physical fields is then $(g_L(t),g_R(t),\Lambda,h)$, and this is the set-up used when gluing two twisted disks together (Figure \ref{WilsonCyl} right). Notice that when computing the partition function, the integral over $h$ just gives $\abs{Z(\Lambda)}$, canceling this same factor in the denominator and giving \eqref{za1}, up to a volume factor $\abs{G}$ which we discard.
\\
A generic amplitude of Wilson lines on this annulus then decomposes as:
\begin{equation}
    \int_C d\Lambda \int_{\frac{LG}{Z(\Lambda)} \times \frac{LG}{Z(\Lambda)}}[\dpi g_L] [\dpi g_R]e^{-S[g_L,\Lambda] + S[g_R,\Lambda]}\left(\int_{Z(\Lambda)} \frac{d h}{\abs{Z(\Lambda)}}\,R_{n m}(g^{-1}_L(t^L_i)\cdot h\cdot g_R(t^R_f))\dots\right), \label{wilsons3} 
\end{equation}
where per example we only inserted a single Wilson line $\mathcal{W}^R_{n m}(t^L_i,t^R_f)$ that begins at the left boundary at time $t^L_i$ and end at the right boundary at time $t^R_f$. Wilson lines that start and end at the same boundary are seen to factor out of the integral over $h$.
\\
This is the more precise way to represent the wormhole-crossing Wilson lines calculated in BF language in the main text, in terms of particle on group variables. In particular, such an integral over twists is secretly implied also present in the path integral \eqref{wormcorr} that results in \eqref{annwl}.\footnote{As a sanity check, notice that the Wilson line inserted in the path integral is a gauge-invariant operator, invariant under independent left and right gauge transformations, because the Haar measure for compact Lie groups is bi-invariant $d h= d (h_L^{-1}\cdot h\cdot h_R)$. Furthermore, we can check the equivalence of rotating either endpoint separately around the annulus: $t_i^L\to t_i^L+\beta_L$ and similarly for $t_f^R$, which pick up the same monodromy $\Lambda$ that can be absorbed into the variable $h \in Z(\Lambda)$ again.}
\\~\\
For coset theories $G/H$, the decomposition \eqref{intspace} becomes:
\begin{align}
\int_{G/H} dM \, \frac{L\left(\frac{G}{H}\right)_M \times L\left(\frac{G}{H}\right)_M}{G_M} \,\, &\simeq \,\, \frac{1}{\left|H\right|} \int_{G} dM \, \frac{L\left(\frac{G}{H}\right)_M \times L\left(\frac{G}{H}\right)_M}{G_M} \nonumber \\
&\simeq \,\, \frac{\left|G\right|}{\left|H\right|}\int_C d\Lambda \,  \frac{L\left(\frac{G}{H}\right)_\Lambda}{Z(\Lambda)} \times \frac{L\left(\frac{G}{H}\right)_\Lambda}{Z(\Lambda)} \times \frac{Z(\Lambda)}{\left|Z(\Lambda)\right|}.\label{cosdecom}
\end{align}
We will see the non-compact analogue of this formula in practice for the Schwarzian limit of Liouville CFT in Appendix \ref{app:liou}.\footnote{For that particular coset, one has that the Liouville volume $V_\phi =\abs{G}/\abs{H} = \left|Z(\Lambda)\right| = V_C$, with $V_C$ defined above in Appendix \ref{app:reg}, so we would get:
\begin{equation}
    \int_C d\Lambda\,  \frac{L\left(\frac{G}{H}\right)_\Lambda}{Z(\Lambda)} \times \frac{L\left(\frac{G}{H}\right)_\Lambda}{Z(\Lambda)} \times Z(\Lambda).\label{liouvilletwist}
\end{equation}}

\subsection{Twists in JT}
\label{app:glueJT}
In BF theory for compact groups, the twist play no significant role. This changes when we consider JT gravity.
\\
In particular, the integration range for the twist variable is different depending on the integration space we define our theory by. There are two natural such choices: Teichm\"uller space $\mathcal{T}$, which is closer in spirit to \eqref{liouvilletwist}, and the moduli space of Riemann surfaces $\mathcal{M}$. In what follows we detail this, from a geometrical perspective.
\\~\\
The moduli space of Riemann surfaces $\mathcal{M}_{g,b}(\bm{\ell})$ of genus $g$ with $b$ geodesic boundaries of fixed lengths $\bm{\ell}$ is related to Teichm\"uller space $\mathcal{T}$ by modding out the mapping class group MCG:\footnote{Teichm\"uller space is just the hyperbolic component of the moduli space of flat $\sltr$ connections. Mathematically, MCG $\simeq \frac{\text{Diff}^+(\Sigma)}{\text{Diff}^+_0(\Sigma)}$, parametrizing the classes of large diffs on the surface.}
\begin{equation}
\frac{\mathcal{T}_{g,b}(\bm{\ell})}{\text{MCG}} \simeq \mathcal{M}_{g,b}(\bm{\ell}).
\end{equation}
This means that surfaces that differ by a Dehn twist are considered equivalent from the perspective of $\mathcal{M}_{g,n}(\bm{\ell})$. Dehn twists are obtained by cutting the surface across a tube, rotating one end by a full $2\pi$-rotation, and regluing. \\
The dimension of the moduli space of a bordered hyperbolic ($\chi <0$) Riemann surface of genus $g$ with $b$ geodesic boundaries is $\dim \mathcal{M}_{g,b}(\bm{\ell}) = 6g-6+2b$.\footnote{Fixing the geodesic lengths to fixed values, means there are 2 real moduli associated to each geodesic boundary. Notice that punctures are treated as geodesic boundaries with $\ell_i \to 0$.} These $6g-6+2b$ degrees of freedom can be thought of as the geodesic lengths $\lambda$ and twists $\tau$ of the $3g-3+b$ independent geodesics associated to some pair-of-pants decomposition of the surface. These are the Fenchel-Nielsen coordinates of the moduli space. 
\\~\\
Both $\mathcal{T}_{g,b}(\bm{\ell})$ and $\mathcal{M}_{g,b}(\bm{\ell})$ are equiped with the Weil-Petersson symplectic measure:
\begin{equation}
  \omega = \sum_{i=1}^{3g-3+b} d\lambda_i \wedge d\tau_i.  
\end{equation}
What distinguishes the moduli space of Riemann surfaces from Teichm\"uller space is the range of integration for these coordinates. In particular, in Teichm\"uller theory, the range of $\tau$ is the entire real axis. Modding by the mapping class group boils down to identifying twists that differ by a full length of the boundary $\tau \sim \tau + 2\pi \lambda$.
\\
This means that if we want to glue surfaces on geodesics, working within the moduli space of Riemann surfaces $\mathcal{M}_{g,b}$, we use a gluing integral of the type:
\begin{equation}
\int \omega = \int d\lambda \wedge d\tau = \int d\lambda \int_0^{2\pi \lambda} d\tau = 2\pi \int_{0}^{\infty} \lambda d\lambda.\label{twistriemann}
\end{equation}
This was used extensively in \cite{sss2}, and is intrinsic to Mirzakhani's recursion relations \cite{Mirzakhani:2006fta,Mirzakhani:2006eta}. 
\\
Working within $\mathcal{T}_{g,b}$ on the other hand, we have a gluing integral of the type:
\begin{equation}
\int \omega = \int d\lambda \wedge d\tau = \int d\lambda \int_{-\infty}^{\infty} d\tau = V_C\int_{0}^{\infty} d\lambda,\label{twists}
\end{equation}
which is the analogue of the similar formula for compact BF theories \eqref{liouvilletwist}. As we demonstrate in section \ref{app:liou}, the volume factor that appears here as a physical interpretation as the Liouville volume $V_\phi=V_C$. Nevertheless we will often neglect it.
\\~\\
Disk amplitudes are not sensitive to this choice of twist range, since the mapping class group in that case is the identity element. Amplitudes of JT gravity on more complicated topologies though, will be significantly different depending on this choice of gluing range.
\\
The example that is most important for this work is the annulus. We see that the different choice of integration contour explain the different known results for the JT gravity annulus partition function: the Liouville or Teichm\"uller-inspired equations \eqref{pfann} or \eqref{Lvpf} in the current work use \eqref{twists}. This should be contrasted to formula (127) of \cite{sss2}, which uses \eqref{twistriemann}.
\\~\\
For higher genus surfaces, we no longer get to choose. Indeed, for $\chi<0$ surfaces it is known that the volume of Teichm\"uller space diverges, whereas the Weil-Petersson volumes of the moduli spaces of Riemann surfaces:
\begin{equation}
    V_{g,b}=\int \exp(\omega),
\end{equation}
are finite.\footnote{One might make sense of gluing in $\mathcal{T}$ by suitably regularizing the IR-divergence, as is done in 2d Liouville CFT. Indeed, the Liouville cylinder amplitude between two FZZT-branes is IR-divergent. It can be interpreted within the Schwarzian double-scaling limit as the JT disk with two local bulk punctures, which is topologically a three-holed sphere. One can make sense of this amplitude by volume-regularization.} In equation \eqref{divchi} below, we demonstrate this explicitly in BF language. The only sensible amplitudes of JT gravity on higher genus surfaces are then obtained by choosing the moduli space of Riemann surfaces $\mathcal{M}$ as integration space. 
\\
This means in particular that when one wants to define JT gravity as a sum over topologies as in \cite{sss2}, one should consider the moduli space of Riemann surfaces $\mathcal{M}$. When one is only interested in the disk and annulus, as we were in the main text of this work, there are two options that are equivalent for what the disk is concerned, but inequivalent on the annulus.

\subsubsection*{Aside: Operator Insertions}
As an aside, let's think about wormhole-crossing Wilson lines in Schwarzian variables when considering the moduli space of Riemann surfaces $\mathcal{M}$ when gluing.\footnote{Wilson lines anchored on the same boundary are easily treated within either $\mathcal{T}$ or $\mathcal{M}$, by gluing with the respective gluing measure. There are no additional complications imposed by the Wilson line itself and the treatment within $\mathcal{M}$ when summing over all topologies was presented in appendix D in \cite{paper5}.} Based on \eqref{wilsons3} a natural guess is:\footnote{The normalization is set because we know we should recover \eqref{twistriemann} as integrand for $\ell \to 0$.}
\begin{equation}
\label{guessbil}
\int_{0}^{2\pi \lambda} d\tau \, \left(\frac{f^{\LL '}_1 f^{\RR '}_2}{\sinh^2 \frac{1}{2}(f^{\LL}_1 - f^{\RR}_2 + \tau)}\right)^\ell.
\end{equation}
Notice though that this operator is not gauge-invariant under left- and right $U(1)$ transformations $f_i \to f_i + c$, due to the integration range.\footnote{This is intuitive: a cross-wormhole line in the double-trumpet geometry is not invariant under a Dehn twist, even though the partition function is.} One then naturally associates to a Wilson line the average over gauge images of \eqref{guessbil}:
\begin{equation}
\frac{1}{\left|\text{MCG}\right|} \sum_{n \in \mathbb{Z}} \int_{0}^{2\pi \lambda} d\tau \, \left(\frac{f^{\LL '}_1 f^{\RR '}_2}{\sinh^2 \frac{1}{2}(f^{\LL}_1 - f^{\RR}_2 + \tau + 2 \pi n \lambda)}\right)^{\ell},
\end{equation}
or
\begin{equation}
\frac{2\pi \lambda}{V_C}\, \int_{-\infty}^{\infty} d\tau \, \left(\frac{f^{\LL '}_1 f^{\RR '}_2}{\sinh^2 \frac{1}{2}(f^{\LL}_1 - f^{\RR}_2 + \tau)}\right)^{\ell}.
\end{equation}
The $2\pi \lambda$ prefactor complicates the computation of correlation functions of this operator, which we leave to future work.

\subsection{Twists in Liouville on the Torus}
\label{app:liou}
Let us stack up the claim that 2d Liouville CFT in the Schwarzian double-scaling limit naturally picks the Teichm\"uller moduli space for JT gravity.\footnote{See also \cite{paper6} for a \emph{bulk} Liouville theory related to JT gravity.}
\\
In particular, we will show that the volume factor $V_\phi$ in \eqref{Lvpf} arises from the zero-mode twist and can be interpreted as the twist variable when gluing using the Teichm\"uller range of the Weil-Petersson measure \eqref{twists}.\footnote{It has indeed been known for a long time that Liouville theory on $\Sigma_{g,n}$ is deeply linked to Teichm\"uller theory on $\Sigma_{g,n}$ \cite{Teschner:2005bz,Teschner:2003at}.}
\\~\\
Liouville on the torus as discussed in section \ref{s:liou} comes with the integration space:\footnote{This structure follows very explicitly from the Gervais-Neveu field redefinition \eqref{GNtrans},\eqref{GNtrans2}.}
\begin{equation}
\label{intLsp}
\int_{\mathbb{R}^+} d \lambda \, \frac{\text{diff }S^1_{\text{L},\lambda} \times \text{diff }S^1_{\text{R},\lambda}}{U(1)} \quad \simeq \quad \int_{\mathbb{R}^+} d \lambda \, \frac{\text{diff }S^1_{\text{L},\lambda}}{U(1)} \times \frac{ \text{diff }S^1_{\text{R},\lambda}}{U(1)} \times \mathbb{R},
\end{equation}
which is indeed conform \eqref{twists}. In terms of field variables, there is on the LHS a left- and right reparametrization $f^{\LL}$ and $f^{\RR}$ with a diagonal $U(1)$ redundancy $f^{\LL, \RR} \to f^{\LL, \RR}+ \alpha$. On the RHS the twist was extracted, which notably ranges over the entire real axis.
\\
This range may seem like a choice, but actually it is not: the Liouville torus amplitude for example is precisely \eqref{Lvpf} which requires the twists to range in $\mathbb{R}$. Indeed, this multiplicative infinity as announced previously is the $V_\phi$ 1d volume of Liouville theory. \\
To see this, consider the region where we have approximate translation invariance, $\phi \sim -\infty$. We have using \eqref{GNtrans} that $\sinh(f^{\LL}-f^{\RR}) \sim e^{f^{\LL}-f^{\RR}}$, and shifting $\phi \to \phi +c$ is identical to shifting $f^{\LL} - f^{\RR} \to f^{\LL} - f^{\RR} + c$, illustrating that the volume factor is the twist factor.\footnote{The transformation \eqref{GNtrans} becomes explicitly:
\begin{equation}
e^{\phi} = - f^{\LL '} e^{-f^{\LL}} f^{\RR '} e^{f^{\RR}} = F^{\LL '} F^{\RR '}, \qquad \text{where } F^{\LL} = e^{-f^{\LL}}, F^{\RR} = e^{f^{\RR}},
\end{equation}
corresponding to the Alekseev-Shatashvili free-field parameterization $\phi = \ln F^{\LL} + \ln F^{\RR}$ to map the chiral boson into the coadjoint orbit action \cite{alekseev1,alekseev2}.}
\\~\\
Under the decomposition \eqref{intLsp}, the Liouville vertex operator insertion becomes: 
\begin{equation}
e^{2\phi} = \frac{f^{\LL '}_1 f^{\RR '}_2}{\sinh^2 \frac{1}{2}(f^{\LL}_1-f^{\RR}_2)} \quad \simeq \quad \int_{-\infty}^{+\infty} d\tau \, \frac{f^{\LL '}_1 f^{\RR '}_2}{\sinh^2 \frac{1}{2}(f^{\LL}_1 - f^{\RR}_2 + \tau)},
\end{equation}
which is the more precise way to state \eqref{wormcorr}. On the LHS we are using field variables with a diagonal redundancy, which has been completely fixed on the RHS. This is to be compared with \eqref{wilsons3}.\\
Notice that the resulting path integral is invariant under a rotation of the Wilson line around the annulus $\tau_{i} \to \tau_{i} + \beta_{i}.$ by redefining $\tau$. In fact ever more is true: before doing the Schwarzian path integrals we can do an shift in twist variables $\tau \to \tau - f^{\LL}_1 + f^{\RR}_2 $. The resulting path integral factorizes into a left- and right-piece, which are then  $\tau_i$-independent. So the amplitude of the single Wilson line stretching the wormhole is expected on general grounds to be independent of $\tau_1$ and $\tau_2$. This was found in the main text via a BF calculation. 
\\
Multiple Liouville vertex operators correspond to: 
\begin{equation}
\int_{-\infty}^{+\infty} d\tau \, \frac{f^{\LL '}_1 f^{\RR '}_2}{\sinh^2 \frac{1}{2}(f^{\LL}_1 - f^{\RR}_2 + \tau)} \frac{f^{\LL '}_3 f^{\RR '}_4}{\sinh^2 \frac{1}{2}(f^{\LL}_3 - f^{\RR}_4 + \tau)} \hdots.
\end{equation}
In the case of two such operators, setting $\tau \to \tau - f^{\LL}_1 + f^{\RR}_2$, one can imagine only dependence on the difference $\tau_3-\tau_4-\tau_1+\tau_2$. This is confirmed via a direct BF calculation of two non-intersection Wilson lines stretching the wormhole.

\subsection{Measure on the Space of Conjugacy Class Elements}
\label{ss:ambiguities}
We distill some formulas relevant for this work from \cite{2dgt1,2dgt2} regarding the precise choice of integration measure on the space of conjugacy class elements (or orbits).
\\~\\
When one usually talks about finite characters, one uses the integration measure on the space of conjugacy class elements, inferred from the Haar measure on the group manifold.\footnote{This choice of measure was taken for example in \cite{quantjt}. For example for $SU(2)$ the measure is $d\theta \sin^2\theta$. For $\sltr$ the measure is $d \lambda \sinh^2 \lambda$. See formula (IV.B) in \cite{quantjt}.} The resulting characters, orthogonal with respect to these measures, are:
\begin{equation}
\text{SU(2)}: \quad \chi_n(\theta)=\frac{\sin n\theta}{\sin \theta}, \qquad\quad \sltr: \quad \chi_\mu(\lambda) = \frac{\cos \mu\lambda}{\sinh \lambda}. \label{ch1}
\end{equation}
And indeed, with the Haar measure inferred from the group manifold, orthogonality can be checked to hold. The point made in \cite{2dgt1} is however, that this integration measure is a \emph{choice}, and depending on the situation a different normalization might be required. 
\\~\\
The Schwarzian amplitudes \eqref{zsch}, to which we want to compare $\slr$ group theoretic amplitudes, are found as the limit of Virasoro CFT amplitudes \cite{schwarzian}, and the same is true for the BF amplitudes used in \cite{origins}. This means we have to choose the measure obtained from the embedding within 2d CFT (and 3d CS).
\\
Combining formulas (4.52) and (4.114) of \cite{2dgt1}, we learn that the measure on the space of conjugacy class elements, inferred from the 2d CFT perspective, is essentially the flat one; the appropriate characters are those where we drop the denominators of \eqref{ch1}. We obtain:
\begin{equation}
\text{SU(2)}: \quad \chi_n(\theta)=\sin n\theta, \qquad\quad \sltr: \quad \chi_\mu(\lambda)=\cos\mu\lambda. \label{ch2}
\end{equation}
These are orthogonal with respect to the flat measures, $d\theta$ and $d\lambda$ respectively. 

\section{Other Euclidean Topologies}
\begin{figure}[h]
\centering
\includegraphics[width=0.8\textwidth]{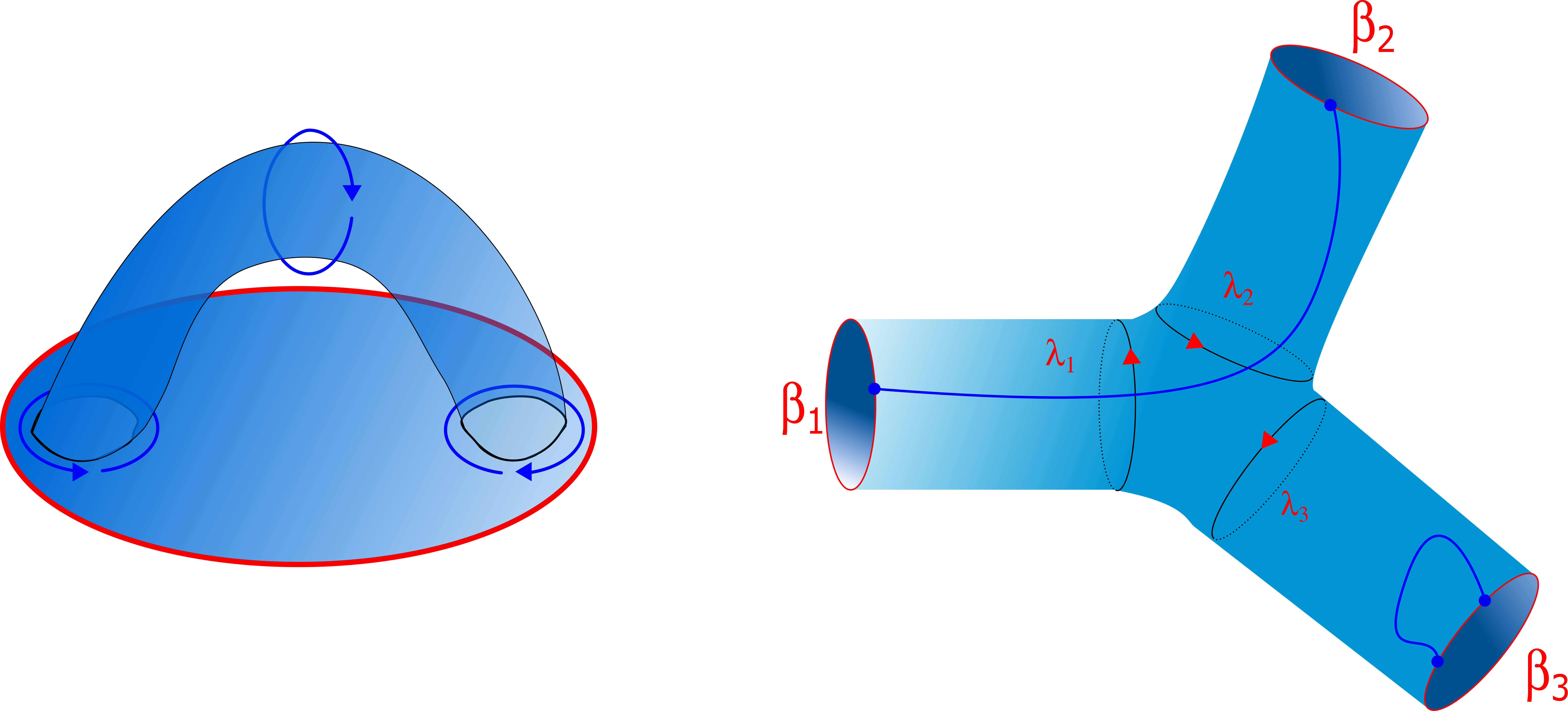}
\caption{Left: Disk with one handle attached. The blue arrows are indicative of a choice of Cauchy slices to do the BF evolution calculation. Right: Multiple Schwarzian boundaries with bilocals in between. Conjugacy class elements gluing the boundary annuli to the three-holed sphere are labeled by $\lambda_i$.}
\label{fig:cauchynonorient}
\end{figure}

As an application of the BF perspective on JT gravity, we show how to calculate amplitudes of generic JT gravity Euclidean manifolds.\footnote{This was studied simultaneously and in more detail in \cite{sss2}, see also \cite{sss}.}\label{app:multi}
\\~\\
Consider a disk with multiple handles attached as for example in Figure \ref{fig:cauchynonorient} left.
The goal here is to explain how to calculate such contributions, within the framework of section \ref{s:slrchar} and \cite{paper3}.
\\~\\
If we naively apply the cutting and gluing rules to such a surfaces by evolving $\slr$ states through the manifold, we obtain for a disk with $h$ handles: 
\begin{equation}
Z_h(\beta)=V_C^h\int \frac{d k} {(k\sinh 2\pi k)^{2h-1}}e^{-\beta k^2}.\label{naive}
\end{equation}
The resulting divergence is directly related to our present choice of integration space / gluing integral as discussed in section \ref{app:glueJT}: the volume of Teichm\"uller space relevant in this setup $\mathcal{T}_{h,1}(\lambda)$ is divergent.\\
We can isolate it by separately evolving through an annulus that includes the Schwarzian boundary, and through the remaining handlebody. The two are then glued together by introducing the resolution of the identity on a circular slice:\footnote{We must also integrate over twists, which gets a volume factor $V_C$ as in \eqref{twists}.}
\begin{equation}
\int d\lambda \ket{\lambda}\bra{\lambda}=\mathbf{1},
\end{equation} 
with $\lambda$ conjugacy class elements. The path integral hence decomposes as:
\begin{equation}
Z_h(\beta) = V_C \int d \lambda\, V_{h,1}(\lambda) \int d k\, \chi_k(\lambda)\, e^{-\beta k^2},\label{handle}
\end{equation}
in terms of the twisted Schwarzian path integral:
\begin{align}
\int d k\ \chi_k(\lambda)\, e^{-\beta k^2}=\int d k\, \cos 2\pi k\lambda\, e^{-\beta k^2}.
\end{align}
One would calculate the volume $V_{h,1}(\lambda)$ of $\mathcal{T}_{h,1}(\lambda)$ as, up to several volume factors:
\begin{equation}
\int d k \frac{\cos 2\pi k \lambda} {(k\sinh 2\pi k)^{2h-1}},
\end{equation}
which is the source of the divergence.\footnote{For compact groups this gives the correct, finite answer \cite{2dgt1}.}
\\~\\
The resolution \cite{dijkgraafwitten}, is to mod by the mapping class group, and considering instead the moduli space of Riemann surfaces $\mathcal{M}_{h,1}(\lambda)$, corresponding to the other possible choice of integration space for JT gravity.
\\
As discussed in section \ref{app:glueJT}, this comes with a change in gluing measure:
\begin{equation}
    V_Cd\lambda\to 2\pi \lambda d\lambda,
\end{equation}
and the volumes of the moduli spaces $V_{h,1}(\lambda)$ in this setup are the finite Weil-Petersson volumes \cite{dijkgraafwitten}, which can be determined recursively \cite{Mirzakhani:2006fta,Mirzakhani:2006eta}. 
\\~\\
This construction can be generalized to arbitrary genus $h$ with any given number of punctures (i.e. defects, coming from character insertions) $n$ and boundaries $b$, where we allow the surfaces to end either on the boundaries, or on Wilson lines, see for example \cite{paper5}. We are led to consider a possibly disconnected surface. 
\\
Within Teichm\"uller theory, each disconnected component comes with a single momentum $k$-integral with weight:\footnote{The Euler character is  $\chi =2-2h -b$.}
\begin{equation}
\label{divchi}
(k\sinh 2\pi k)^{\chi}.
\end{equation}
This diverges when $\chi < 0$, and to get a sensible answer one should instead consider JT gravity with contour $\mathcal{M}$. If however $\chi \geq 0$ for all components, meaning that there are only disks or annuli, both gluing integrals give sensible and in general (when there are annuli) different answers.
\\
As an example, consider three Schwarzian boundaries with no handles attached (Figure \ref{fig:cauchynonorient} right). The Teichm\"uller result diverges:
\begin{align}
Z(\beta_1,\beta_2,\beta_3) =\int \frac{d k}{k\sinh 2\pi k} e^{-k^2(\beta_1+\beta_2+\beta_3)},
\end{align}
and to make sense of this amplitude we should treat surfaces as in $\mathcal{M}$ instead of $\mathcal{T}$. One cuts off the annuli at the three boundaries and glues these to a three-holed sphere. The amplitude of the latter is the Weil-Petersson volume $V_0(\lambda_1,\lambda_2,\lambda_3)$, in terms of the conjugacy classes of the three gluing cycles. Using that $V_{0,3}=1$, since the moduli space is just a point for the three-boundary sphere with fixed boundary lengths, we are left with the product of three integrals of the type:
\begin{equation}
\int d\lambda_i \lambda_i e^{- \frac{\lambda_i^2 \pi^2}{\beta_i}}\left(\frac{\pi}{\beta_i}\right)^{1/2} = \sqrt{\frac{1}{4\pi^3}}\sqrt{\beta_i}.
\end{equation}
This results in the known three-boundary amplitude $Z(\beta_1,\beta_2,\beta_3)\sim \sqrt{\beta_1\beta_2\beta_3}$ \cite{Moore:1991ir}.
\\
It is in principle possible to calculate generic correlation functions on these multi-boundary manifolds. As an example, consider the Wilson line stretching between boundary 1 and boundary 2 in figure \ref{fig:cauchynonorient} right. We find: 
\begin{equation}
\frac{1}{Z}\int d k \frac{\Gamma(\ell)^2\Gamma(\ell\pm 2ik)}{\Gamma(2\ell)}e^{-k^2 (\beta_1+\beta_2+\beta_3)}.
\end{equation}
Notice that the geometry of the manifold minus the Wilson line is topologically an annulus, which comes with a flat measure such that we get a finite answer when gluing within $\mathcal{T}$.

\section{Edge States in Chern-Simons Theories}
\label{app:CS}
BF theory is defined as the dimensional reduction of 3d CS. The goal in this Appendix is to repeat the discussion of section \ref{sect:BFedge} for CS. By comparing famous CS formulas with some of the BF formulas obtained in section \ref{sect:BFedge} we provide with an alternative proof of the latter. 

\subsection{Edge Dynamics from the Path Integral}
\begin{figure}[h]
\centering
\includegraphics[width=0.5\textwidth]{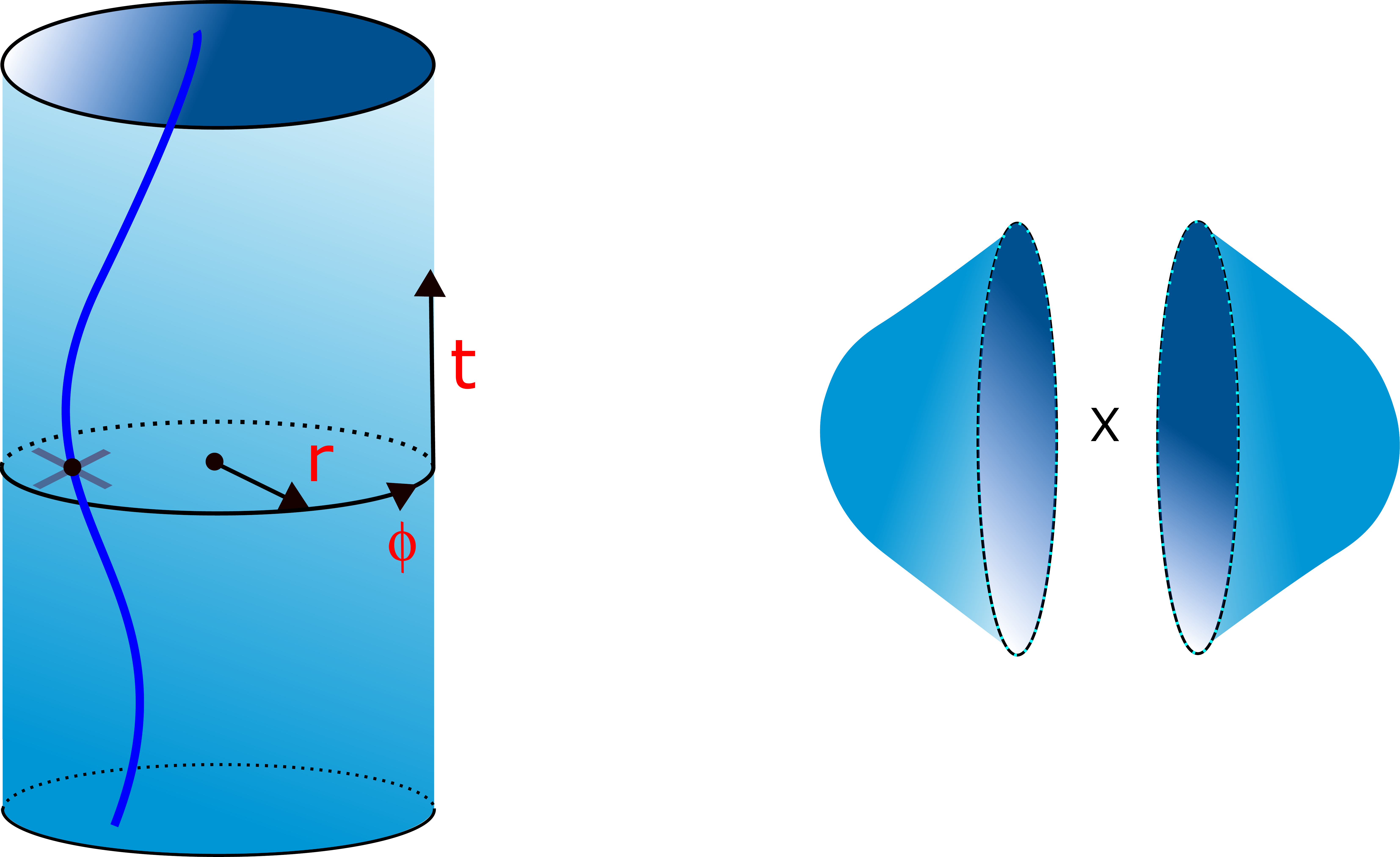}
\caption{Left: Chern-Simons cylinder amplitude and coordinatization, with a Wilson line piercing the fixed timeslice. Right: gluing two Chern-Simons theories on $D \times \mathbb{R}$ into a boundaryless Chern-Simons theory on $S^2 \times \mathbb{R}$.}
\label{fig:disk}
\end{figure}

We first review how Chern-Simons on a manifold with boundary, leads to a Wess-Zumino-Witten 2d CFT on the boundary \cite{jones,mooreseibergelitzurschwimmer}, in parallel to the BF argument of Section \ref{sect:bf}. Focusing on a manifold with cylindrical topology, we write
\begin{align}
S[A]&=\int_{\mathcal{M}} d^3x \epsilon^{\mu\nu\sigma}\Tr\left[A_\mu \partial_\nu A_\sigma +\frac{2}{3}A_\mu A_\nu A_\sigma\right]\\
&=\int_{\mathcal{M}} d t d r d \phi \Tr\left[A_r \partial_t A_\phi -A_\phi \partial_t A_r+2 A_t F_{\phi r}\right]-\int_{\partial \mathcal{M}} d t d \phi \Tr\left[A_t A_\phi\right].
\label{CSaction1}
\end{align}
The background-dependence is only in the orientation of the chosen coordinate axes which we choose $\epsilon^{t r \phi}=1$. We parameterize the spatial disk $D$ as in Figure \ref{fig:disk}.
Variation results in the boundary condition $A_\phi \atbdy \sim A_t\atbdy$. Rescaling the coordinates is a symmetry of the problem hence we can bring the proportionality factor to $\pm 1$. Changing sign corresponds to changing orientation and with our ordering of the coordinates, only the $+$-sign leads to a positive Hamiltonian: CS on a manifold with boundary is only consistent with the boundary conditions:
\begin{equation}
\left(A_t - A_\phi\right) \atbdy = A_{\bar{z}} \atbdy=0.\label{bc}
\end{equation} 
Path integration over the Lagrange multiplier $A_t$ results in 
\begin{align}
A_\phi &= g^{-1}\partial_\phi g\label{aphi}\\
A_r&=g^{-1}\partial_ r g,\label{ar}
\end{align}
with $g$ a $G$-valued field, in general twisted in the $\phi$-direction: $g(\phi+2\pi)=U_\lambda g(\phi)$, with $U_\lambda$ determined by a possible Wilson line insertion in irrep $\lambda$ in the $t$-direction. Bulk values of $g$ are redundant and only its boundary profile is a physical degree of freedom. Moreover there is a global $G$ redundancy in \eqref{aphi}, \eqref{ar} under $g\to V g$ with $V$ constant.\footnote{This results in the equivalence $U\sim V^{-1}UV$ hence the space of all inequivalent holonomies $U$ is isomorphic to the space of conjugacy class elements $\lambda$.} The path integral over $A$ in \eqref{gluecs} is reduced to a path integral over boundary configurations $g$.\footnote{Depending on the topology, in general there may or may not also be an integral over conjugacy class elements $\lambda$.}
\\~\\
Making the substitution $g(\phi)\to \Lambda(\phi) g(\phi)$ with $\Lambda(\phi+2\pi)^{-1}\Lambda(\phi)=U_\lambda$ untwists $g(\phi)$. Using partial integration combined with the boundary conditions $A_\phi \atbdy \pm A_t\atbdy$, the CS action \eqref{CSaction1} becomes a right-moving chiral WZW model, or a (right-moving) affine coadjoint orbit action:
\begin{equation}
\boxed{S[g,\lambda]=\int_{\partial \mathcal{M}} d t d \phi \Tr\left((\gphi+\lambda)\gt-(\gphi+\lambda )^2\right) + \Gamma_{WZ}.}\label{CSotherorbit}
\end{equation}
Let's now take two such Chern-Simons theories on spatial disks, and glue them into a single $S^2$ along the equator (figure \ref{fig:disk} right) \cite{wong,Fliss:2017wop}. The correct way to split the Chern-Simons Lorentzian path integral is by the introduction of a functional delta constraint:
\begin{align}
Z= \int [\dpi A_L]\exp(iS[A_L])\delta(A_L\atbdy-A_R\atbdy)\int [\dpi A_R]\exp(iS[A_R]).\label{gluecs}
\end{align}
The CS path integral on $S^2$ hence decomposes as:
\begin{equation}
Z_{S^2} = \int[\dpi g_L]\exp(iS[g_L])\delta(g_L-g_R)\int[\dpi g_R]\exp(-iS[g_R]) = 1,\label{diskpathint}
\end{equation}
the final equality being true because the Hilbert space of CS on $S^2$ is just the vacuum \cite{jones}.
Upon taking the $t$-dimensional reduction, the chiral WZW model \eqref{CSotherorbit} reduces precisely to the twisted particle on group \eqref{twistedpogaction} and \eqref{diskpathint} goes to \eqref{split}, upon renaming $\phi \to t$. Notice again that the two actions will cancel upon gluing. The left action is minus the right one, or $k\to -k$.\footnote{This sign flip can be undone by considering a coordinate system more natural for the left observer, in which the boundary condition \eqref{bc} destroys the holomorphic polarization instead and we obtain a left moving affine coadjoint orbit action.}
\\~\\
The argument in the functional delta in \eqref{gluecs} becomes the WZW current density upon path integrating out $A_t$: $\delta(\mj_L-\mj_R)$. The theory associated with the submanifold $R$ only is obtained from \eqref{diskpathint} by dropping all reference to $L$ and is just the chiral WZW model:
\begin{equation}
Z_R=\int [\dpi g_R]\exp\left( i S[g_R,\lambda_R]\right) = \int[\dpi \mj_R]\int_{A_t \atbdy = A_\phi \atbdy = \mj_R} [\dpi A_R]\exp(iS[\mj_R,A_R])\label{edgedynamicsCS},
\end{equation}
which can also be interpreted as path integrating over all boundary sources $\mj_R$ with a suitable action. In terms of Hilbert spaces this means there is a extended Hilbert space construction that accounts for edge states on the dividing surface, with again the gluing condition $\delta(\mj_L-\mj_R)$ acting as a Gupta-Bleuler constraint, projecting onto the physical subspace. \\
Explicitly, the Gupta-Bleuler constraint selects just one state in the extended Hilbert space associated with the entangling surface, to be written as an Ishibashi state \cite{wong,Fliss:2017wop}:
\begin{equation}
\ket{\lambda} =\frac{1}{\sqrt{\tensor{S}{_\lambda^0}}}\sum_n \ket{\lambda,n} \otimes \ket{\lambda,n}.\label{ishi}
\end{equation} 
of the left- and right sectors of the 2d WZW CFT on the entangling surface. In the BF limit, this becomes the factorization property \eqref{statefactor} or \eqref{statesfactorjt}, as only primaries survive.

\subsection{Two-Boundary Models}
\begin{figure}[h]
\centering
\includegraphics[width=0.5\textwidth]{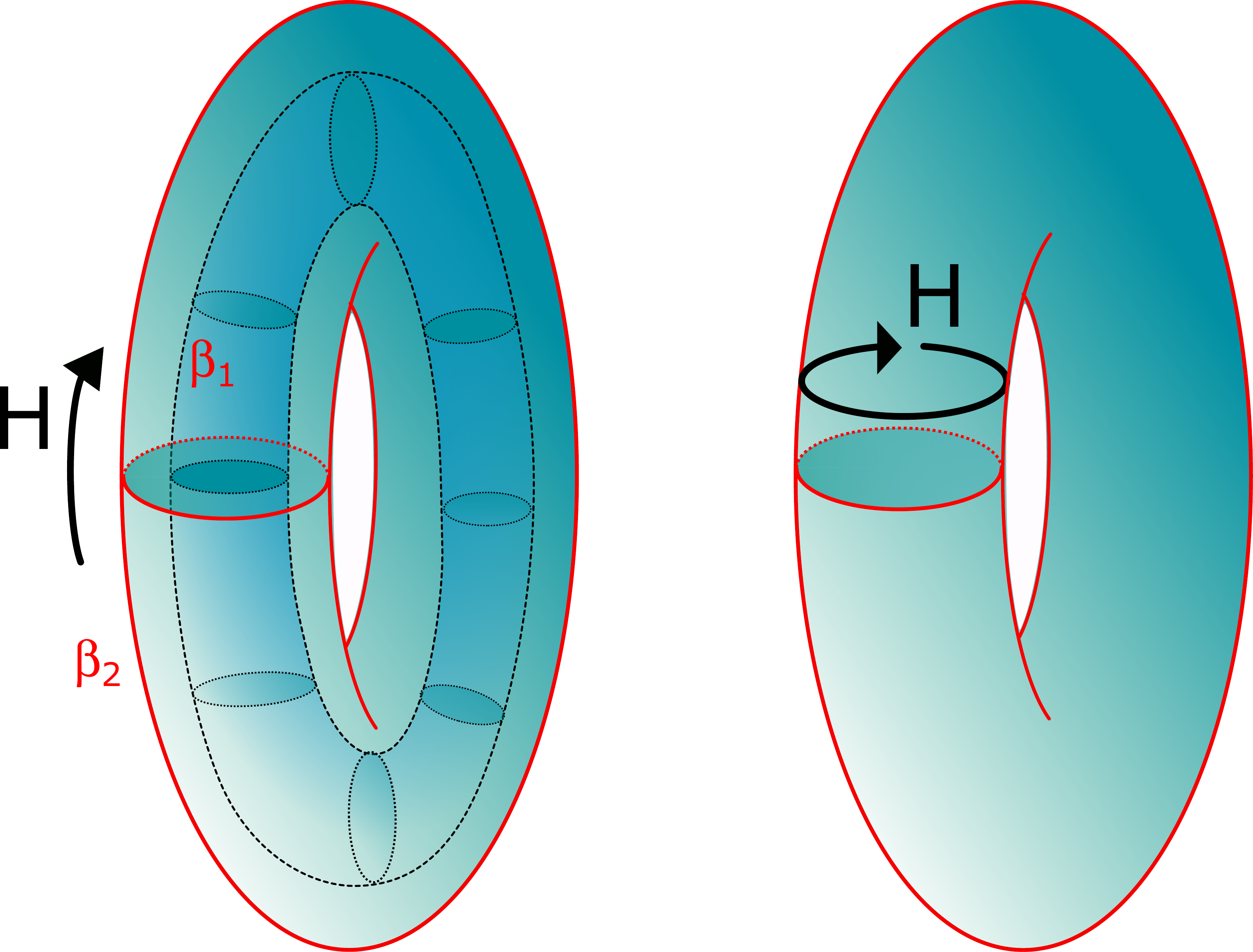}
\caption{Left: Annulus partition sum $Z$ with (vertical) length $\beta_1$ for the inner tube, and $\beta_2$ for the outer tube.  Right: Limit where $\beta_1 \to 0$, leading to the exterior of the solid torus or, alternatively, the interior of the $S$-dual surface.}
\label{fig:CStorus}
\end{figure}

As an application of the above consider the annulus path integral:
\begin{equation}
Z=\sum_\lambda \int [\dpi g_L]\exp(iS[g_L,\lambda])\int [\dpi g_R]\exp(-iS[g_R,\lambda]).\label{modeltwo}
\end{equation}
Since the $\phi$-direction is non-contractible in the annulus $A$, the holonomy of the connection can be arbitrary and hence the path integral includes a sum over conjugacy class elements.\footnote{The holonomy on the inner boundary equals the holonomy on the outer boundary because $g_L$ and $g_R$ are derived from a single field $g$ with an $r$-independent holonomy. Similarly, the global redundancy in the parameterization is the diagonal $g_{L,R}\to V g_{L,R}$.}
\\
The Euclidean configuration associated with this setup in $A\cross S^1$ whose boundaries are two tori. The boundaries break topological invariance, as made explicit by the dependence of the theory on a choice of modular parameter $\beta_i$ on both tori (Figure \ref{fig:CStorus} left):
\begin{equation}
Z=\sum_\lambda \chi_\lambda(\beta_1)\chi_\lambda(\beta_2), \qquad \chi_\lambda(\beta) = \Tr_\lambda e^{-\beta L_0}. \label{zac}
\end{equation}
This is equivalent to the statement that the spectrum of the theory consists of the states $\ket{\lambda,m}\otimes \ket{\lambda,n}$ and the Hamiltonian is $\beta_1 H_1 +\beta_2 H_2$. This should be compared to \eqref{zib}.
\\~\\
In the special (thermal) case that $\beta_1=\beta_2$, \eqref{zac} is just the partition function of a non-chiral WZW model: the fields $g_L$, $g_R$ and $\lambda$ can be recombined into a single non-chiral field $g$ as described for example in \cite{paper3} and applied to Liouville in section \ref{s:liou}. Notice in particular that this expression is modular invariant. 
\\~\\
The case that $\beta_1=0$ is closely related to the construction of the thermofield double state from the single-sided Hilbert space, and corresponds geometrically to (the exterior of) a solid torus (Figure \ref{fig:CStorus} right). The amplitude \eqref{zac} can then be interpreted as $\bra{\text{TFD}}\ket{\text{TFD}}$ by analyzing the relation between Rindler (or modular or one-sided) time $\tau$ and Kruskal (or global) time $t$. At the asymptotic boundary, we have $t=\frac{\beta}{2\pi}\tau$. At the horizon, there is an infinite redshift between the Rindler time frame and the Kruskal time frame, and the associated edge degrees of freedom seem frozen on the horizon according to the modular time frame. This is the analogue of the discussion around Figure \ref{annulusredshift}. From \eqref{zac} we immediately write down the purification:
\begin{equation}
\boxed{\ket{\text{TFD}}=\int_\oplus d\lambda\sum_{m,n}\ket{\lambda,m,n} \otimes \ket{\lambda,m,n}\,e^{-\frac{\beta}{2}L_0(\lambda, m)}.}\label{tfdfac}
\end{equation}
Using the factorization property \eqref{ishi}, the thermofield double can hence be rewriten into the form:
\begin{equation}
\ket{\text{TFD}}=\int d\lambda \sqrt{ \tensor{S}{_\lambda^0} } \sum_m \ket{\lambda,m,m} e^{-\frac{\beta}{2}L_0(\lambda,m)}.\label{tfdsann}
\end{equation}
In the BF limit, the CFT Hamiltonian goes to the Casimir $L_0(\lambda)\to \cas_\lambda$ as shown by the Sugawara construction and we recover the factorized \eqref{tfd} and non-factorized \eqref{diska} thermofield double states respectively.
\\~\\
One can match the norm of the thermofield double with the torus partition function: 
\begin{equation}
Z = \bra{\text{TFD}}\ket{\text{TFD}}=\int d\lambda \tensor{S}{_0^\lambda}\chi_\lambda (\beta)=\chi_0(S\cdot \beta),
\end{equation}
with the $S$-transform reflecting that the Hamiltonian now generates evolution along the $A$-cycle of the torus. Indeed, the $\beta_1 \to 0$ limit yields the \emph{exterior} of the original torus as the partition function. The latter is then related to a usual torus precisely by a modular $S$-transform \cite{jones}.

\section{Some Representation Theory of $\sltr$}
\label{s:rep}
We review some of the representation theory of $\sltr$ that is used in the main text. We will be mainly concerned with the parabolic basis which paves the way for the representation theory of $\slr$ in Appendix \ref{s:repsemi}. The emphasis here is on the continuous series irreps for which we derive the matrix elements and the Plancherel measure in a down-to-earth manner. This section is largely based on \cite{VK}. 
\\~\\
Group elements of $\sltr$ can be represented as the set of matrices:
\begin{equation}g=
\left(\begin{array}{cc}
a & b \\
c & d \\
\end{array}\right), \qquad ad-bc = 1.
\end{equation}
The (self-adjoint) generators of the group are the traceless matrices $J_a$:
\begin{eqnarray}
\label{gener}
iJ_+=\left(\begin{array}{cc}
0 & 1 \\
0 & 0 \\
\end{array}\right) , \quad iJ_- = \left(\begin{array}{cc}
0 & 0 \\
1 & 0 \\
\end{array}\right), \quad iJ_0 = \frac{1}{2}\left(\begin{array}{cc}
-1 & 0 \\
0 & 1 \\
\end{array}\right),
\end{eqnarray}
satisfying the $\mathfrak{sl}(2,\mathbb{R})$ algebra:
\begin{equation}
\label{alge}
\comm{J_0}{J_\pm} = \pm i J_\pm, \quad \comm{J_+}{J_-} = 2iJ_0.
\end{equation}
The Casimir is: $\cas=J_0^2+\frac{1}{2}\left\{J_+,J_-\right\}$. A set of basis functions of $\sltr$ is obtained by diagonalizing the Casimir and one of the generators $J_a$. Suppose we label the eigenvalues of the Casimir as $\cas=j(j+1)$. For each fixed $j$, a spin $j$ representation is defined as a basis for the corresponding eigenspace of $J_a$. Labeling the eigenvalues of the diagonalized generator of choice as $J_a=\nu$, we end up with the orthonormal states $\ket{j\nu}$. 
\\
To make this more explicit we must specify a realization of the algebra or the group. We will be discussing functions on the real line $x\in\mathbb{R}$ with the usual inner product. A spin $j$ representation is obtained by defining the action of the group element $g$ on the basis functions $f^j_\nu(x)$ as:
\begin{equation}
\label{repgroup}
f^j_\nu(x) \to (g \cdot f^j_\nu)(x) = \left|bx+d\right|^{2j}f^j_\nu\left(\frac{ax+c}{bx+d}\right).
\end{equation}
Infinitesimally, using $g = 1 + i \epsilon^a J_a$, we observe that this is the Borel-Weil realization of the $\mathfrak{sl}(2,\mathbb{R})$ algebra:
\begin{align}
\label{BW}
iJ_- &= \partial_x, \nonumber \\
iJ_0 &= -x\partial_x + j, \nonumber \\
iJ_+ &= -x^2\partial_x + 2j x.
\end{align}
This also confirms that \eqref{repgroup} is a spin $j$ representation: using \eqref{BW} the Casimir is immediately calculated to be $\mathcal{C} = j(j+1)$.
\\
Representation matrices are as always the Fourier components of transformed states:
\begin{equation}
\bra{j\nu}g\ket{l\mu} =\delta_{j,l}\int d x f^*_\nu(x) (g\cdot f_\mu)(x).\label{matel}
\end{equation}
The above can be viewed as introducing a complete set of states $\ket{x}$ with 
\begin{equation}
\bra{x}g\ket{j\mu}=(g\cdot f^j_\mu)(x).
\end{equation}
One immediately verifies that these satisfy the composition property:
\begin{align}
\label{compogroup}
\bra{j\nu}g_1g_2\ket{j\mu} &= \int dx f^*_\nu(x) (g_1g_2\cdot f_\mu)(x), \qquad\quad\qquad\qquad g_i = \left(\begin{array}{cc}
a_i & b_i \\
c_i & d_i \\
\end{array}\right),  \nonumber \\
&= \int dx f^*_\nu(x) \,g_1 \cdot \left(\left|b_2x+d_2\right|^{2j}f_\mu\left(\frac{a_2x+c_2}{b_2x+d_2}\right)\right) \nonumber \\
&= \int dx f^*_\nu(x) \left|bx+d\right|^{2j} f_\mu\left(\frac{ax+c}{bx+d}\right)
,\qquad g_1 g_2 = \left(\begin{array}{cc}
a & b \\
c & d \\
\end{array}\right),
\end{align}
indeed demonstrating the defining property of a representation: $R(g_1) \cdot R(g_2) = R(g_1 \cdot g_2)$. From the definition of the adjoint action $g^\dagger$:
\begin{equation}
\label{adjg}
\bra{j\nu}g \ket{j\mu} = \int dx f^*_\nu(x) (g\cdot f_\mu)(x) \equiv \int dx (g^\dagger \cdot f_\nu(x))^* f_\mu(x),
\end{equation}
we obtain:
\begin{equation}
(g^\dagger \cdot f_\nu)(x) = \left|-bx+a\right|^{-2j-2}f_\nu\left(\frac{dx-c}{-bx+a}\right),
\end{equation}
such that the adjoint action is obtained by acting with the inverse group element $g^{-1}$.

\subsection{Mixed Parabolic Basis}
In the harmonic analysis on $\sltr$ two sets of unitary irreducible representations of $\sltr$ appear: the discrete ones with $j=\ell$ and $2\ell \in \mathbb{N}$, and the continuous ones $j=-\frac{1}{2}+i k$ with $k\in \mathbb{R}$. The goal of this section is to find explicit formulas for the associated matrix elements and Plancherel measure. With one eye on $\slr$ we choose to focus on only the continuous irreps here, and we choose to construct the matrix elements in the mixed parabolic basis.\footnote{We are free in the choice of generator $J_a$ to diagonalize, the Plancherel measure nor any of the physics is affected by this choice.}
\\~\\
Suppose one chooses to diagonalize $J_-$ or equivalently the subgroup $h_-(t)=\exp i t J_-$ with $t\in\mathbb{R}$. A basis of the spin $j$ representation is then the plane wave basis $f^k_\nu(x) = e^{i \nu x}$:
\begin{equation}
(h_-(t)\cdot f^k_{\nu})(x) = f^k_{\nu}(x+t) = e^{i \nu t} f_{j\nu}(x),
\end{equation}
with $J^- = \nu\in \mathbb{R}$. We will denote the associated state by $\ket{\nu_-}$, suppressing the $j$ index, such that
\begin{equation}
\bra{x}\ket{\nu_-}=e^{i\nu x}.
\end{equation}
Alternatively one may choose to diagonalize $J_+$ or equivalently the subgroup $h_+(t)=\exp it J_+$. A basis of the irrep is now formed by $f^k_{\nu}(x) = \left|x\right|^{2ik-1} e^{i \frac{\nu}{x}}$ with $J^+ = \nu$. Denoting the associated states by $\ket{\nu_+}$ we obtain:
\begin{equation}
\label{Lbasis}
\bra{x}\ket{\nu_+} = \left|x\right|^{2ik-1} e^{i \frac{\nu}{x}}.
\end{equation}
One can transform $J_-$ eigenstates into $J_+$ eigenstates by applying the group element
\begin{eqnarray}
\mathbf{s} = \left(\begin{array}{cc}
0 & 1 \\
-1 & 0 \\
\end{array}\right),
\end{eqnarray}
as $\mathbf{s}$ transforms $h_-$ into $h_+$: $\mathbf{s} \cdot h_-\cdot \mathbf{s}^{-1}=h_+$. And indeed, from the property $\bra{x}\mathbf{s}\ket{\nu_-}=\bra{x}\ket{-\nu_+}$ we find: 
\begin{equation}
\mathbf{s} \ket{\nu_-}=\ket{-\nu_+}.\label{equivalence}
\end{equation}
This property will prove pivotal in the construction that follows.
\\~\\
Mixed parabolic matrix elements are defined as
\begin{equation}
\bra{\nu_-}g\ket{\lambda_+},
\end{equation}
and form a basis of wavefunctions for the continuous spectrum of quantum mechanics on $\sltr$. Indeed, ordinary matrix elements for example in the basis $\ket{\nu_-}$ are orthogonal:
\begin{equation}
\int d g \bra{k \nu_-}g\ket{k \mu_-}\bra{k' \nu'_-}g\ket{k' \mu'_-}^*=\frac{\delta(k-k')}{\rho(k)}\delta(\nu-\nu')\delta(\mu-\mu'),
\end{equation}
with $d g$ the Haar measure and $\rho(k)$ the Plancherel measure. Using the property \eqref{equivalence} and the invariance of the Haar measure under $g\to g\cdot \mathbf{s}$ proves that mixes parabolic matrix elements are orthogonal in precisely the same way:
\begin{equation}
\int d g \bra{k \nu_-}g\ket{k \mu_+}\bra{k' \nu'_-}g\ket{k' \mu'_+}^*=\frac{\delta(k-k')}{\rho(k)}\delta(\nu-\nu')\delta(\mu-\mu'),\label{plandef}
\end{equation}
The same argument can be used to show that the $3j$-symbols in the $J_+$ basis are the same as those calculated in the $J_-$ basis, modulo some sign changes.

\subsection{Matrix Elements}
\begin{figure}[h]
\centering
\includegraphics[width=0.3\textwidth]{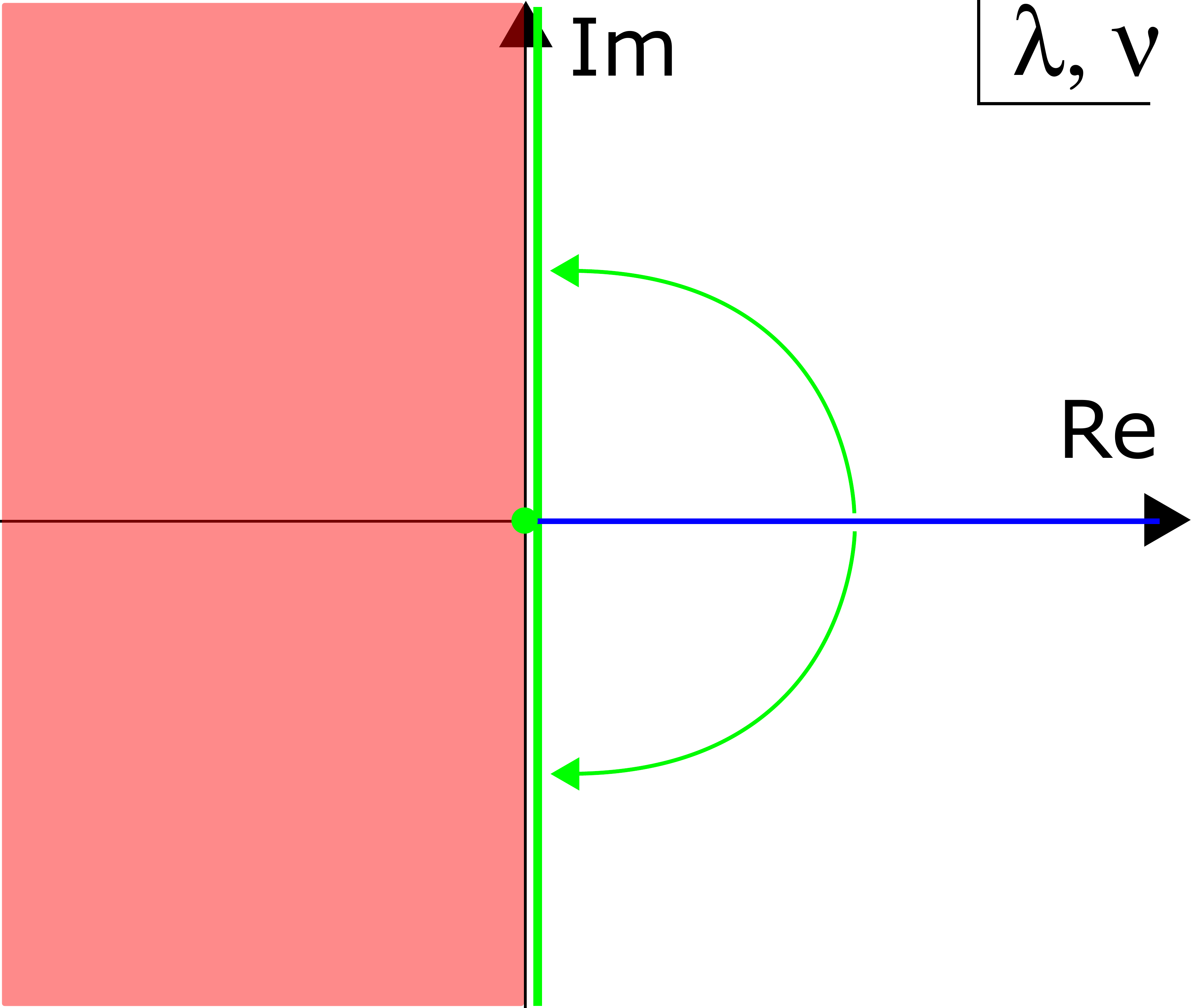}
\caption{Analytic continuation of the integral representation \eqref{BInt} of the modified Bessel function $K_{2ik}(x)$. We need $\Re(\nu,\lambda)>0$ for convergence.}
\label{BesselAnalytic}
\end{figure}

Let us then continue to compute the matrix elements explicitly. As a first instructive example consider the $g = \mathbf{1}$ matrix elements or the overlap $\bra{\nu_-}\ket{\lambda_+}$. From the definition \eqref{matel} we find:
\begin{equation}
\label{overla}
\bra{\nu_-}\ket{\lambda_+}= \int_{-\infty}^{+\infty} dx \abs{x}^{2ik-1} e^{-i\nu x } e^{i \frac{\lambda}{x}}=\int_0^\infty d x x ^{2ik-1}e^{-i\nu x}e^{i\frac{\lambda}{x}}+\int_0^\infty d x x ^{2ik-1}e^{i\nu x}e^{-i\frac{\lambda}{x}}. 
\end{equation}
To evaluate these integrals we use the integral representation of the modified Bessel function of the second kind:
\begin{equation}
\label{BInt}
\int_{0}^{+\infty}d x \,x^{2ik-1} e^{-\nu x}e^{-\frac{\lambda}{x}} =  \left(\frac{\lambda}{\nu}\right)^{ik} K_{2ik}\left(\sqrt{\nu\lambda}\right), \quad \nu,\lambda > 0
\end{equation}
This values of $\nu$ and $\lambda$ can be taken to the imaginary axis, analytically continuing from the positive real axis as in Figure \ref{BesselAnalytic}. Taking $\nu\to e^{i\pi/2}\nu$ and $\lambda\to e^{-i\pi/2}\lambda$ results in:
\begin{equation}
\int_{0}^{+\infty}d x \, x^{2ik-1} e^{-i\nu x}e^{i\frac{\lambda}{x}} = e^{\pi k}\left(\frac{\lambda}{\nu}\right)^{ik} K_{2ik}\left(\sqrt{\nu\lambda}\right), \quad \lambda,\nu > 0.
\end{equation}
Similarly by taking $\nu\to e^{-i\pi/2}\nu$ and $\lambda\to e^{i\pi/2}\lambda$ to rotate in the other direction, we find:
\begin{equation}
\int_{0}^{+\infty}d x \, x^{2ik-1} e^{i\nu x}e^{-i\frac{\lambda}{x}} = e^{-\pi k}\left(\frac{\lambda}{\nu}\right)^{ik} K_{2ik}\left(\sqrt{\nu\lambda}\right), \quad \lambda,\nu > 0.
\end{equation}
Combining both we obtain:\footnote{This formula appears in \cite{VK}. It should be replaced by a Bessel-J function in case $\lambda\nu<0$. We leave this implicit. }
\begin{equation}
\bra{\nu_-}\ket{\lambda_+}=\cosh\left(\pi k\right) \left(\frac{\lambda}{\nu}\right)^{ik} K_{2ik}\left(\sqrt{\nu\lambda}\right).\label{overlap}
\end{equation}
\\
More generically we are interested in computing $\bra{\nu_-}g\ket{\lambda_+}$. For this, we choose to parameterize the $\sltr$ group element by its Gauss decomposition:
\begin{eqnarray}
\label{Gaussfull}
g(\phi,\gamma_-,\gamma_+)=h_-(\gamma_-) \cdot d(\phi) \cdot h_+(\gamma_+) = \left(\begin{array}{cc}
1 & 0 \\
\gamma_- & 1 \\
\end{array}\right)\left(\begin{array}{cc}
e^{-\phi} & 0 \\
0 & e^{\phi} \\
\end{array}\right)\left(\begin{array}{cc}
1 & \gamma_+ \\
0 & 1 \\
\end{array}\right),
\end{eqnarray}
where
\begin{equation}
d(\phi)=\left(\begin{array}{cc}
e^{-\phi} & 0 \\
0 & e^{\phi} \\
\end{array}\right).
\end{equation}
This covers the Poincar\'e patch of $\sltr$ with metric:
\begin{equation}
ds^2 = \text{Tr}(g^{-1}dg)^2 = d\phi^2 + e^{-2\phi}d\gamma_-d\gamma_+.
\end{equation}
Since $\ket{\nu_\pm}$ diagonalizes $J_\pm$ we find:
\begin{equation}
\bra{\nu_-}g(\phi,\gamma_-,\gamma_+)\ket{\lambda_+}=e^{i\gamma_-\nu}e^{i\gamma_+\lambda}\bra{\nu_-}d(\phi)\ket{\lambda_+}=e^\phi \bra{\nu e^\phi _-}\ket{\lambda e^\phi_+}e^{i\gamma_-\nu}e^{i\gamma_+\lambda},
\end{equation}
where the second equality follows from a change of integration variables $x \to x e^{-\phi}$ in \eqref{matel}. Inserting \eqref{overlap} now directly results in the relevant matrix elements: 
\begin{align}
\label{matelexplicit}
\boxed{
R_{k,\nu\lambda}(g) = \left\langle \nu_-\right|g(\phi,\gamma_-,\gamma_+) \left|\lambda_+\right\rangle = \cosh\left(\pi k\right) \left(\frac{\lambda}{\nu}\right)^{ik} e^{\phi} K_{2ik}(\sqrt{\nu\lambda}e^{\phi})e^{i\nu\gamma_- + i \lambda \gamma_+}.}
\end{align}

\subsection{Plancherel measure}
Finally, we would like to read off the Plancherel measure using \eqref{plandef} and the orthogonality relation of the Bessel functions. To do this, we must make a detour on the coordinatization of the $\sltr$ manifold as the Gauss parameterization \eqref{Gaussfull} does not cover the entire $\sltr$ manifold but only the Poincar\'e patch. Any integral over the full group manifold (such as \eqref{plandef}) is a sum of four terms. The whole $\sltr$ group is covered by four patches of the form \cite{Forgacs:1989ac}:
\begin{equation}
g = h_- \cdot d \cdot h_+ \cdot \omega, \qquad \omega = \pm \left(\begin{array}{cc}
0 & 1 \\
-1 & 0 \\
\end{array}\right) = \pm \mathbf{s}\quad \text{or} \quad \omega = \pm \mathbf{1}.
\end{equation}
These patches give 2 by 2 the same result as an overall sign of $\omega$ gives the same matrix elements.\footnote{As we are in fact studying PSL$(2,\mathbb{R})$.} This means the group integral splits in two a priori distinct pieces: one over $g(\phi,\gamma_-,\gamma_+)$ and one over $g(\phi,\gamma_-,\gamma_+)\cdot \mathbf{s}$. We demonstrate in the next subsection that an elementary substitution is sufficient to show that the second term equals the first one, both for the grand orthogonality as for the $3j$-integral. We end up with:
\begin{equation}
\label{plmeasu}
\boxed{
\rho(k)=\frac{k\sinh 2\pi k}{\cosh^2 \pi k} = k\tanh \pi k},
\end{equation}
which is indeed known to be the Plancherel measure on $\sltr$. Of course, this result can be obtained in any basis of interest. Most discussions utilize the hyperbolic basis to deduce this (see e.g. \cite{Kitaev:2017hnr} for a recent account).

\subsection{Covering of the $\sltr$ manifold by Gauss Patches}
To go to the $\omega = \mathbf{s}$ patch, we set $e^{\phi} \to -e^{\phi}/\gamma_+$, $\gamma_+ \to - 1/\gamma_+$ and $\gamma_- \to \gamma_- + e^{2\phi}/\gamma_+$ in the matrix element \eqref{matelexplicit} \cite{Basu:1980cx}, and we obtain:
\begin{align}
R_{k,\nu\lambda}(g \cdot \mathbf{s}) = \cosh\left(\pi k\right) \left(\frac{\lambda}{\nu}\right)^{ik} \frac{e^{\phi}}{\left|\gamma_+\right|} K_{2ik}\left(\sqrt{\nu\lambda}\frac{e^{\phi}}{\left|\gamma_+\right|}\right)e^{i\nu \left(\gamma_- + \frac{e^{2\phi}}{\gamma_+}\right)- i \lambda \frac{1}{\gamma_+}},
\end{align}
from which we read off the contribution to orthonormality of the $\omega = \mathbf{s}$ patch to be:
\begin{align}
&\int dg R_{k,\nu\lambda}(g \cdot \mathbf{s}) R_{k,\nu'\lambda'}^{*}(g \cdot \mathbf{s}) \nonumber \\
&= \delta(\nu-\nu') \int_{0}^{+\infty} \frac{d\gamma_+}{\gamma_+^2} \int_{-\infty}^{+\infty} d\phi \cosh\left(\pi k_1\right) \cosh\left(\pi k_2 \right)\left(\frac{\lambda}{\nu}\right)^{ik_1} \left(\frac{\lambda'}{\nu'}\right)^{-ik_2} \nonumber \\
&\qquad \times K_{2ik_1}\left(\sqrt{\nu\lambda}\frac{e^{\phi}}{\left|\gamma_+\right|}\right) K_{2ik_2}\left(\sqrt{\nu'\lambda'}\frac{e^{\phi}}{\left|\gamma_+\right|}\right)e^{i (\lambda -\lambda') \frac{1}{\gamma_+}} \nonumber \\
&= \delta(\nu-\nu')\delta(\lambda-\lambda') \int_{-\infty}^{+\infty} d\phi \cosh\left(\pi k_1\right) \cosh\left(\pi k_2 \right)\left(\frac{\lambda}{\nu}\right)^{i(k_1-k_2)}K_{2ik_1}(\sqrt{\nu\lambda}e^{\phi}) K_{2ik_2}(\sqrt{\nu\lambda}e^{\phi}) \nonumber \\
&= \frac{\delta(\nu-\nu')\delta(\lambda-\lambda')\delta(k_1-k_2)}{\rho(k)}, \qquad \quad \rho(k) = k\tanh \pi k,
\end{align}
where in the first equality we did the $\gamma_-$-integral, in the second one we first shifted $\phi \to \phi + \ln\left|\gamma_+\right|$, used $\gamma_+ \to 1/\gamma_+$ and did the $\gamma_+$-integral, while in the third equality we did the final $\phi$-integral. \\
This is just the same answer as the Gauss $\omega = \mathbf{1}$ patch, and the only effect of considering all four patches is a quadrupling of the result, leading to the Plancherel measure \eqref{plmeasu}.

\section{Some Representation Theory of $\slr$}
\label{s:repsemi}
We present the representation theory of $\slr$. It is very closely related to that of $\sltr$ itself, and large parts of it can be found in the available literature \cite{vilenkin, VK, Dijkgraaf:1991ba}.
\\~\\
The semigroup $\slr$ is defined as the set of positive $\sltr$ matrices with the usual matrix operations:
\begin{equation}
\left(\begin{array}{cc}
a & b \\
c & d \\
\end{array}\right), \qquad ad-bc = 1, \quad a,b,c,d > 0.
\end{equation}
In spite of the lack of an inverse, hence the name \emph{semi}group, it is possible to set up a meaningful representation theory in the sense that
\begin{equation}
R(g_1\cdot g_2) = R(g_1)\cdot R(g_2).
\end{equation}
It has an action on $L^2(\mathbb{R}^+)$ in the same way as \eqref{repgroup}, but restricted to $x>0$:
\begin{equation}
f_j(x) \to (g \cdot f_j)(x) = \left|bx+d\right|^{2j}f_j\left(\frac{ax+c}{bx+d}\right),
\end{equation}
Due to the positivity of all matrix entries, this operation is internal in $\mathbb{R}^+$ and is well-defined. 
The $\mathfrak{sl}(2,\mathbb{R})$ algebra is still relevant.
\\~\\
A matrix element in a representation $j$ is defined as the overlap:
\begin{equation}
R_{ab}(g) \equiv \left\langle j a\right| g \left|j b\right\rangle = \int_{0}^{+\infty} dx f_{ja}^*(x) (g \cdot f_{jb}(x)) .\label{rab}
\end{equation}

\subsection{Matrix Elements}
The matrix elements of the subsemigroup $\slr$ can be found as a subset of the $\sltr$ matrix elements when diagonalizing the $J^0$ generator. This basis is called the \emph{hyperbolic} basis. To describe it, it's convenient to briefly return to $\sltr$. The eigenfunctions ($x>0$)
\begin{align}
\left\langle x\right|\left.s\right\rangle = \frac{1}{\sqrt{2\pi}}x^{is-1/2}, \qquad \left\langle s\right|\left.x\right\rangle = \frac{1}{\sqrt{2\pi}}x^{-is-1/2},\label{hypbas}
\end{align}
are a basis on $\mathbb{R}^+$:
\begin{align}
\label{basis}
\int_{0}^{+\infty} \frac{dx}{x}\, x^{is}x^{-is'} = 2\pi \delta(s-s'), \qquad \int_{-\infty}^{+\infty} ds \, x^{is-1/2} x'^{-is-1/2} = 2\pi \delta(x-x'),
\end{align}
with parameter $s$ related to the $J^0$-eigenvalue by \eqref{BW}. An analogous basis is constructed on $\mathbb{R}^-$. 
Defining the four matrix elements
\begin{equation}
K^{\pm \pm}_{s_1 s_2}(g) \equiv \left\langle s_1,\pm\right| g \left|s_2,\pm\right\rangle,\label{kdef}
\end{equation}
linking basis functions in the $x<0$ $(-)$ or $x>0$ $(+)$ sector with one another, we can write the matrix elements of $\sltr$ on $L^2(\mathbb{R})$ in the hyperbolic basis as a $2\times 2$ matrix of matrix elements:
\begin{equation}
\mathbf{K}(g) = \left(\begin{array}{cc}
K^{++} & K^{+-} \\
K^{-+} & K^{--} \\
\end{array}\right).
\end{equation}
This matrix composes under group transformations using matrix multiplication: $\mathbf{K}(g_1 \cdot g_2) = \mathbf{K}(g_1) \cdot \mathbf{K}(g_2)$. Specifying now to $\slr$, the matrix elements in the hyperbolic basis of $\slr$ are just $K^{++}$.\footnote{Also their $q$-deformed variants are known, which reduce to these in the classical limit again \cite{Ip}.} Indeed, for the subsemigroup elements $g_1$, $g_2$ the composition law of $\slr$ irrep matrices implies the composition law of $\slr$ irrep matrices: $K^{++}(g_1 \cdot g_2) = K^{++}(g_1) \cdot K^{++}(g_2)$. 
This matrix element can be explicitly computed by evaluating the defining integral:
\begin{equation}
K^{++}_{s_1 s_2}(g) = \left\langle s_1\right| g \left|s_2\right\rangle = \int_{0}^{+\infty} dx x^{-is_1-1/2}(g \cdot x^{is_2-1/2}).
\end{equation}
The Gauss decomposition of a generic $\slr$ matrix is given by:
\begin{eqnarray}
\label{Gauss}
g=e^{i\gamma_- J^-}e^{2i\phi J^0}e^{i \gamma_+ J^+} = \left(\begin{array}{cc}
1 & 0 \\
\gamma_- & 1 \\
\end{array}\right)\left(\begin{array}{cc}
e^{-\phi} & 0 \\
0 & e^{\phi} \\
\end{array}\right)\left(\begin{array}{cc}
1 & \gamma_+ \\
0 & 1 \\
\end{array}\right), \quad \gamma_-,\gamma_+ >0,
\end{eqnarray}
and provides a complete covering of the $\slr$ manifold. It has corresponding metric
\begin{equation}
ds^2 = \frac{1}{2} \text{Tr}\left[(g^{-1}dg)^2\right] = d\phi^2 + e^{-2\phi}d\gamma_-d\gamma_+ , \quad \gamma_-,\gamma_+ >0.
\end{equation}
Note here that for $g\in\slr$, though no inverse exists in the semigroup, $g^{-1}$ is well-defined because $\slr$ is a subregion of $\slr$. This is crucial, as otherwise the construction of a $\slr$ BF theory would not be valid. 
\\
For each of the three constituents of \eqref{Gauss}, one obtains the matrix elements $(j=-\frac{1}{2}+ik)$:
\begin{align}
\label{matgenexpl}
K^{++}_{s_1 s_2}(\phi) &= e^{2i(k-s_2)\phi}\delta(s_1-s_2), \nonumber \\
K^{++}_{s_1 s_2}(\gamma_-) &= \frac{1}{2\pi}\frac{\Gamma(-is_1+1/2)\Gamma(is_1-is_2)}{\Gamma(-is_2+1/2)}\gamma_-^{is_2-is_1}, \\
K^{++}_{s_1 s_2}(\gamma_+) &= \frac{1}{2\pi}\frac{\Gamma(is_2-is_1)\Gamma(is_1+1/2-2ik)}{\Gamma(is_2+1/2-2ik)}\gamma_+^{is_1-is_2}. \nonumber
\end{align}
The generic matrix element can then be readily computed as
\begin{equation}
\label{matgen}
K^{++}_{s_1 s_2}(g) = \int_{-\infty}^{+\infty} d\alpha d\beta \, K^{++}_{s_1 \alpha}(\gamma_-) K^{++}_{\alpha \beta}(\phi) K^{++}_{\beta s_2}(\gamma_+).
\end{equation} 
The orthogonal wavefunctions are then obtained as
\begin{equation}
\psi^k_{s_1 s_2}(g)=\sqrt{k\sinh 2\pi k}\, K^{++}_{s_1 s_2}(g),
\end{equation}
and the Plancherel measure is deduced as
\begin{equation}
\label{plmeasup}
\boxed{\rho(k) = k \sinh 2\pi k.}
\end{equation}

\subsection{Unitarity of the Matrix Elements}
\label{app:uni}
We can use the explicit expressions \eqref{matgenexpl} and \eqref{matgen} to prove that the continuous representation $K_{++}(g)$ is unitary. We compute:
\begin{align}
\int &ds \, K^{++}_{s_1 s}(g) \, K^{++}_{s_2 s}(g)^* \nonumber \\
&= \int_{-\infty}^{+\infty} ds d\alpha d\beta K^{++}_{s_1 \alpha}(\gamma_-)K^{++}_ {\alpha\alpha}(\phi)K^{++}_{\alpha s}(\gamma_+) K^{++}_{s_2 \beta}(\gamma_-)^*K^{++}_{\beta \beta}(\phi)^*K^{++}_{\beta s}(\gamma_+) ^* \nonumber \\
&= \int_{-\infty}^{+\infty} ds\, d\alpha\, d\beta\, \int_{0}^{+\infty} dx\, du\, dy\, dv \, x^{-is_1-1/2}(x+\gamma_-)^{i\alpha-1/2}y^{is_2-1/2}(y+\gamma_-)^{-i\beta-1/2} \nonumber \\
&\times u^{-i\alpha-1/2}(u+\gamma_+)^{is-1/2}v^{i\beta-1/2}(v+\gamma_+)^{-is-1/2}e^{2i(k-\alpha)\phi} e^{-2i(k-\beta)\phi}, 
\end{align}
Using successively
\begin{alignat}{2}
\int_{-\infty}^{+\infty}ds (u+\gamma_+)^{is-1/2}(v+\gamma_+)^{-is-1/2} &= \delta(u-v), \qquad  \int_{0}^{+\infty}\frac{du}{u} u^{i(\beta-\alpha)} &&= \delta(\alpha-\beta), \nonumber \\
\int_{-\infty}^{+\infty}d\alpha (x+\gamma_-)^{i\alpha-1/2}(y+\gamma_-)^{-i\alpha-1/2} &= \delta(x-y), \qquad \int_{0}^{+\infty} \frac{dx}{x} x^{i(s_2-s_1)} &&= \delta(s_1-s_2),
\end{alignat}
we find
\begin{align}
\int_{-\infty}^{+\infty} ds \, K^{++}_{s_1 s}(g) \, K^{++}_{s_2 s}(g)^*= \delta(s_1-s_2),
\end{align}
identifying the inverse representation matrix with the adjoint:
\begin{equation}
K^{++}_{\alpha,\beta}(g)^{-1}= K^{++}_{\beta,\alpha}(g)^*.
\end{equation}

\subsection{Gravitational Matrix Elements}
Gravitational matrix elements are associated with the parabolic states $\ket{\mathfrak{i}_\pm}$ defined as before to satisfy $J_\pm\ket{\mathfrak{i}_\pm} = \pm i\ket{\mathfrak{i}_\pm}$. In the coset slicing we are interested in obtaining $\bra{s}g(\phi,\gamma_-)\ket{\mathfrak{i}_+}$. In the Schwarzian slicing we are interested in obtaining $\bra{\mathfrak{i}_-}g(\phi)\ket{\mathfrak{i}_+}$. In the coordinate basis we obtain a damped exponential:
\begin{equation}
\label{gravbas}
\bra{\mathfrak{i}_-}\ket{x}=e^{-x}, \qquad \bra{x}\ket{\mathfrak{i}_+}=x^{2ik-1}e^{-\frac{1}{x}}.
\end{equation}
Notice that neither the damped $e^{-\nu x}$ nor the oscillating exponentials $e^{i\mu x}$ are orthogonal on $\mathbb{R}^+$. This is because $J_+$ as defined in \eqref{BW} is not self-adjoint on $\mathbb{R}^+$, so its eigenfunctions are not necessarily orthogonal and the vectors $\ket{\mu_+}$ do not form a basis, in sharp contrast with the situation in $\sltr$ above. $J^0$ on the other hand \emph{is} self-adjoint on $\mathbb{R}^+$, and leads to the hyperbolic basis \eqref{hypbas} we constructed above. 
\\~\\
These states can be decomposed in the hyperbolic basis using the Cahen-Mellin integral:
\begin{equation}
e^{-y} = \frac{1}{2\pi i}\int_{c-i\infty}^{c+i\infty}ds \, \Gamma(s) y^{-s}, \qquad y>0.\label{mellin}
\end{equation}
Using \eqref{hypbas} and \eqref{gravbas}, one finds for the overlap:
\begin{equation}
\bra{\mathfrak{i}_-}\ket{s} = \frac{1}{\sqrt{2\pi}}\Gamma\Big(is+\frac{1}{2}\Big), \qquad \bra{s}\ket{\mathfrak{i}_+} = \frac{1}{\sqrt{2\pi}}\Gamma\Big(is+\frac{1}{2}-2ik\Big).
\end{equation}
This transition is the same as that linking Minkowski eigenmodes to Rindler modes.\footnote{The Mellin transform and its inverse are explicitly:
\begin{align}
f(x) &=\int_{-\infty}^{+\infty}ds F(s) x^{-is-1/2}, \\
F(s) &= \int_{0}^{+\infty}dx f(x) x^{is-1/2},
\end{align}
which follows from \eqref{basis}. The Mellin transform links a function on $\mathbb{R}^+$ to a function on $\mathbb{R}$.
}
The matrix element of the middle Cartan element $e^{2i\phi J^0}$ is called the Whittaker function (or coefficient). The elementary basis functions $f_\nu $ and $f_\lambda$ are called the Whittaker vectors in the mathematics literature \cite{Jacquet,Schiffmann,Hashizume1,Hashizume2}. 
\\
The matrix element between these states is easily found as
\begin{equation}
R_{k}(g) = \left\langle \mathfrak{i}_-\right|g(\phi) \left|\mathfrak{i}_+\right\rangle =\frac{1}{2} \int_0^\infty \frac{dx}{x} x^{ik} e^{\phi} e^{-2\phi i k}e^{- x}e^{-\frac{e^{2\phi}}{x}}= e^{\phi} K_{2ik}\left(e^{\phi}\right).\label{nulambda}
\end{equation}
These are orthogonal with the Plancherel measure $k\sinh 2\pi k$ as expected.
An addition theorem can be found by inserting a complete set of intermediate states in the \emph{hyperbolic} basis:
\begin{equation}
\label{addtheo}
\left\langle \mathfrak{i}_-\right| g_1\cdot g_2 \left|\mathfrak{i}_+\right\rangle = \int_{-\infty}^{+\infty} ds \left\langle \mathfrak{i}_-\right| g_1 \left|s\right\rangle\left\langle s\right| g_2 \left|\mathfrak{i}_+\right\rangle.
\end{equation}
All intermediate channel are always treated in the hyperbolic basis. 

\section{Schwarzian Bilocals from $\sltr$ BF}
\label{app:kaplan}
This Appendix concerns the holographic evaluation of a bulk crossing Wilson line $\mo^\ell(\tau_1,\tau_2)$. In particular this is a Wilson line in the lowest weight state of a discrete $j=\ell$ representation of $\slr$.\footnote{This representation is identical to a discrete representation of $\sltr$.} The goal is to prove formula \eqref{kaplan} of the main text.
\\~\\
In BF theory \cite{paper2,paper3}, a Wilson line evaluates upon path integrating out $\chi$ to: 
\begin{equation}
\mathcal{P}e^{\int_{z_i}^{z_f} A(z) dz} = g(z_f) g^{-1}(z_i).
\end{equation}
Specifying to a boundary-anchored Wilson line, we set $z_i = t_i$ and $z_f = t_f$. In gravity, the boundary is subject to the gravitational constraints \eqref{gravcon}, such that $g(t_f)$ is an implicit function of $f(t_f)$. We aim to make this explicit. In the defining $2\times2$ representation, this is solving the matrix equation:
\begin{equation}
iJ^- - \frac{T(t)}{2}  iJ^+ = \left(\begin{array}{cc}
0 & 0 \\
1 & 0 \\
\end{array}\right) 
 - \frac{T(t)}{2} \left(\begin{array}{cc}
0 & 1 \\
0 & 0 \\
\end{array}\right)
 = g \partial_t g^{-1}, \quad g^{-1} =\left(\begin{array}{cc}
A & B \\
C & D \\
\end{array}\right),
\end{equation}
with solution determined by
\begin{align}
A'' + \frac{1}{2}T(t) A =0, &\quad B =A', \\
C'' + \frac{1}{2} T(t) C = 0, &\quad D = C',
\end{align}
and the $\sltr$ constraint $AC'-A'C = 1$. The Fuchsian differential equation is Hill's equation, familiar from the study of Virasoro coadjoint orbits \cite{Balog:1997zz}. $A$ and $C$ are the two linearly independent solutions to Hill's equation, with the $\sltr$ constraint playing the role of the Wronskian condition. Up to permutations (obtained by performing a M\"obius transformation), there is a unique solution to this system:
\begin{equation}
A=\frac{1}{\sqrt{f'}},\qquad C=\frac{f}{\sqrt{f'}},
\end{equation}
with  $f$ the solution to $\left\{f,t\right\} = T(t)$. Parametrizing $g^{-1}$ using the Gauss parameterization
\begin{eqnarray}
g^{-1}=e^{i\gamma_L J^-}e^{2i\phi J^0}e^{i \gamma_R J^+}  = \left(\begin{array}{cc}
1 & 0 \\
\gamma_- & 1 \\
\end{array}\right)\left(\begin{array}{cc}
e^{-\phi} & 0 \\
0 & e^{\phi} \\
\end{array}\right)\left(\begin{array}{cc}
1 & \gamma_+ \\
0 & 1 \\
\end{array}\right),
\end{eqnarray}
we identify:
\begin{align}
\gamma_- = f, \qquad e^{-\phi}=\frac{1}{\sqrt{f'}}, \qquad \gamma_+ = -\frac{1}{2}\frac{f''}{f'}.\label{solution}
\end{align}
We are interested in evaluating the Wilson line in the lowest weight representation of a discrete irrep $\ell$ of $\slr$. This is easy in the Borel-Weil realization 
\begin{align}
\label{BWb}
i\hat{J}^- &= \partial_x, \nonumber \\
i\hat{J}^0 &= -x\partial_x - j, \\
i\hat{J}^+ &= -x^2\partial_x - 2j x, \nonumber
\end{align}
where we know the lowest weight state $\ket{\ell,0}$ to be of the form \cite{Fitzpatrick:2016mtp}:
\begin{equation}
\bra{x}\ket{\ell,0} = \frac{1}{x^{2\ell}}, \quad \bra{\ell,0}\ket{x} = \delta(x),
\end{equation}
and the generators \eqref{BWb} exponentiate to one-parameter subgroups of $\sltr$, acting as:
\begin{alignat}{3}
J^-&: f(x) && \quad \rightarrow \quad f(x+c), \qquad &&c\in\mathbb{R},\\
J^0&: f(x) &&\quad \rightarrow \quad e^{-2\ell\phi}f(e^{-2\phi} x), \qquad &&\phi\in\mathbb{R},\\
J^+&: f(x) &&\quad \rightarrow \quad (bx+1)^{-2\ell}f\left(\frac{x}{(bx+1}\right) ,  \qquad &&b\in\mathbb{R},
\end{alignat}
corresponding to translation, scaling and special conformal transformations respectively. \\
Hence the Wilson line can be written as:
\begin{equation}
\mo^\ell(\tau_1,\tau_2)=\bra{\ell,0}g(t_f) g^{-1}(t_i)\ket{\ell,0}=\int d x\, \delta(x) \left( g(t_f) g^{-1}(t_i)\cdot \frac{1}{x^{2\ell}} \right),
\end{equation}
with the differential operator
\begin{equation}
\label{sol}
g^{-1} = e^{\gamma_-  \partial_x}e^{2\phi (-x\partial_x + \ell)}e^{\gamma_+ (-x^2\partial_x - 2\ell x)},
\end{equation}
with parameters \eqref{solution}. Explicitly, the wavefunction transforms under the action of $g(z_f)g^{-1}(z_i)$ as:
\begin{align}
\frac{1}{x^{2\ell}} &\qquad \underset{g^{-1}(t_i)}{\longrightarrow} \qquad  \frac{{f'_1(t_i)}^\ell}{(x+f_1(t_i))^{2\ell}} \qquad \underset{g(t_f)}{\longrightarrow} \qquad \frac{({f'_1(t_i)}{f'_2(t_f)})^\ell}{f'_2(t_f) x+(f_1(t_i)-f_2(t_f))^{2\ell}(1+\gamma_+x))^{2\ell}},
\end{align}
with $f_1$ and $f_2$ possibly different functions associated with the respective holographic boundaries on which $t_i$ and $t_f$ are located.
Setting $x=0$, and using the notation of \eqref{kaplan} we obtain:
\begin{equation}
\left(\frac{\dot{F^{\LL}}(t_1) \dot{F^{\RR}}(t_2)}{(F^{\LL}(t_1)-F^{\RR}(t_2))^2}\right)^{\ell}.
\end{equation}


\end{document}